\documentclass[longauth]{aa}  
\usepackage{graphicx}
\usepackage{txfonts}
\usepackage{cancel}
\usepackage{newtxtext,newtxmath}
\usepackage[T1]{fontenc}
\usepackage{ae,aecompl}
\usepackage{amsmath}	
\usepackage{amssymb}	
\usepackage{hyperref}
\usepackage[normalem]{ulem}
\usepackage{placeins} 
\hypersetup{
    colorlinks=true,
    citecolor=blue,
    linkcolor=blue,
    urlcolor=blue,
}
\usepackage{comment}

\begin{document} 

   \title{The PLATO Input Catalog of targets (tPIC) \\
   for the first Long-duration Observation Phase field}

\author{
M.~Montalto\inst{1},
G.~Piotto\inst{2,3}\fnmsep\thanks{Corresponding author: \email{giampaolo.piotto@unipd.it}}, 
P.~M.~Marrese\inst{4,5},
L.~Prisinzano\inst{6},
S.~Marinoni\inst{4,5},
V.~Granata\inst{7,3},
J.~Cabrera\inst{8},
V.~Nascimbeni\inst{3},
S.~Desidera\inst{3},
V.~Adibekyan\inst{9},
S.~Ortolani\inst{2},
E.~Alei\inst{10},
C.~Aerts\inst{11},
G.~Altavilla\inst{4,5},
K.~Belkacem \inst{12},
S.~Benatti\inst{6},
A.~B\"orner\inst{8},
M.~Deleuil\inst{13},
M.~Fabrizio\inst{4,5},
L.~Gizon\inst{14},
M.~J.~Goupil\inst{15},
M.~G\"unther \inst{16},
A.~M.~Heras\inst{16},
D.~Magrin\inst{3},
L.~Malavolta\inst{2,3},
J.~M.~Mas-Hesse\inst{17},
H.~P.~Osborn\inst{18},
I.~Pagano\inst{1},
C.~Paproth\inst{8},
D.~Pollacco\inst{19},
R.~Ragazzoni\inst{3,2},
G.~Ramsay\inst{20},
H.~Rauer\inst{8},
S.~Udry\inst{21}
}

\institute{
Istituto Nazionale di Astrofisica -- Osservatorio Astrofisico di Catania, Via S. Sofia 78, 95123, Catania, Italy
\and
Dipartimento di Fisica e Astronomia ``Galileo Galilei'', Universit\`a degli Studi di Padova, Vicolo dell'Osservatorio 3, 35122 Padova, Italy
\and
Istituto Nazionale di Astrofisica -- Osservatorio Astronomico di Padova, Vicolo dell'Osservatorio 5, 35122 Padova, Italy
\and
Istituto Nazionale di Astrofisica -- Osservatorio Astronomico di Roma, Via Frascati 33, 00078 Monte Porzio Catone, Roma, Italy
\and
Space Science Data Center - ASI, Via del Politecnico snc, 00133 Roma, Italy
\and
Istituto Nazionale di Astrofisica -- Osservatorio Astronomico di Palermo, Piazza del Parlamento 1, 90134 Palermo, Italy
\and
Centro di Ateneo di Studi e Attività Spaziali ``Giuseppe Colombo'' (CISAS), Universit\`a degli Studi di Padova, Via Venezia 1, 35131 Padova, Italy
\and
Deutsches  Zentrum  f\"ur  Luft-  und  Raumfahrt  (DLR),  Institut  f\"ur  Optische  Sensorsysteme,  Rutherfordstra{\ss}e  2,  12489  Berlin-Adlershof, Germany
\and
Instituto de Astrof\'isica e Ci\^encias do Espa\c{c}o, Universidade do Porto, CAUP, Rua das Estrelas, 4150-762 Porto, Portugal  \& Departamento de F\'{\i}sica e Astronomia, Faculdade de Ci\^encias, Universidade do Porto, Rua do Campo  Alegre, 4169-007 Porto, Portugal
\and
NASA Goddard Space Flight Center, 8800 Goddard Rd, Greenbelt, 20771, MD, USA (NPP Fellow)
\and
Institute of Astronomy, KU Leuven, Celestijnenlaan 200D, 3001, Leuven, Belgium
\and
LIRA, Observatoire de Paris, Universit\'e PSL, CNRS, Sorbonne Universit\'e, Universit\'e Paris Cit\'e, CY Cergy Paris Universit\'e, 92190 Meudon, France
\and
Aix-Marseille Universit\'e, CNRS, CNES, Laboratoire d’Astrophysique de Marseille, Technop\^{o}le de Marseille-Etoile, 38, rue Fr\'ed\'eric Joliot-Curie, 13388 Marseille cedex 13, France
\and
Max-Planck-Institut f\"ur Sonnensystemforschung, Justus-von-Liebig-Weg~3, 37077~G\"ottingen, Germany 
\and
LESIA, CNRS UMR 8109, Universit\'e Pierre et Marie Curie, Universit\'e Denis Diderot, Observatoire de Paris, 92195 Meudon, France
\and
European Space Agency (ESA), European Space Research and Technology Centre (ESTEC), Keplerlaan 1, 2201 AZ Noordwijk, The Netherlands
\and
Centro de Astrobiolog\'{\i}a (CSIC--INTA), 28692 Villanueva de la Ca\~nada, Madrid, Spain
\and
 Center for Space and Habitability, University of Bern, Gesellschaftsstrasse 6, 3012, Bern, Switzerland
\and
Department of Physics, University of Warwick, Gibbet Hill Road, Coventry CV4 7AL,   
United Kingdom
\and
Armagh Observatory \& Planetarium, College Hill, Armagh, BT61 9DG, United Kingdom
\and 
Observatoire de Gen\`eve, Universit\'e de Gen\`eve, Chemin Pegasi 51, 1290 Sauverny, Switzerland 
}

\date{Received; accepted}
 
  \abstract
   {   
The ESA PLAnetary Transits and Oscillations of stars (PLATO) mission is designed to detect terrestrial planets in the habitable zones of solar-type stars. Owing to telemetry constraints, the selection of PLATO targets must be performed in advance.
   }
   {
We present the first public release of the PLATO Input Catalog of targets (tPIC2.2), which provides the list of stars that will be observed during PLATO’s first Long-duration Observation Phase field at South (LOPS2) as part of its core science program.    
   }
   {
   We used astrometric and photometric data from Gaia data release 3 (DR3), together with three-dimensional maps of the local interstellar medium, to identify stars belonging to the PLATO stellar samples in accordance with the mission requirements.
   }
   {
The tPIC comprises 217\,741 stars, including 202\,315 FGK dwarfs and subgiants, 15\,037 M dwarfs, and 789 known planet-host stars. The median distances of the samples are 512\,pc for FGK stars and 133 \,pc for M dwarfs. We estimated interstellar reddening for almost all targets and developed an algorithm to infer fundamental stellar parameters (effective temperature, radius, and mass) homogeneously from astrometric and photometric observables.
}
   {
   The tPIC fulfills all the science requirements of the PLATO mission. It also includes a list of stars that host known exoplanets, confirmed or still candidate,  located within the LOPS2 field.}

             
   \titlerunning{The PLATO Input Catalog for its first Long Pointing field}
   \authorrunning{Montalto et al.} 
   \maketitle

\section{Introduction}

The PLATO mission \citep{rauer2025} is the third medium-class mission of the European Space Agency (ESA) Cosmic Vision program. Its primary objectives are the detection and characterization of terrestrial exoplanets with periods extending  to the habitable zone of bright solar-type stars, together with a detailed investigation of their host stars. The PLATO mission is optimized to detect terrestrial planets on long orbital periods, including those located within the habitable zones of their parent star, with the ultimate goal of identifying Earth analogs under potentially favorable conditions for the emergence of life.

To achieve these goals, PLATO will search for small decreases in stellar brightness caused by planetary transits across the disks of their host star. The mission is designed to continuously observe two Long-duration Observation Phase (LOP) fields for at least two years each, while retaining the capability to observe additional fields for shorter intervals of up to three months during the so-called ``step-and-stare'' observation phase (SOP).

The PLATO instrument is equipped with an array of 26 cameras, where the term ``camera'' denotes the complete optical and detection subsystem, including the telescope optics, focal plane assembly, and all associated ancillary components (e.g., baffle and electronics). Each camera is based on refractive optical elements in the 20 cm class, corresponding to an equivalent final aperture of 12 cm, for a total circular field of view of about 1\,037 square degrees \cite{ragazzoni2016}.

The 24 cameras (all observing in white light) are arranged in four groups of six cameras each, co-aligned with four slightly overlapping fields of view. Therefore, the signal-to-noise ratio (S/N) of each star depends on its precise location within the overall PLATO field of view.
The combined field of view of the camera system amounts to $\sim 2\,132$ square degrees \citep[][]{ragazzoni2016,magrin2018}.
The remaining two cameras (one observing in blue and one in red) are optimized for fast monitoring of very bright stars, their color measurements \citep[e.g.,][]{grenfell2020}, and fine guidance and navigation. Their focal plane is equipped with frame-transfer detectors for a field of view somewhat larger than 600 square degrees. Under nominal conditions, the geometry of the patches covered by the entire set of cameras on the sky remains symmetric for 90-degree rotations around the line of sight. This configuration maintains the same field-of-view coverage throughout a pointing and allows continuous in-flight calibration among the different cameras.

Compliance with the performance specifications is assessed using models that account for performance degradation over the mission lifetime. This degradation is expected to be moderate and is primarily driven by a reduction in optical throughput. The impact is mitigated by using radiation-hardened glasses for the optical elements most exposed to cosmic radiation \citep{magrin2016,corso2018}, as well as by adopting a conservative end-of-life (EOL) scenario in which up to two camera detectors or associated electrical subsystems are assumed to fail.

\noindent
Due to limitations in the spacecraft-to-ground telemetry rate, the downlink of full-frame charge-coupled device (CCD) images at a cadence suitable for planet detection is not feasible. Therefore, targets must be preselected. For a subset of targets, customized pixel sub-arrays (``imagettes'') are downlinked to the ground segment. For the remaining selected targets, centroid and photometric extraction are performed on board, and only the derived photometric time series are transmitted to the ground.

This operational concept has been successfully implemented in previous space missions. Target selection catalogs were developed for CoRoT (Exo-Dat; \citealt{deleuil2009}),  {\it Kepler} (KIC; \citealt{brown2011}), K2 (EPIC; \citealt{huber2016}) and TESS (TIC; \citealt{stassun2018,stassun2019}) to identify and prioritize targets for exoplanet detection. 

We present the PLATO Input Catalog of targets (tPIC2.2, hereafter tPIC), which is provided through the PLATO archive
\footnote{\url{https://pax.esac.esa.int/plato}})
and corresponds to the first long-operation phase field in the south (LOPS2). This catalog constitutes the reference database of stellar targets for the core science of the PLATO mission.

The target-selection strategy adopted in this work follows the methodology originally described by \citet{montalto2021} for the Gaia DR2-based all-sky catalog (asPIC 1.1.0). The asPIC represents the first large-scale implementation of the PLATO target selection strategy and provides an all-sky catalog of candidate PLATO targets based on Gaia DR2 astrometry and photometry, including stellar classification, de-reddening, and identification of the main PLATO target samples. The present work extends and refines that approach using data from the Gaia 3 data release (DR3) and updated calibrations.

The first official PIC version, named PIC2.1.0.1, based on Gaia DR3 data and restricted to the LOPS2 footprint \citep{nascimbeni2025} was released to the PLATO Consortium on February 21, 2025. \citet{prisinzano2025} presents a dedicated analysis of the selection and characterization of M dwarfs (sample P4) within PIC2.1.0.1.
The complete PLATO Input Catalog (PIC2.2; hereafter PIC) also includes additional stellar samples, namely the calibration sample \citep[cPIC,][]{heller2026}, the fine guidance sample \citep[fgPIC,][]{heller2026}, and the science calibration and validation sample \citep[scvPIC,][]{zwintz2026}.  
The description of the assembly of the single sub-catalogs and the content of the complete PIC  are outside the scope of this paper and will be described in Marrese et al. (in prep.).

Establishing the main parameters of PLATO's host stars is crucial for determining the properties of transiting planets. The uncertainty in the transiting planet's radius is directly related to the uncertainty in its host-star radius.  The properties of planet hosts are also essential for modeling stars using asteroseismology.
In addition, a planet's habitability depends, among other factors, on the properties of its host star. Moreover, knowledge of the values of stellar parameters permits the statistical study of the frequency of planets as a function of stellar type, metallicity, environment, and other factors. \citep[e.g.,][]{howard2012,bryson2020}.
   
\noindent
The construction of the tPIC requires the identification (within the selected observing field) of stars meeting the required spectral-type and luminosity-class criteria. In addition, photometric noise constraints must be taken into account to assess the capability to detect planetary transits  around the selected targets. Consequently, magnitude limits for the stellar samples are defined in accordance with the mission performance requirements, and the level of flux contamination from neighboring sources is evaluated for each target.
The PLATO pixel scale is 15 arcsec pix$^{-1}$ on axis, intermediate between that of  {\it Kepler} (4 arcsec pix$^{-1}$; \citealt{borucki2010}) and TESS (21 arcsec pix$^{-1}$; \citealt{ricker2014}).
The point spread function (PSF), while showing moderate variations across the field of view and depending on the thermal state of the cameras, is expected to extend over an area of a few pixels \citep[see also][]{gullieuszik2016,umbriaco2018}. As a consequence, particularly in crowded fields, multiple sources can be combined within a single photometric aperture, potentially leading to false-positive transit-like signals \citep[e.g.,][]{santerne2015,fressin2013}.

The present work is based on Gaia DR3\citep{gaiadr3}, which enables accurate distance determinations and, consequently, reliable estimates of absolute stellar luminosities for the majority of PLATO target candidates. Future Gaia data releases will be used to prepare the PIC for possible new LOP fields (other than LOPS2)  that PLATO might observe.

\noindent
Following the adoption of the mission by ESA in June 2017, 
PLATO entered the development phase with the objective of being ready for launch in early 2027. At this stage of mission development, the selection of the first LOPS2 has been finalized \citep{nascimbeni2025}.

The present paper focuses on the LOPS2 field. The corresponding catalog was made publicly available to the scientific community by the time of the first ESA Guest Observer (GO) call\footnote{\url{https://www.cosmos.esa.int/web/plato/ao-1}} (April 7–May 21, 2026). The catalog will be updated regularly to include new confirmed or candidate planet hosts as they become available.
Section~\ref{sec:stellar_samples} presents a brief overview of the mission requirements defining the PLATO stellar samples and the target selection criteria.
Section~\ref{sec:stellar_photom} describes the corrections applied to Gaia DR3 photometry, followed by the computation of the visible magnitude and the PLATO magnitude, and a detailed computation of the noise-to-signal ratio and the scientific ranking of the tPIC targets.
We describe the treatment of interstellar extinction adopted in this analysis, the determination of stellar parameters, and the distribution in radius, mass, and temperature of the tPIC sources in Sect.~\ref{sec:stellar_parameters}.  Section~\ref{sec:confirmed_candidate_planet_hosts} describes a dedicated list of confirmed and candidate exoplanet host stars included in the tPIC.  Sect.~\ref{sec:comparisons} provides a comparison with previously published results. Section~\ref{sec:summary} summarizes the main results of this work, including stellar counts for the different tPIC samples. Section~\ref{sec:access} provides information on access to PIC2.2 and accompanying documents describing the data model.  In Appendix~\ref{sec:StellarLibraries}, we present the libraries of synthetic stellar spectra used to derive the calibration relations presented in this paper. Appendix~\ref{sec:PLATOresponseFunction} presents the PLATO response functions and Appendix~\ref{sec:PLATOmagnitude} presents a detailed description and the equations used to calculate the PLATO magnitude. Section~\ref{sec:flags} describes the flags introduced in the tPIC to associate the single targets with the appropriate subsample, and to describe their spectral class. 

\begin{table*}
\caption{Summary of scientific requirements for PLATO stellar samples (cumulative over all LOPs fields).
\label{tab:samplesrequirements}}
\centering
\begin{tabular}{lccccc}
 \hline
 \hline
  & P1 & P2 & P4 & P5 & color sample \\  
\hline 
Stars & $\ge$15 000 (goal 20 000) & $\ge$1000 & $\ge$5000 & $\ge$245 000 & 300 \\
Spectral Type & Dwarf and  &  Dwarf and &  M Dwarfs &  Dwarf and  & Anywhere in \\
 & subgiants F5-K7 & subgiants F5-K7 &  & subgiants F5-late K & the HR diagram \\
Limit {\it V} (mag) & 11 & 8.5 & 16 & 13 &  \ldots \\
Random noise (ppm in 1 h) & $<$50 & $<$50 &   \ldots & \ldots & \ldots \\
Wavelength (nm)& 500--1000 & 500--1000 & 500--1000 & 500--1000 & Blue (500--675)\\
 & & & & & and\\
 & & & & & red (675--1000)\\
 & & & & & spectral bands\\
\hline 
\end{tabular}
\end{table*}

\section{ tPIC stellar samples }
\label{sec:stellar_samples}

\subsection{Sample definition}
\label{sec:samples_definition}

Targets for the PLATO mission include four main stellar samples, named P1, P2, P4, and P5. For historical reasons, sample P3 was eliminated, but the numbering of the PLATO samples was left unchanged). These samples will be observed in the wavelength range between 500 nm and 1000 nm.  
An additional sample,  located in the central part of LOPS2 and called the color sample, will be observed in two broad blue and red spectral bands spanning the 500--675 and 675--1000 nm wavelength ranges, respectively \citep{corso2018}. 

The science requirements for the PLATO stellar samples, cumulative over all LOPs fields, according to the ESA Science Requirements Document \citep[SciRD;][]{rauer2025}, are listed below and summarized in Table~\ref{tab:samplesrequirements}.
Defining the PLATO samples requires knowledge of the visual apparent magnitude, $V$, because the SciRD adopts this magnitude as a reference to identify targets that are sufficiently bright for ground-based spectroscopic follow-up. 
We selected the targets using $V$ magnitudes obtained from our own calibrated transformation of Gaia DR3 \citep{gaiadr3}
magnitudes and colors into the Johnson $V$ band (see also Sect.~\ref{sec:vphot}). This choice ensures maximum homogeneity of $V$ band photometry.
We describe the calibration procedure in Appendix~\ref{sec:JohnsonVSyntheticMagnitudes}.\\

\noindent {Stellar sample 1 (P1):}

\begin{itemize}
    \item The total number of targets in P1 (cumulative over all LOPs fields) shall be at least 15 000 dwarf and subgiant stars of spectral types from F5 to K7, with a goal of 20\,000.
    \item The dynamic range of P1 shall be {\it V} $\le$ 11~mag.
     \item The random noise level for P1 shall be below 50 ppm in 1 h.
    \item In P1, the proportion of brighter targets ({\it V} $\le$ 10.5~mag) shall be maximized.
    \item P1 shall be observed during two LOPs.
\end{itemize}

\noindent Stellar sample 2 (P2):
\begin{itemize}
\item The total number of targets (cumulative over all LOP fields) shall be at least 1\,000 dwarf and subgiant stars of spectral types from F5 to K7.
\item The dynamic range shall be $V \le 8.5$~mag.
\item The random noise level shall be below 50 ppm in 1 h.
\item P2 shall be observed during two LOPs.
\end{itemize}

\noindent Stellar sample 4 (P4):
\begin{itemize}
\item The total number of targets (cumulative over all sky fields) shall be at least 5\,000 cool late-type dwarfs (M dwarfs) monitored during a LOP.
\item The dynamic range shall be $V \le 16$~mag.
\end{itemize}

\noindent {Stellar sample 5 (P5):}
\begin{itemize}
\item The total number of targets (cumulative over all LOP
fields) shall be at least 245\,000 dwarf and subgiant stars of spectral types
from F5 to late K.
\item The dynamic range  shall be $V \le 13$~mag.
\item P5 shall be observed during two LOPs.
\end{itemize}

\noindent Color sample\\

The color sample is the subset of stars in the PIC selected by the PLATO Mission Consortium for observation with the fast cameras. The payload design requirement was to provide color information for at least 150 bright stars in the sky field per pointing. The instrument can observe 325  targets per pointing, of which 25 are reserved for the Guest Observer program, in addition to the stars observed with the fast cameras for fine guidance \citep{Griessbach2021}. The stars in the color sample will be observed at a cadence of 2.5 s with imagettes, and the photometry will be extracted on the ground. Broadband color information is useful for planet validation \citep{leger2009}, planetary atmosphere studies \citep{grenfell2020,Carone2026}, and complementary science. In particular, color information may help break degeneracies in asteroseismology models of pulsators with a convective core that experience heat-driven coherent oscillation modes \citep[e.g.,][Chapter 6]{Aerts2010}. Targets in the color sample will be selected according to the core-mission science program from the prime sample stars \citep{nascimbeni2026} and from P2 stars of interest for asteroseismology of solar-like oscillations, rotation analysis, and activity monitoring (flares; disentangling spots versus faculae).

\subsection{Selection criteria for P1, P2, P4, and P5}
\label{sec:selection_criteria}

The PLATO target samples P1, P2, and P5
include
dwarf and subgiant stars with spectral type between F5 and
late K,
while P4 comprises M dwarfs.
According to the definitions reported in Table~5 of \citet[]{pecaut2013}, spectral type F5 corresponds to $T_\textrm{eff} = 6\,550\,$ K, while spectral type M0 corresponds to $T_\textrm{eff} = 3\,850\,$K. Dwarfs and subgiants have $\log g > 3.5$.

\subsubsection{Selection from Gaia DR3}

To translate these physical definitions into observational selection criteria, we first identified the regions occupied by these stellar populations in the Gaia DR3 color–magnitude diagram (CMD). Although Gaia provides the photometric quantities used for the final target selection, the locations of stars with given effective temperature and surface gravity cannot be determined directly from the observed CMD. We therefore used Galactic population simulations generated with TRILEGALv1.6 (TRIdimensional modeL of thE GALaxy; \citealt{girardi2005}), which provide intrinsic stellar parameters that allowed us to map the PLATO target populations in the Gaia CMD. 

We ran ten simulations, each covering an area of ten square degrees. We placed the centers of the simulated fields at the centers of the LOPN1 and LOPS2 fields \citep{nascimbeni2022,nascimbeni2025} and at four surrounding locations separated by 10 degrees from the centers of these fields in Galactic longitude or latitude.

We kept all TRILEGAL parameters at their default values, including a 30\% binary fraction.  We determined the intrinsic visual magnitude, $V_0$, using our adopted calibration relation (Sect.~\ref{sec:vphot}). We calculated reddening using our adopted extinction map  and converted it to the extinction coefficient in the $V$ band ($A_V$)  as described in Sect.~\ref{sec:monochromatic_extinction}. We computed the apparent magnitude, $V$, by  adding the derived $A_V$ to the intrinsic magnitude.
 We only considered stars with $V \le 13$~mag for FGK dwarfs and subgiants, and $V \le 16$~mag for M dwarfs and we dereddened the CMD accordingly to the procedure described in Section~\ref{sec:stellar_parameters}.
\noindent
We then defined the boundaries for the FGK and M samples in the CMD. We defined blue and bright selection boundaries for FGK dwarfs and subgiants and a separation boundary between FGK and M dwarfs. 
We assumed that all boundaries could be described by linear models.

\begin{figure*}[!t]
	\centering
	\includegraphics[width=\columnwidth]{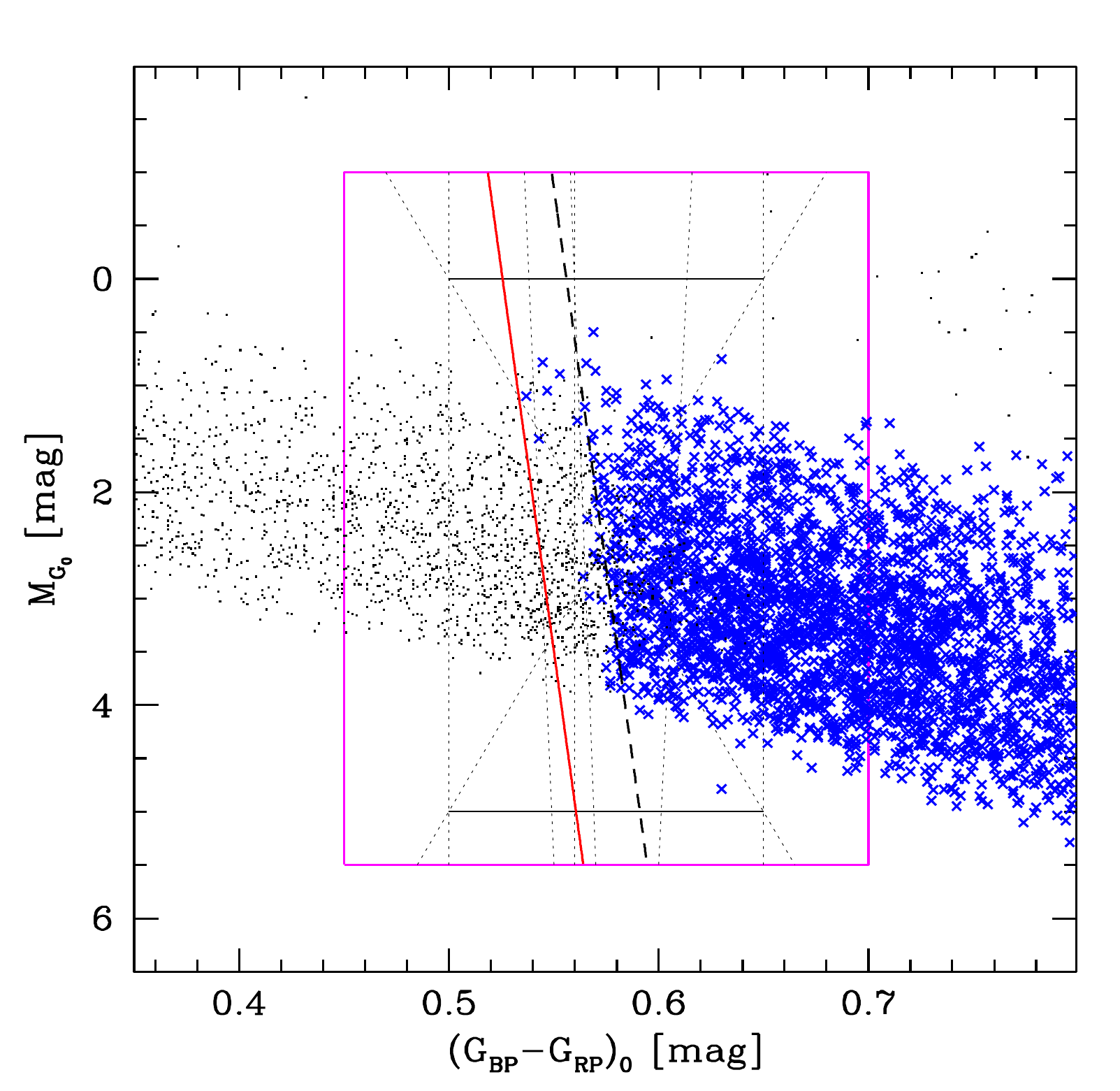}
    \includegraphics[width=\columnwidth]{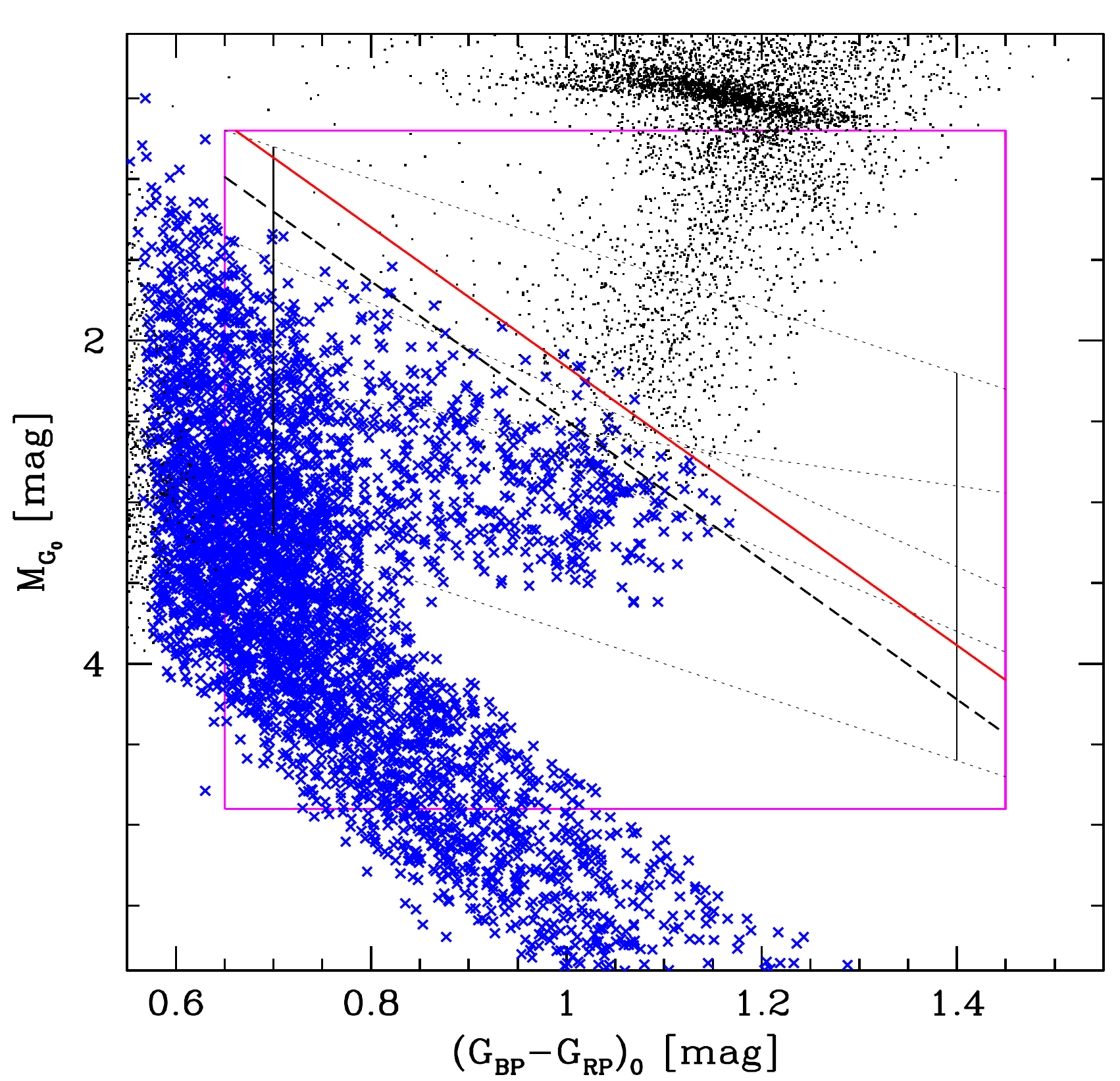}
    \includegraphics[width=\columnwidth]{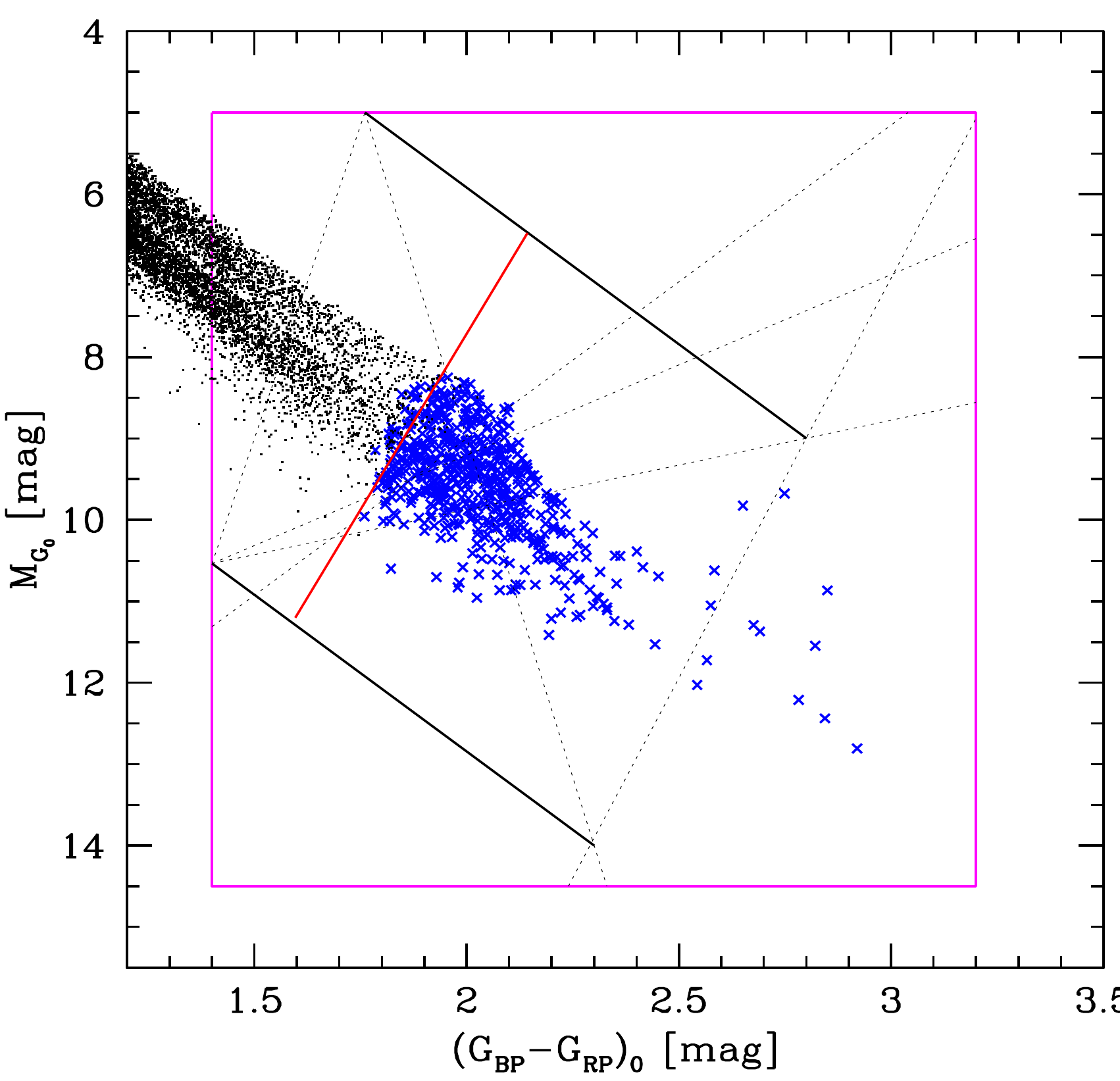}
    \includegraphics[width=\columnwidth]{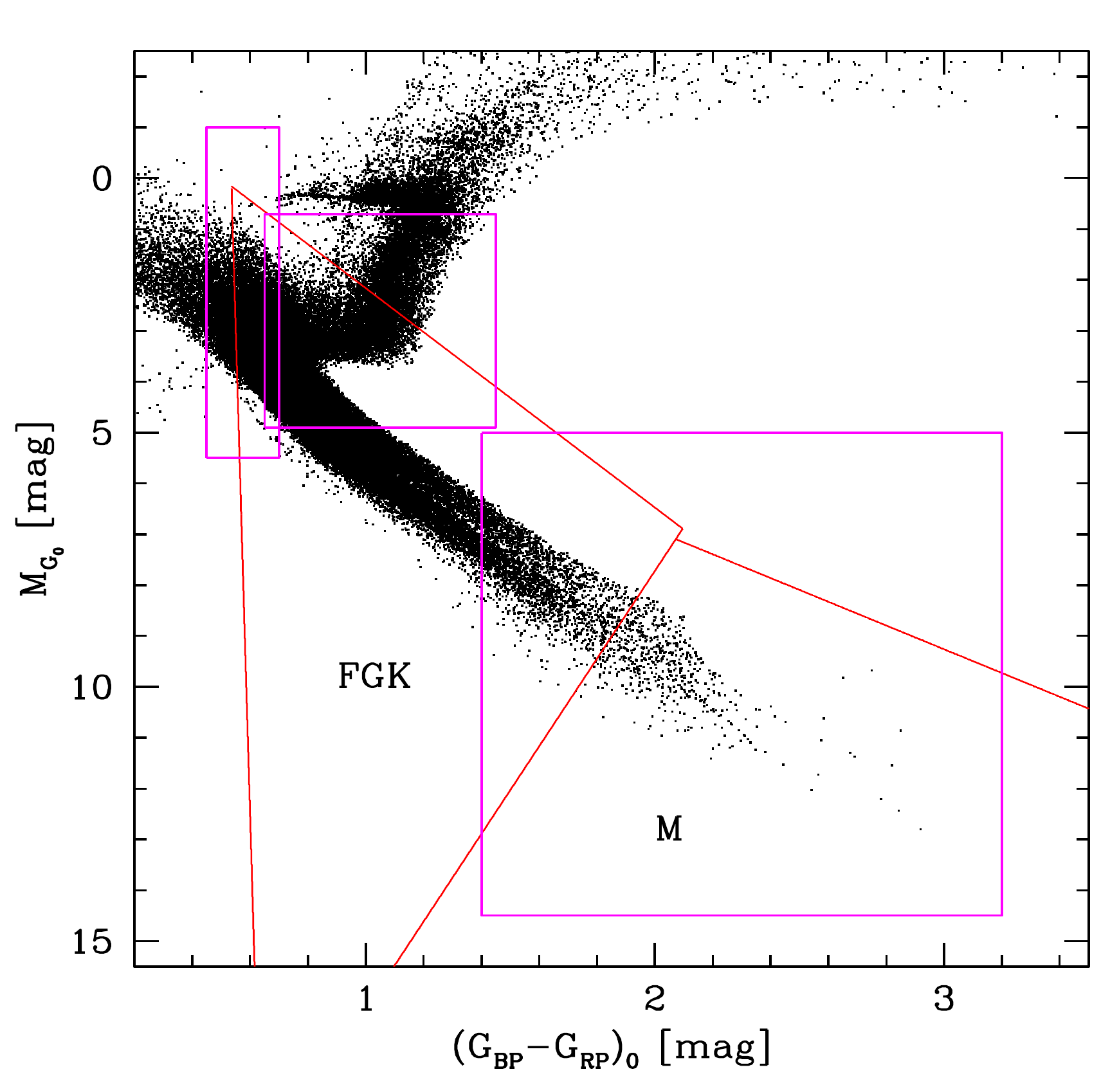}
	\caption{
    {Top left:} Selection region used to define the blue
    separation boundary (magenta rectangle) for FGK dwarfs and subgiants in the P1, P2, and P5 samples. Constraints for the linear separation boundaries (solid black lines), samples of separation boundaries (dotted lines), the best theoretical separation boundary (dashed line), and the final adopted separation boundary (solid red line). Blue crosses represent FGK dwarfs and subgiants, while black dots represent contaminants; both have $V \le 13$~mag (see Sect. \ref{sec:selection_criteria}). The final adopted separation boundary differs from the best theoretical boundary because it is a more conservative boundary and accounts for observational errors.{\it Top right:}
    Selection region used to define the bright
    separation boundary (magenta rectangle) for FGK dwarfs and subgiants in the P1, P2, and P5 samples. Symbols and separation lines are as in the top-left panel.    
    {\it Bottom left: } Selection region used to determine the blue
    separation boundary (magenta rectangle) for M dwarfs in P4.  Blue crosses represent M dwarfs with $V \le 16$~mag, while black dots represent contaminants with $V \le 16$~mag  (see Sect. \ref{sec:selection_criteria}).  The best theoretical  and final adopted separation boundaries are both represented by the solid red line in this case. Other sample lines are as in the top-left panel.
    {\it Bottom-right: }
    Final adopted selection regions for FGK dwarfs and subgiants and for M dwarfs, as indicated by the labels. Black points denote simulated stars with $V\le16$~mag. Magenta rectangles display the selection regions shown in the other panels.
    }
	\label{fig:boundary_FGK_blue}
\end{figure*}

To define the blue selection boundary of FGK stars, we considered only the targets and contaminants located within the magenta rectangle shown in Fig.~\ref{fig:boundary_FGK_blue}. We defined targets from the TRILEGAL simulations as stars with $3850,{\rm K}<T_{\rm eff}<6550,{\rm K}$, $\log g > 3.5$, and $V\le13$ mag, while contaminants comprised all remaining simulated stars with $V\le13$ mag. 
We searched for the line that best separates targets from contaminants. We constructed candidate separation boundaries as straight lines intersecting the two horizontal solid black lines in Fig.~\ref{fig:boundary_FGK_blue}, corresponding to $M_{G,0}=0$ mag and $M_{G,0}=5$ mag. We allowed the intersection points to move along these horizontal lines within the interval $0.5 \le (G_{\rm BP}-G_{\rm RP})_0 \le 0.65$ mag, with a step of 0.001 mag, generating 22,500 candidate boundaries. We show examples with dotted lines in Fig.~\ref{fig:boundary_FGK_blue}.
For each candidate boundary, we computed the metric
$S=N_{\rm targ}-N_{\rm cont}$,
where N$_{\rm targ}$ and N$_{\rm cont}$ are the numbers of targets and contaminants, respectively, located on the red side of the boundary within the magenta rectangle. Maximizing $S$ simultaneously maximizes the number of selected targets and minimizes the number of contaminants. We adopted the boundary that yields the highest value of $S$ as the best theoretical separator.

We accounted for observational uncertainties by fixing the slope of this boundary and perturbed the intrinsic colors and magnitudes of the simulated stars according to the expected errors in reddening, distance modulus, and Gaia photometry. We generated 10,000 realizations and determined the intercept of the optimal boundary in each realization. From the resulting distribution, we computed the mean intercept and its dispersion, $\sigma_b$. We defined the final adopted boundary using an intercept equal to the mean value plus $5\sigma_b$, resulting in the relation
M$_{\rm G,0}$=192.308($\,$G$_{\rm BP}$-G$_{\rm RP})_0$-102.88$\,{\rm mag}.$
We adopted this more conservative boundary, shown as the solid red line in Fig.~\ref{fig:boundary_FGK_blue}, to account for observational uncertainties and to avoid rejecting potentially valid targets.

\noindent

We set the bright limit of the FGK dwarfs and subgiants selection region for P1, P2, and P5 following a similar approach to that illustrated in Fig.~\ref{fig:boundary_FGK_blue}.
The final separation boundary is given by the equation $M\rm_{G,0} =4.314 \times (G\rm_{BP}-G\rm_{RP})_0-2.153$~mag. We selected  the P4 sample following the procedure detailed in \citep{prisinzano2025}.

\citet{prisinzano2025} describes the P4 sample included in version 2.1 of the PLATO Input Catalog. We used the same approach as for P4 in tPIC, with differences arising only from the calculation of the noise-to-signal ratio (NSR) based on updated satellite performance. 
In the following, we summarize the main steps and boundary definitions.
We set the blue limit of the M dwarf selection region following a similar approach, as illustrated in the bottom-left panel of Fig.~\ref{fig:boundary_FGK_blue}. In this case, we did not perturb the best theoretical boundary because it separates FGK from M dwarfs, and a larger tolerance for the M dwarf region would therefore be unfavorable for the FGK region. The adopted final separation boundary corresponds to the best separation boundary and is given by the equation $M\rm_{G,0}=-8.620 \times (G\rm_{BP} -G \rm_{RP} )_0 +24.960$~mag.

Finally, we set the bright limit of the M dwarf selection region to that adopted in asPIC1.0 \citep{montalto2021}, which corresponds to the best regression line of a 10 Myr solar-metallicity isochrone from the Padova database \citep{bressan2012}. The separation boundary is given by \citep{prisinzano2025}: 
$M \rm_{G,0} =2.334 \times (G\rm_{BP} -G \rm_{RP} )_0 +2.259$~mag.

We defined the final separation limits for the FGK and M dwarf samples (Fig~\ref{fig:boundary_FGK_blue}) from the combined set of boundaries described above. Together with the criteria illustrated in Sect.~\ref{sec:stellar_samples}, these limits define the selection conditions for the different stellar samples:\\

\begin{equation*}
    \textrm{P1 sample}=\begin{cases}
    M_{G,0} \leq 192.308\,(G_{\rm BP}-G_{\rm RP})_0-102.880 \\
    M_{G,0} \geq 4.314\,(G_{\rm BP}-G_{\rm RP})_0-2.153\\
     M_{G,0} \leq -8.62\,(G_{\rm BP}-G_{\rm RP})_0+24.96\\
    V\leq11\textrm{ mag}\\
    \textrm{NSR}_\textrm{sys} < 50\,\textrm{ppm}\,\textrm{h}^{-1}
    \end{cases}
\end{equation*}

\begin{equation*}
    \textrm{P2 sample}=\begin{cases}
    M_{G,0} \leq 192.308\,(G_{\rm BP}-G_{\rm RP})_0-102.880 \\
    M_{G,0} \geq 4.314\,(G_{\rm BP}-G_{\rm RP})_0-2.153\\
     M_{G,0} \leq -8.62\,(G_{\rm BP}-G_{\rm RP})_0+24.96\\
    V\leq8.5\textrm{ mag}\\
    \textrm{NSR}_\textrm{sys} < 50\,\textrm{ppm}\,\textrm{h}^{-1}
    \end{cases}
\end{equation*}

\begin{equation*}
    \textrm{P4 sample}=\begin{cases}
    M_{G,0} \geq 2.334\,(G_{\rm BP}-G_{\rm RP})_0+2.259\\
     M_{G,0} > -8.62\,(G_{\rm BP}-G_{\rm RP})_0+24.96\\
    V\leq16\textrm{ mag}\\
    \end{cases}
\end{equation*}

\begin{equation*}
    \textrm{P5 sample}=\begin{cases}
    M_{G,0} \leq 192.308\,(G_{\rm BP}-G_{\rm RP})_0-102.880 \\
    M_{G,0} \geq 4.314\,(G_{\rm BP}-G_{\rm RP})_0-2.153\\
    M_{G,0} \leq -8.62\,(G_{\rm BP}-G_{\rm RP})_0+24.96\\
    [(V < 11 \, \textrm{ mag} \,\, AND \,\, \textrm{NSR}_\textrm{sys} \geq 50 \,\textrm{ppm}\,\textrm{h}^{-1})\\ 
    \textrm{OR}\\  
    11 \textrm{mag} <  \, V\leq 13  \textrm{ mag}]\\
    \end{cases}.
\end{equation*}

\noindent
The quantities in the above relations are the intrinsic color, $(G_{\rm RP}-G_{\rm BP})_0$ in the Gaia bands, the Gaia intrinsic absolute magnitude ($M_{{\rm G},0}$), the apparent visual magnitude ($V$), and the NSR$_{\textrm{sys}}$, which  accounts for random and systematic errors.

As described in Sect.~\ref{sec:noise_to_signal_ratios}, the PLATO Performance Team (PPT) provided the NSR of each star in tPIC.
To calculate the NSR, the PPT uses the PIC list of photometric contaminants which includes all Gaia sources in the LOPS2 field and is complemented at the bright end by \textit{Hipparcos2} and Tycho2TDSCmerge.  
The \texttt{PICcontaminant} table is the starting point for selecting the P1, P2, P4, and P5 targets. It contains all Gaia DR3 objects in the LOPS2 field and is complemented at the bright end using \textit{Hipparcos2} \citep{leeuwen2007} and Tycho-2 \citep{hog2000}. The caseFlag is obtained for all contaminants using Gaia, Hipparcos/Tycho-2, or Johnson magnitudes.
Stars with caseFlag$ = 1, 8, $ or $14$ (F, G, and K dwarfs or subgiants, respectively) are selected as a P1, P2,  or P5 candidate.
If it has caseFlag$= 2$ (M dwarf), it is selected as a P4 candidate.
There is a single object in the tPIC that is not in Gaia: it is a Hipparcos \citep{1997ESASP} star with a planet (not P1, P2, P4, or P5).
Further information on PIC contaminants is provided in Marrese et al. (in prep.).

The tPIC includes stars that may fall within CCD gaps or slightly outside the nominal PLATO field of view, as indicated by a null value of the appropriate number-of-cameras-coverage flags). We included these stars to account for small uncertainties in the pointing and field rotation. For these stars, we used an extrapolated value of the NSR, assuming that each star was observed by six cameras. 

\subsubsection{Selection  based on Hipparcos photometry}

When {\it Gaia} photometry was not available, we used Hipparcos photometry and defined the following inequalities following a similar procedure to that described in Sect.~\ref{sec:selection_criteria}:

\begin{equation*}
    \textrm{P1 sample}=\begin{cases}
    M_{Hp0} \geq -98.21429\,(B_{\rm T}-V_{\rm T})_0+44.95477 \\
    M_{Hp,0} \geq 3.42857\,(B_{\rm T}-V_{\rm T})_0-0.62125\\
     (B_{\rm T}-V_{\rm T})_0 \leq1.49 \textrm{ mag}\\
    V\leq11\textrm{ mag}\\
    \textrm{NSR}_\textrm{sys} < 50\,\textrm{ppm}\,\textrm{h}^{-1}
    \end{cases}
\end{equation*}

\begin{equation*}
    \textrm{P2 sample}=\begin{cases}
 M_{Hp0} \geq -98.21429\,(B_{\rm T}-V_{\rm T})_0+44.95477 \\
    M_{Hp,0} \geq 3.42857\,(B_{\rm T}-V_{\rm T})_0-0.62125\\
     (B_{\rm T}-V_{\rm T})_0 \leq1.49 \textrm{ mag}\\   
    V\leq8.5\textrm{ mag}\\
    \textrm{NSR}_\textrm{sys} < 50\,\textrm{ppm}\,\textrm{h}^{-1}
    \end{cases}
\end{equation*}

\begin{equation*}
    \textrm{P5 sample}=\begin{cases}
M_{Hp0} \geq -98.21429\,(B_{\rm T}-V_{\rm T})_0+44.95477 \\
    M_{Hp,0} \geq 3.42857\,(B_{\rm T}-V_{\rm T})_0-0.62125\\
     (B_{\rm T}-V_{\rm T})_0 \leq1.49 \textrm{ mag}\\
    [(V < 11 \, \textrm{ mag} \,\, AND \,\, \textrm{NSR}_\textrm{sys} \geq 50 \,\textrm{ppm}\,\textrm{h}^{-1})\\ 
    \textrm{OR}\\  
    11 < \textrm{mag} \, V\leq 13  \textrm{ mag} ]\\
    \end{cases}.
\end{equation*}

\subsubsection{Selection based on Johnson photometry}

When {\it Gaia} and Hipparcos photometries were not available, we used Johnson photometry with the following inequalities:

\begin{equation*}
    \textrm{P1 sample}=\begin{cases}
    M_{V,0} \geq -117.02128\,(B-V)_0+52.45523 \\
    M_{V,0} \geq 3.2\,(B-V)_0-0.080025 \\
    (B-V)_0\leq1.29 \textrm{ mag}\\
    V\leq11\textrm{ mag}\\
    \textrm{NSR}_\textrm{sys} < 50\,\textrm{ppm}\,\textrm{h}^{-1}
    \end{cases}
\end{equation*}

\begin{equation*}
    \textrm{P2 sample}=\begin{cases}
    M_{V,0} \geq -117.02128\,(B-V)_0+52.45523 \\
    M_{V,0} \geq 3.2\,(B-V)_0-0.080025 \\
    (B-V)_0\leq1.29 \textrm{ mag}\\
    V\leq8.5\textrm{ mag}\\
    \textrm{NSR}_\textrm{sys} < 50\,\textrm{ppm}\,\textrm{h}^{-1}
    \end{cases}
\end{equation*}

\begin{equation*}
    \textrm{P5 sample}=\begin{cases}
    M_{V,0} \geq -117.02128\,(B-V)_0+52.45523 \\
    M_{V,0} \geq 3.2\,(B-V)_0-0.080025 \\
    (B-V)_0\leq1.29 \textrm{ mag}\\
    [(V < 11 \, \textrm{ mag} \,\, AND \,\, \textrm{NSR}_\textrm{sys} \geq 50 \,\textrm{ppm}\,\textrm{h}^{-1})\\ 
    \textrm{OR}\\  
    11 \textrm{mag} < \, V\leq 13  \textrm{ mag]}\\
    \end{cases}.
\end{equation*}

We did not select the P4 sample using Hipparcos or Johnson photometry because the corresponding  photometric bands are too blue to properly characterize M dwarfs.
In the tPIC 2.2 version, we did not use the Johnson photometry as input because all tPIC targets were either Gaia or Hipparcos/Tycho objects; for completeness, we include the approach here.\\

\section{Stellar photometry}
\label{sec:stellar_photom}

\subsection{Corrections applied to Gaia DR3 photometry}
\label{sec:correction_gaia_dr3}

Gaia DR3 photometry is affected by systematic effects documented in the literature \citep{riello2021}. In particular, the correction reported in Sec. 8.3 of \citet{riello2021} for systematics caused by the use of the default color in the image parameter determination was already applied in Gaia DR3 and was therefore not considered here. However, as reported in the corrigendum to\citet{brown2021}, the correction was erroneously applied to 20 million sources, mostly faint, in Gaia DR3. We ignored this issue in our analysis because these objects are extremely faint sources in Gaia DR3 and therefore did not have an impact on our final selection.  We applied the saturation correction described in Appendix C.1 of \citet{riello2021}. We also applied the correction described in Sect. 8.4 of \citet{riello2021}, which applies to \textit{G}-band photometry of blue and bright sources ($G_{\rm BP}-G_{\rm RP}<-0.1$ and $8 < G < 13$~mag). 

\subsection{Visible magnitude}
\label{sec:vphot}

We calculated the visible magnitude from Gaia photometry or, when Gaia photometry was not available, from Hipparcos photometry.  We used several synthetic stellar libraries to calibrate the $(G-V)_0$ versus $(G_{\rm BP}-G_{\rm RP})_0$ relation, as detailed in Appendix~\ref{sec:StellarLibraries}.
We adopted a calibration relation in the form of a polynomial:

\begin{equation}
\label{eq:pv0_bv0}
    (G-V)_0=\sum_{i=1}^{i=6} b_i(G_{\rm BP}-G_{\rm RP})_0^i,
\end{equation}

\noindent
where the best-fit coefficients defined in Eq.~\ref{eq:pv0_bv0} are given in Table~\ref{tab:best_fit_coefficients_dwarfs_gv_bprp}. 
 The first column gives the stellar type (dwarfs or giants) to which each relationship should be applied. The last column gives the color range over which each relation is valid.
Figure\,\ref{fig:GV_BPRP_GIANTS} illustrates the calibration relation for giant stars \citep[see][for the corresponding figure for dwarf stars]{prisinzano2025}.

\noindent
When using Hipparcos photometry, we instead adopted the following relation\footnote{ \url{http://cdsarc.cds.unistra.fr/ftp/cats/I/239/version_cd/docs/vol1/sect1_03.pdf}} from the Hipparcos manual:

\begin{equation}
V_0=(V_T)_0-0.09\,(B_T-V_T)_0,
\end{equation}

\noindent
valid for $-0.2<(B_T-V_T)_0<1.8$~mag.


When using Gaia photometry, we obtained the uncertainty in the intrinsic visible photometry ($\delta V_0$) by error propagation using the following equation:

\begin{equation}
    \delta V_0=\delta G_0+\Big|\sum_{i=1}^{i=6} i \times (-b_i) \times (G_{\rm BP}-G_{\rm RP})_0^{i-1} \times \delta (G_{\rm BP}-G_{\rm RP})_0\Big|,
\end{equation}

\noindent where

\begin{equation}
    \delta G_0=\sqrt{(\delta G)^2+(\delta A_G)^2},
\end{equation}

\noindent and

\begin{equation}
    \delta (G_{\rm BP}-G_{\rm RP})_0=\sqrt{(\delta G_{\rm BP})^2+(\delta G_{\rm RP})^2+(\delta E(G_{\rm BP}-G_{\rm RP}))^2}.
\end{equation}

\noindent Here, $\delta G$, $\delta G_{\rm BP}$, and $\delta G_{\rm RP}$ are the uncertainties in the Gaia DR3
photometry and $\delta A_G$, and $\delta E(G_{\rm BP}-G_{\rm RP})$ are the uncertainties  in the $G$-band extinction and in the $(G_{\rm BP}-G_{\rm RP})$ reddening, respectively.

\noindent
We obtained the uncertainty in the apparent visual magnitude ($\delta V$) from
\begin{equation}
    \delta V=\sqrt{(\delta V_0)^2+(\delta A_V)^2},
    \label{eq:apparent_V}
\end{equation}
where $\delta A_V$ is the uncertainty in the visible band extinction.
When using Hipparcos photometry, we obtained the uncertainty in the intrinsic visible photometry ($\delta V_0$) by error propagation using the following equation:
\begin{equation}
    \delta V_0=\delta (V_T)_0+0.09\,\delta (B_T-V_T)_0.
\end{equation}

\noindent

\subsection{PLATO magnitude}
\label{sec:plato_magnitude}

We derived the PLATO magnitude for the tPIC targets using the following procedure. First, we derived the PLATO response function, as detailed in Appendix~\ref{sec:PLATOresponseFunction}. Next, for the different libraries of stellar spectra described in Appendix~\ref{sec:StellarLibraries}, we computed synthetic magnitudes in the PLATO passbands and in the Gaia DR3 $G$, $G_{\rm BP}$, $G_{\rm RP}$, Hipparcos, Tycho-2, and Johnson bands, as detailed in Appendix~\ref{sec:PLATOmagnitude}. 

Finally, we interpolated the following polynomial to 
the $(P-G)_0$ versus $(G_{\rm BP}-G_{\rm RP})_0$ colors:

\begin{equation}
\label{eq:pg0_bprp0}
    (P-G)_0=\sum_{i=1}^{i=6} b_i[(G_{\rm BP}-G_{\rm RP})_0]^i,
\end{equation}

\noindent
both for the normal cameras (N-CAMs) and the fast cameras (F-CAM$\rm_b$ and F-CAM$\rm_r$), and for both dwarf and giant stars.
 Table~\ref{tab:best_fit_coefficients_dwarfs_ncams} lists the best-fit coefficients defined in Eq.~\ref{eq:pg0_bprp0}. Similar relations were also derived for Hipparcos and Johnson photometry (Table~ \ref{tab:best_fit_coefficients_hipparcos} and Table~\ref{tab:best_fit_johnson}).
 The first two columns of Table~\ref{tab:best_fit_coefficients_dwarfs_ncams}-Table~\ref{tab:best_fit_johnson} 
 indicate the camera type (N-CAMs, F-CAMb, or F-CAMr) and the stellar type (dwarfs or giants) to which each relation applies.
 The last column indicates the color range over which the relation is valid.
\\

\section{Extinction and stellar parameters}\label{sec:stellar_parameters} 
The determination of extinction and stellar temperature is an iterative process, described below.

\subsection{Determination of monochromatic extinction}\label{sec:monochromatic_extinction}

We calculated interstellar extinction following an approach similar to that adopted for asPIC1.0 \citep{montalto2021}.
We used the 3D interstellar extinction map of \citet{lallement2022}, which provides the extinction density at 550 nm ($A_{550}$) within a volume of $6\times 6\times0.8$ kpc volume around the Sun, sampled every 10 pc along each dimension. We calculated the extinction uncertainties using the map of \citet{vergely2022}, which covers a larger volume  ($10\times10\times0.8$ kpc) but is sampled every  20 pc.
To obtain the extinction $A_{550}$ and its uncertainty, we numerically integrated the extinction maps from the Sun's position to the target  position. We performed the integration on a 5 pc discretized grid for the \citet{lallement2018} map and on a 10 pc discretized grid for the \citet{lallement2022} error map.

\subsection{Conversion between monochromatic extinction and extinction in other photometric bands}
\label{sec:conversion_monochromatic_extinction}

We converted the monochromatic extinction, $A_{550}$,  into extinction in other photometric bands, in particular the Gaia bands ($A\rm_G$, $A\rm_{BP}$, $A\rm_{RP}$), the visible band ($A\rm_V$), the 2MASS \citep{skrutskie2006} $Ks$ band ($A\rm_{Ks}$), and the PLATO bands ($A\rm_P$, $A\rm_{FCAMb}$, $A\rm_{FCAMr}$). This conversion depends on both the temperature of the source and the $A_{550}$. To calculate the conversion factors, we considered a set of 31 solar-metallicity dwarf ($\log g = 4.5$) stellar spectra from the MARCS library (Model Atmospheres with a Radiative and Convective Scheme; \citealt{gustafsson2008}), with temperatures between
2\,500 K and 8\,000 K. For each model,
we considered a grid of 2\,000 monochromatic extinction values between $A_{550}=0.01$ mag and $A_{550}=20$ mag in steps of 0.01 mag. For each value of the monochromatic extinction, we reddened the stellar spectrum using the extinction law of \citet{fitzpatrick2019}, integrated the flux over each photometric band, and calculated the conversion factor $A_{X}/A_{550}$, where $X$ corresponds to one of the photometric bands specified above. We interpolated the resulting tables in temperature using a cubic spline and linearly interpolated in extinction to derive the conversion factor
$A_{X}/A_{550}$ for any arbitrary value of the monochromatic extinction between $0<A_{550}\le20$~mag and temperature between 
$2\,500\, \rm{K} \le T \rm_{eff}\le 8\,000$ K.

\subsubsection{Gaia photometry}

We simultaneously determined the effective temperature and extinction in all  photometric bands considered using an iterative procedure. This procedure used a calibration relation between the intrinsic color and the effective temperature. 
To construct this relation, we considered a set of FGK dwarfs from the sample of \citet[]{casagrande2010} and M dwarfs from the sample of \citet[]{mann2015}. We matched the FGK dwarfs with Gaia DR3 and considered only stars within 70 pc that satisfied our
quality flags (see Sect.~\ref{sec:quality_flag}), resulting in a sample of 147 stars. Similarly, we considered a set of 179 M dwarfs from \citet{mann2015}. We fitted a polynomial relation between the effective temperature (T$\rm_{eff}$) of the selected stars and their intrinsic  Gaia DR3 color $(G\rm_{BP}-G\rm_{RP})_0$. Because the selected stars are all very close to the Sun, we assumed that interstellar extinction was negligible when constructing the effective-temperature-intrinsic-color relation.
The resulting relation is

\begin{table*}
	\centering
    \caption{Best-fit coefficients for the  $T_{\rm eff}$ vs. $(G_{\rm BP}-G_{\rm RP})_0$ relationship represented by Eq.~\ref{eq:teff_color}.}
	\begin{tabular}{cccccc}
	\hline
    \hline
   $c_0$ & $c_1$ & $c_2$ & $c_3$ & $c_4$ & $c_5$  \\
    (K) & (K mag$^{-1}$) &  (K mag$^{-2}$)&  (K mag$^{-3}$)&  (K mag$^{-4}$) & (K mag$^{-5}$) \\
    \hline
    9\,649.1817 & $-$7\,175.80969 & 3\,642.30312 & $-$1\,020.37499 & 146.20008 &  $-$8.30455 \\
    \hline
    \end{tabular}   
   \label{tab:teff_color}
\end{table*}

\noindent

\begin{equation}
\label{eq:teff_color}
 T_{\rm eff}=c_0+\sum_{i=1}^{i=5} c_i[(G_{\rm BP}-G_{\rm RP})_0]^i.
\end{equation}

\begin{figure}[!t]
	\centering
	\includegraphics[width=\columnwidth]{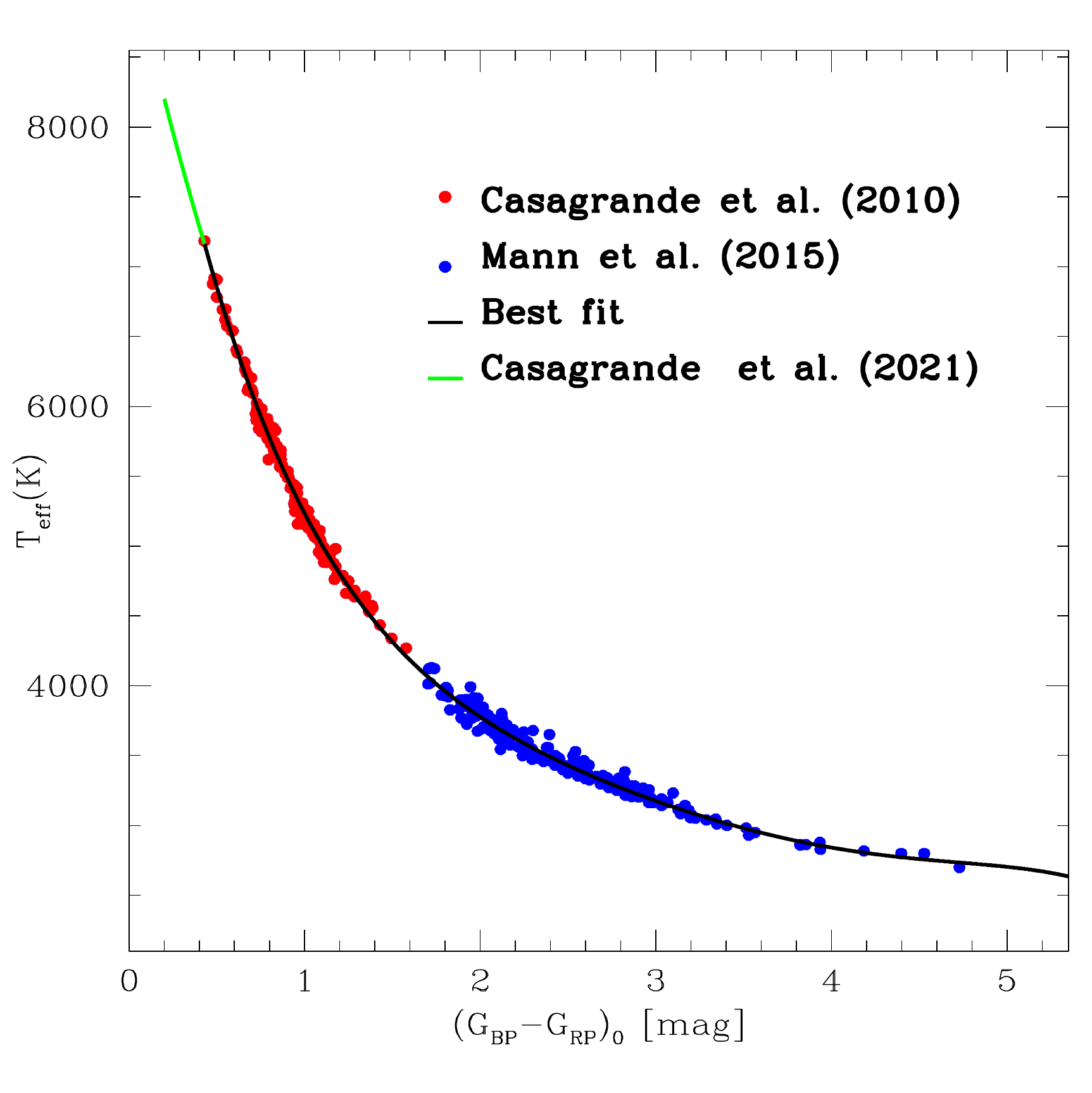}
	\caption{
    Best-fit relationship between effective temperature and color (Eq.~\ref{eq:teff_color}, solid black line). Red points represent the sample of \citet{casagrande2010}, while blue points represent the sample of \citet{mann2015}. The green line represents Eq.~1 of \citet{casagrande2021}, for which we imposed $\log g = 4.438068$ and
    [Fe/H] = 0.
	}
	\label{fig:teff_color}
\end{figure}

\noindent
Table~\ref{tab:teff_color} lists the best-fit coefficients and the validity color range is $0.43 \,\textrm{mag}  \le (G \rm_{BP} -G \rm_{RP} ) _0\le 5$~mag. 

Fig.\ref{fig:teff_color} shows the best-fit relation. 

To extend the color range over which Eq.\,(\ref{eq:teff_color}) is applicable, we constructed the relation by imposing continuity at $(G \rm_{BP} -G \rm_{RP} )_0=0.43$~mag with Eq.~1 provided by \citet{casagrande2021}. We imposed $\log g = 4.438068$ and [Fe/H] = 0 in Eq.~1 of \citet{casagrande2021}, together with the coefficients reported in Table~A1 in Appendix~A of \citet{casagrande2021}.
We used Eq.~\ref{eq:teff_color} to extend the effective-temperature-intrinsic-color relationship down to $(G\rm_{BP}-G\rm_{RP})_0=0.2$~mag.
Initially, we calculated the effective temperature of the tPIC targets using Eq.~\ref{eq:teff_color}, assuming
$(G \rm_{BP}-G\rm_{RP})_0= (G\rm_{BP}-G\rm_{RP})$. We then estimated the monochromatic extinction from \citet{lallement2022} (Sect.~\ref{sec:monochromatic_extinction}) and converted it into extinction in all photometric bands (Sect.~\ref{sec:conversion_monochromatic_extinction}). We then estimated the intrinsic color as
$(G\rm_{BP}-G\rm_{RP})_0= (G\rm_{BP}-G\rm_{RP})-E(G\rm_{BP}-G\rm_{RP})$. We iterated this procedure three times, which was sufficient to obtain convergence in the effective temperature ($\rm \Delta T_{\rm eff} \lesssim $10 K). 

\subsubsection{Hipparcos photometry}

When the Gaia color was not available, we instead used the Hipparcos $(B\rm_T-V\rm_T)_0$ color.
In this case, we adopted the following calibration relation from \citet{casagrande2010}:

\begin{equation}
T_{\rm eff}=\frac{5040~\textrm{K}}{0.5839+0.4\,(B_T-V_T)_0-0.0067\,(B_T-V_T)^2_0},
\end{equation}

\noindent
which is valid in the range $0.19 < (B_T-V_T)_0 < 1.49$~mag. In the above equation, we neglected the dependence on metallicity by assuming [Fe/H] = 0. We adopted an iterative procedure similar to that described for Gaia photometry
to simultaneously estimate the intrinsic color $(B_T-V_T)_0$ and the effective temperature.

\subsubsection{Johnson photometry}

When neither the Gaia color nor the Hipparcos color was available, we used the Johnson $(B-V)_0$ color.
 In this case, we adopted the following calibration relation from \citet{casagrande2010}:

\begin{equation}
T_{\rm eff}=\frac{5040~\textrm{K}}{0.5665+0.4809\,(B-V)_0-0.0060\,(B-V)^2_0},
\end{equation}

\noindent
which is valid in the range $0.18<(B-V)_0<1.29$~mag. In the above equation, we neglected the dependence on the metallicity assuming [Fe/H] = 0. We adopted an iterative procedure similar to that described for Gaia photometry
to simultaneously estimate the intrinsic color $(B-V)_0$ and the effective temperature.

\subsection{Bolometric correction}
\label{sec:bolometric_correction}

\subsubsection{\it{Gaia} photometry}

We calculated the bolometric correction, $BC_{\rm G}$, for the $G$-band magnitude. We interpolated using a spline in the tables provided by the Gaia DR3  team 
\footnote{\url{https://www.cosmos.esa.int/web/gaia/dr3-bolometric-correction-tool}}, imposing $\log g = 4.5$, [Fe/H] = 0, and [$\alpha$/Fe] = 0.
The bolometric correction is valid over the effective-temperature range $2500\, {\rm K} \le \rm T_{\rm eff} \le 20000$ K. The value of $BC\rm_{G}$ is based on the assumption of $M \rm_{G,\odot} = 4.66$ mag, that is, $BC \rm_{G,\odot} = +0.08$ mag 
\citep{creevey2023}.

To assess the impact of adopting fixed values of $\log g$, [Fe/H], and [$\alpha$/Fe], we computed $BC_G$ over the FGK temperature range, varying $\log g$ between 4.0 and 5.0, [Fe/H] between $-0.5$ and $+0.5$, and [$\alpha$/Fe] between 0.0 and 0.4. We find that the dependence on [$\alpha$/Fe] is negligible, while the dependence on $\log g$ is generally below 0.01 mag. The largest effect is due to metallicity, producing variations in $BC_G$ of up to $\sim0.10$ mag at the cool end of the FGK range and typically below 0.03 mag for solar-type stars. These variations correspond to uncertainties of less than about 5\% in the inferred stellar radius and typically below 1.5\% for solar-type stars.

\subsubsection{Hipparcos and Johnson photometry}

We calculated the bolometric correction for the Hipparcos \textit{$V\rm_{T}$} and the Johnson $V$ magnitudes using the following equation from \citet{eker2020}:

\begin{equation}
BC=a_0+\sum_{i=1}^{i=4} a_i\, [\log\,(T\rm_{eff})]^i.
\label{eq:BC}
\end{equation}

\noindent
Table~\ref{tab:BC} reports the coefficients $a_i$ of Eq.~\ref{eq:BC}.

\begin{table*}
	\centering
    \caption{Coefficients of the bolometric-correction formula (Eq.~\ref{eq:BC}) for the Hipparcos \textit{V$\rm_{T}$} and the Johnson \textit{V} bands.}
	\begin{tabular}{ccccc}
	\hline
    \hline
    $a_0$ & $a_1$ & $a_2$ & $a_3$ & $a_4$ \\ 
      (mag) & (mag) &  (mag) & (mag) &  (mag)\\ \hline
     $-$2360.69565 &  2109.00655 &  $-$701.96628 & 103.30304 & $-$5.68559 \\
    \hline
    \end{tabular}
   \label{tab:BC}
\end{table*}

\subsection{Absolute magnitude and absolute luminosity}

We calculated the intrinsic absolute magnitude, $M_\textrm{G,0}$, and its uncertainty, $\delta M_\textrm{G,0}$, as follows:

\begin{equation}
    M_\textrm{G,0} = G_0 - 5\,\log_{10}\,d + 5,
\end{equation}

\begin{equation}
    \delta M_\textrm{G,0}= \sqrt{\delta G^2+\Big(\frac{5\,\delta d}{\ln(10)\,d}\Big)^2+
    \delta (A_G)^2}.   
\end{equation}

\noindent
We calculated the luminosity, $L$, for all tPIC targets (excluding the P4 sample) as

\begin{equation}
    \frac{L}{L_\odot}=10^{-0.4\,(M_\textrm{G,0} + BC_\textrm{G} - M_{\textrm{BOL}_{\odot}})},
\end{equation}

\noindent
where $d$ is the geometric distance of the star, in parsecs, from  \citet{baijon2021},
$\delta d$ is the uncertainty in the distance,;
$\delta G$ is the uncertainty in the $G$ band photometry; $\delta A_G$ is the uncertainty in the extinction in the $G$ band, $M_{\textrm{BOL},\odot} = 4.74$~mag; and $L_\odot =3.828 \times 10^{26}$\,W. We used similar relations for Hipparcos or Johnson astrometry and photometry.\\
If $G$ photometry was unavailable, we used Hipparcos or Johnson photometry; if the distance was also unavailable, we used only apparent magnitudes. These situations and the corresponding assumptions are codified by the caseFlag. In the tPIC, there is only one Hipparcos star for which we calculated the distance using as the inverse of the parallax (in this case, taken from the Hipparcos catalogue itself). In addition, there are 16 pure planet-host targets for which Gaia does not provide proper motions and parallax (i.e., no Bailer-Jones distances were available); for these stars, the caseFlag is 21 (see Table \ref{tab:CaseFlag}).

\subsection{Stellar radii}
\label{sec:stellar_radii}

For FGK dwarfs and subgiants, we calculated the stellar radius, $R$,  from the Stefan-Boltzmann law:

\begin{equation}
    \frac{R}{R_\odot}=\left( \frac{T_\textrm{eff}}{T_{\textrm{eff},\odot}} \right)^{-2}\, \sqrt{\frac{L}{L_\odot}}.
\end{equation}

\noindent
For M dwarfs with  $ 4 \le M_{K_{s,0}}\le 10$~mag, we determined the stellar radius from Eq.~4 of \citet{mann2015}: 

\begin{equation}
\frac{R}{R_{\odot}}=1.9515-0.3520\,(M_{K_{s,0}})+0.01680\,(M_{K_{s,0}})^2,
\end{equation}

\noindent
where $M_{K_{s,0}}$ is the intrinsic absolute magnitude in the $K_s$ band,

\begin{equation}
    M_{K_{s,0}} = K_{s,0} - 5\,\log_{10}\,d + 5 .
\end{equation}
We retrieved the apparent $K_s$ magnitudes used to derive $M_{K_{s,0}}$ from the 2MASS point source catalog \citep{skrutskie2006}.

\subsection{Stellar masses}
\label{sec:stellar_masses}

We determined the stellar mass using the following empirical relation from \citet{moya2018}, which was calibrated on a sample of stars with accurately measured parameters from asteroseismology, eclipsing binaries, and interferometry, and expressed as a function of effective temperature and luminosity:

\begin{equation}
    \log \left ( \frac{M}{M_\odot} \right ) = -0.119 + 2.14 \times 10^{-5}\, T_\textrm{eff} + 0.1837 \times \log \left ( {\frac{L}{L_\odot}} \right ).
\end{equation}

\noindent
The relation is valid over the ranges $4780\, \textrm{K} \le T_\textrm{eff} \le 10\,990\, \textrm{K}$ and $-0.717 \le \log \left ( L / L_\odot \right )  \le 2.01$. 
For late K dwarfs with $T_{\rm eff}<4780$ K and $\rm 4 \le M_{K_{s,0}}\le10$~mag and for M dwarfs with 
$\rm 4 \le M_{K_{s,0}}\le10$~mag, we adopted the empirical mass--$M_{K_s}$ relation of \citet{mann2015}, which was calibrated using nearby cool dwarfs with accurately determined stellar parameters. The use of the absolute $K_s$ magnitude minimizes the impact of extinction and reduces the sensitivity to metallicity effects.

For late K dwarfs with $T_{\rm eff}<4780$ K and $\rm 4 \le M_{K_{s,0}}\le10$~mag and for M dwarfs with 
$\rm 4 \le M_{K_{s,0}}\le10$~mag, we determined the stellar mass using Eq.~10 of \citet{mann2015}:

\begin{align}
\frac{M}{M_{\odot}}=0.5858 &+ 0.3872\,(M_{K_{s,0}}) 
-0.1217\,(M_{K_{s,0}})^2 + \\ \nonumber
& +0.0106\,(M_{K_{s,0}})^3-2.7262\times10^{-4}(M_{K_{s,0}})^4.
\end{align}

\noindent
Finally, for late K dwarfs with $3\,700 \,\textrm{K} \le T\rm_{eff} < 4\,780$ K for which
Eq.~10 of \citet{mann2015} was not applicable, we used spline interpolation to compute the stellar mass from the effective temperature using a 1 Gyr, solar-metallicity isochrone from the Padova database \citet{girardi2005}.

We used this isochrone-based interpolation only to bridge the limited temperature interval in which neither of the adopted empirical mass relations was applicable. Since late-K main-sequence stars evolve slowly, the mass--$T_{\rm eff}$ relation in this regime is only weakly dependent on age. The assumption of a 1 Gyr, solar-metallicity isochrone therefore provides a smooth transition between the Mann et al. (2015) relation for cool dwarfs and the Moya et al. (2018) relation for FGK stars. We note that this approximation may be less accurate for the small number of evolved or slightly evolved stars in this temperature range.

\subsection{Uncertainties}
\label{sec:uncertainty}

We determined the uncertainties in the stellar parameters using Monte Carlo simulations, perturbing all observed quantities accordingly to their associated uncertainties, and assuming Gaussian distributions. We perturbed the magnitudes, derived from the color-effective temperature relation, using a Gaussian uncertainty of 200~K. We performed 100 simulations for each star and calculated the uncertainties in the effective temperature, mass and radius as the half interval between the $16 \rm^{th}$ and $84 \rm^{th}$ percentiles of the cumulative distributions of the simulated values.

\subsection{Noise to Signal Ratio}
\label{sec:noise_to_signal_ratios}

The noise-to-signal ratios (NSRs) for each source in the tPIC were provided by the PPT following \citet{borner2024} and \citet{cabrera2026}. The NSR depends on the instrument models, parameters, and observational conditions, including the exact pointing and the pointing rotation angle. Changes in these models and parameters result in different NSR values. In tPIC we report  the beginning-of-life (BOL) NSR  for N-CAMs considering both random and systematic noise (tPIC column name: BOLrandomSysNSRNCAM\_T). Further details are provided by\citet{borner2024} and \citet{cabrera2026}.

\subsection{tPIC scientific ranking}
\label{sec:scientific_ranking}

We ranked the tPIC targets according to the following metric: 
\begin{equation}
\mathcal{M} = \frac{\textrm{80 ppm}}{\textrm{NSR}}\left ( \frac{R_\star}{R_\odot} \right ) ^ {-2}\left ( \frac{T_\mathrm{eff}}{5778\textrm{ K}} \right ) ^{-1}, 
\end{equation}
where the NSR is expressed in ppm, $R_{\star}$ is in solar radii and $T_\textrm{eff}$ is in Kelvin. For stars with a missing stellar radius, or effective temperature, or with NSR = 0, we set the ranking to the average value of the corresponding stellar sample (P1, P2, P3, or P4). For all stars with planets (Sect.~\ref{sec:confirmed_candidate_planet_hosts}), we set the scientific ranking to one. Further details are provided in \citet{nascimbeni2026}.\\

\subsection{Distribution of stellar parameters} 
\label{sec:distributions}

\begin{figure*}[!t]
	\centering
	\includegraphics[width=0.66\columnwidth]{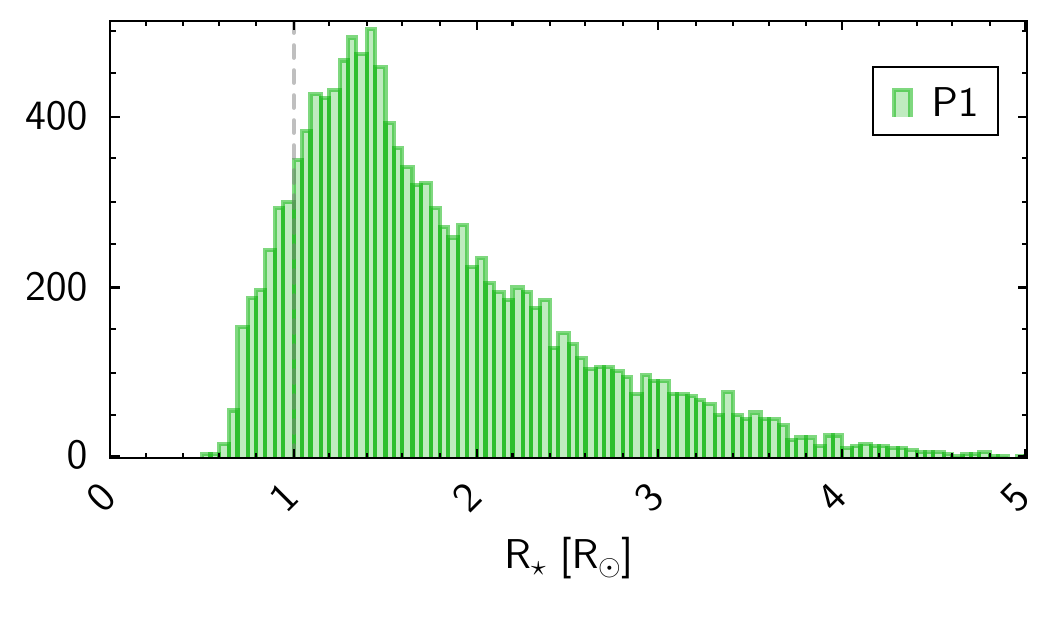}
	\includegraphics[width=0.66\columnwidth]{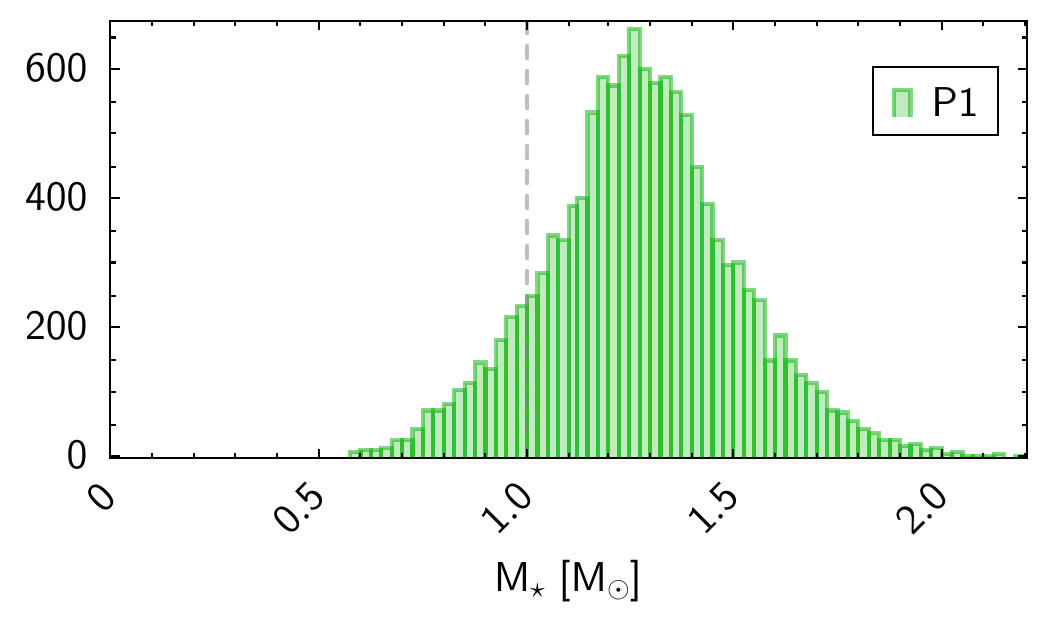}
	\includegraphics[width=0.66\columnwidth]{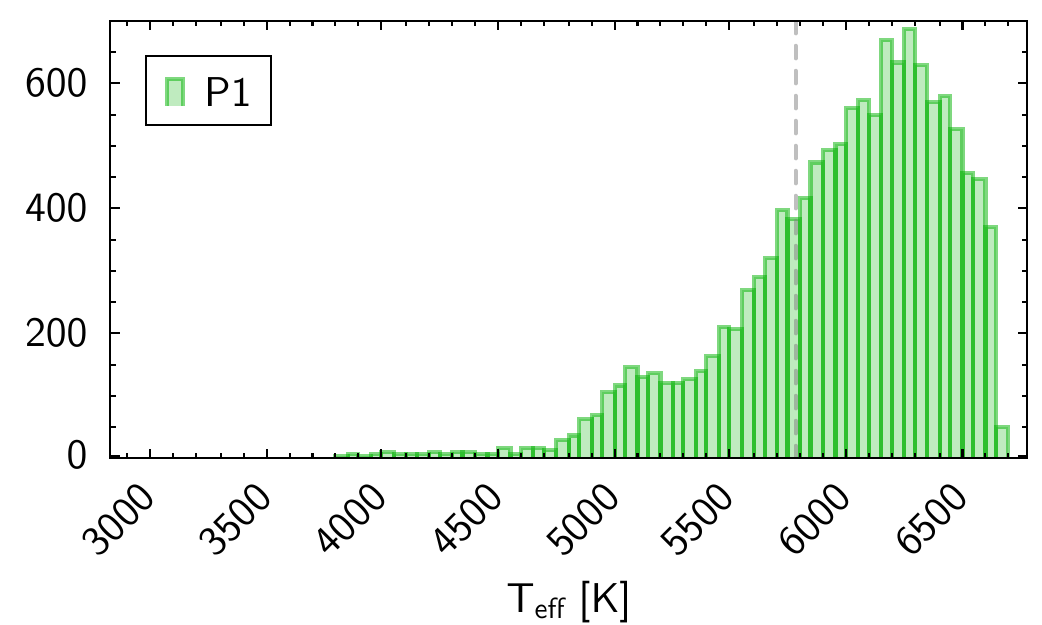}
    \includegraphics[width=0.66\columnwidth]{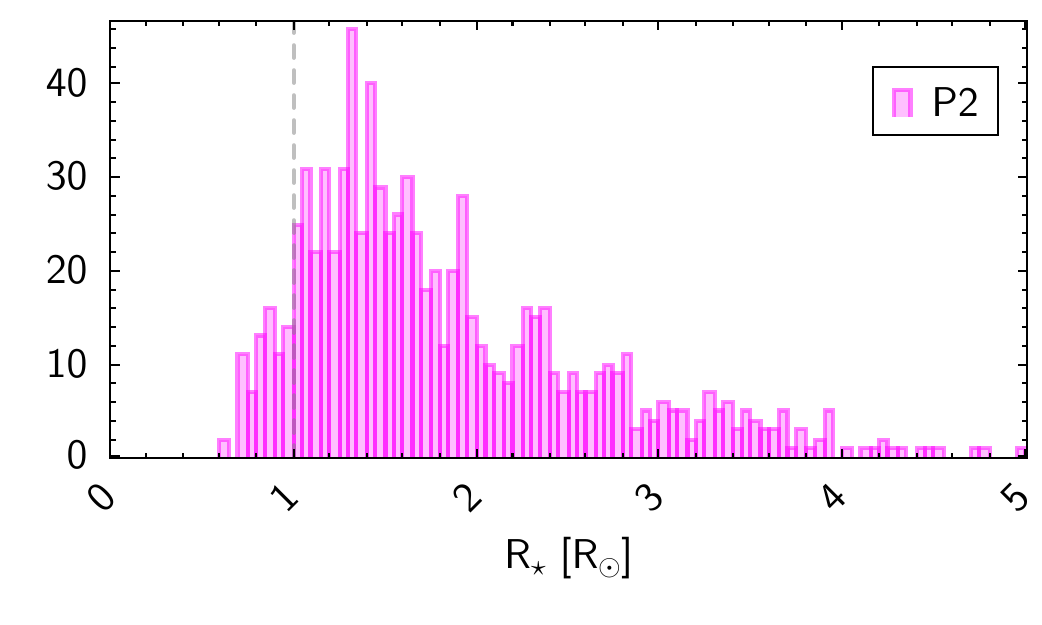}
	\includegraphics[width=0.66\columnwidth]{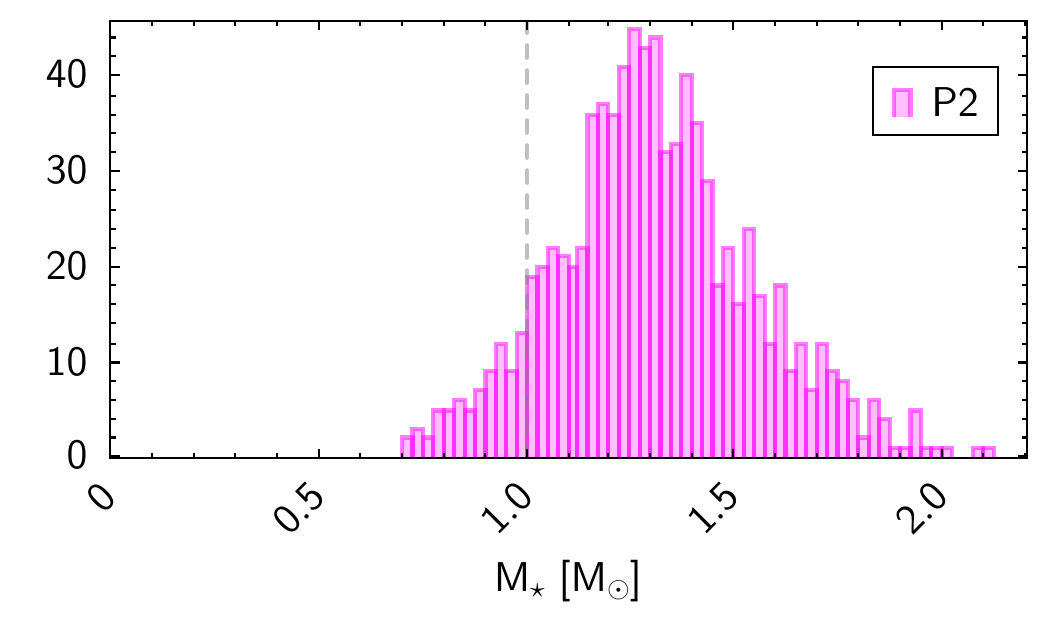}
	\includegraphics[width=0.66\columnwidth]{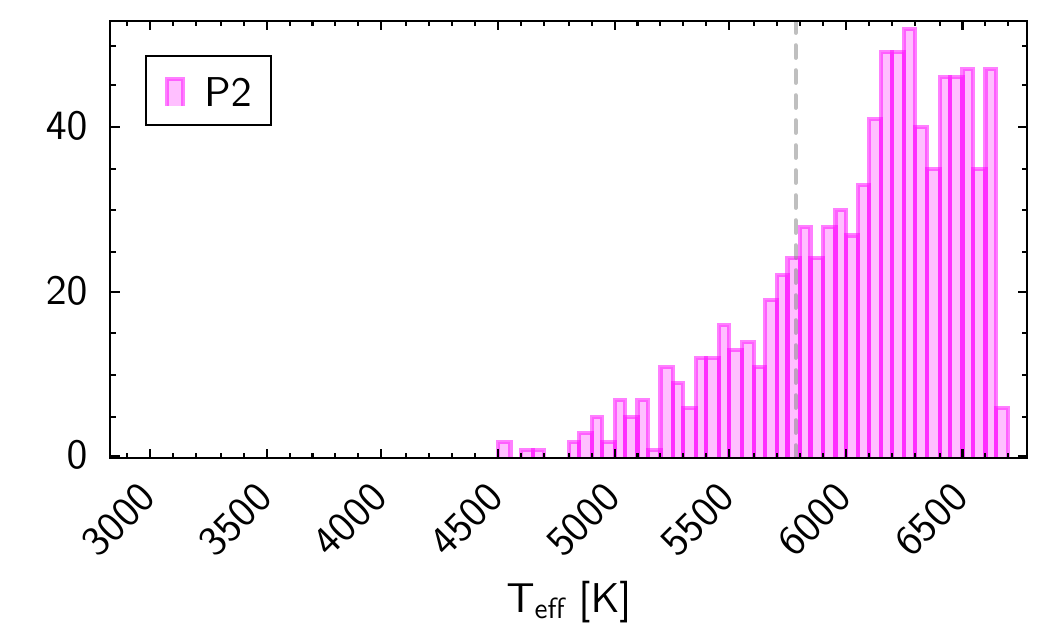}
    \includegraphics[width=0.66\columnwidth]{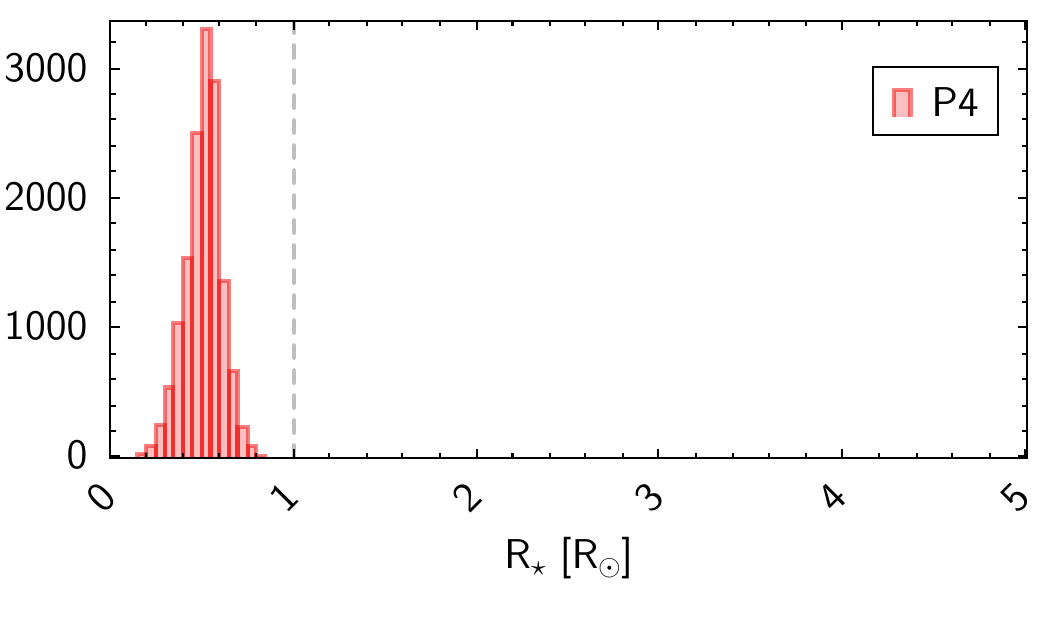}
	\includegraphics[width=0.66\columnwidth]{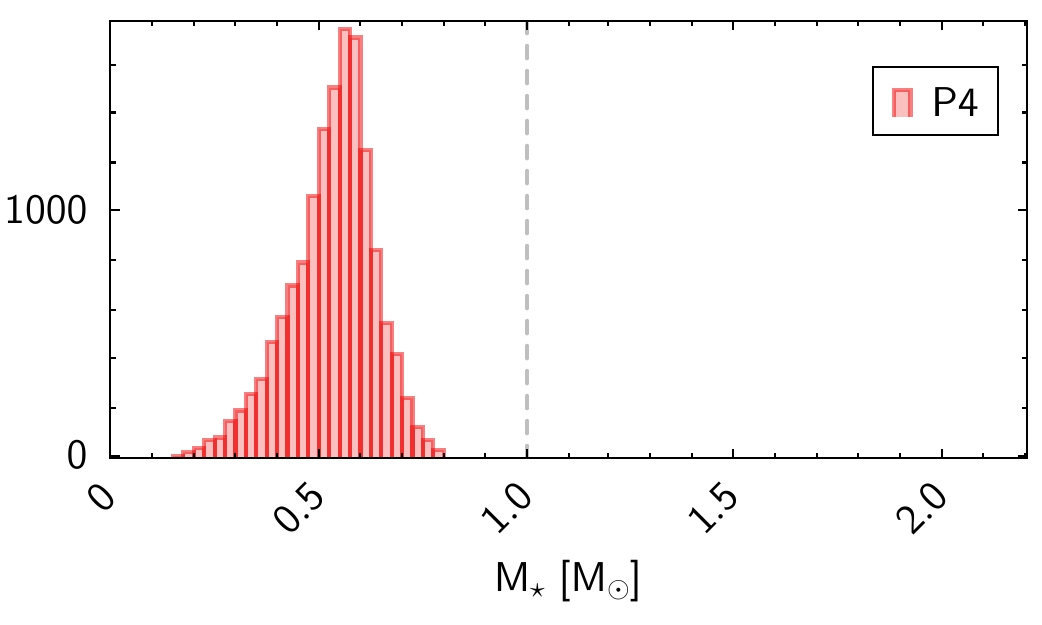}
	\includegraphics[width=0.66\columnwidth]{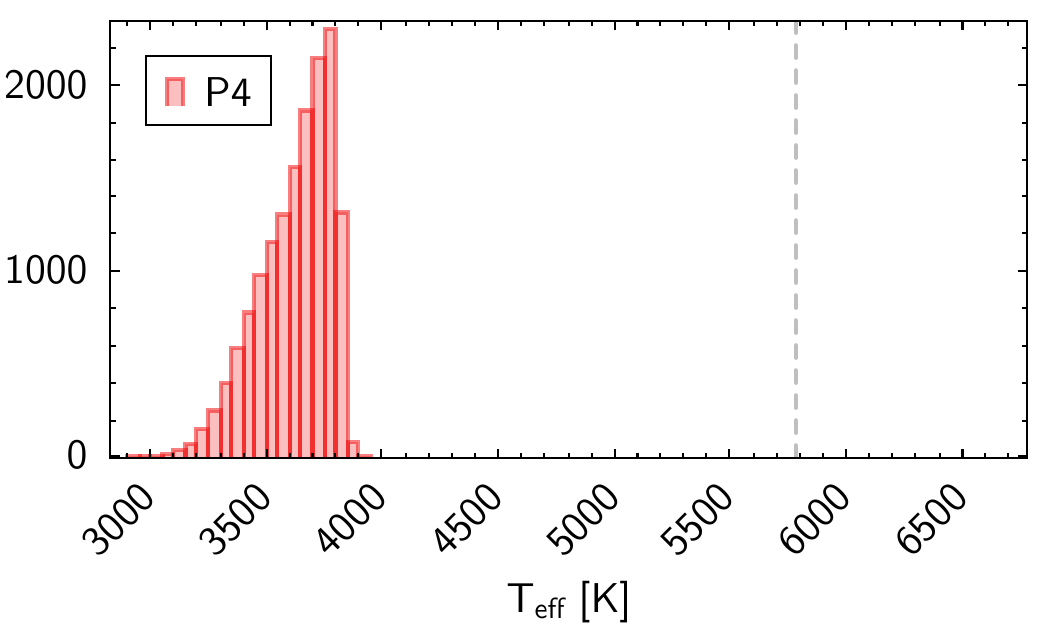}
    \includegraphics[width=0.66\columnwidth]{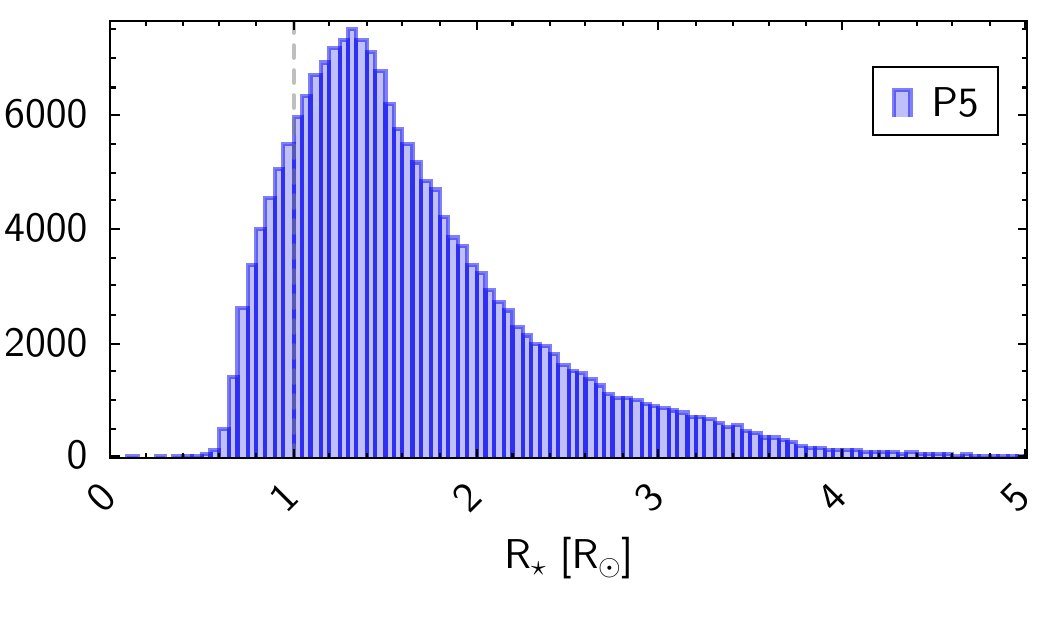}
	\includegraphics[width=0.66\columnwidth]{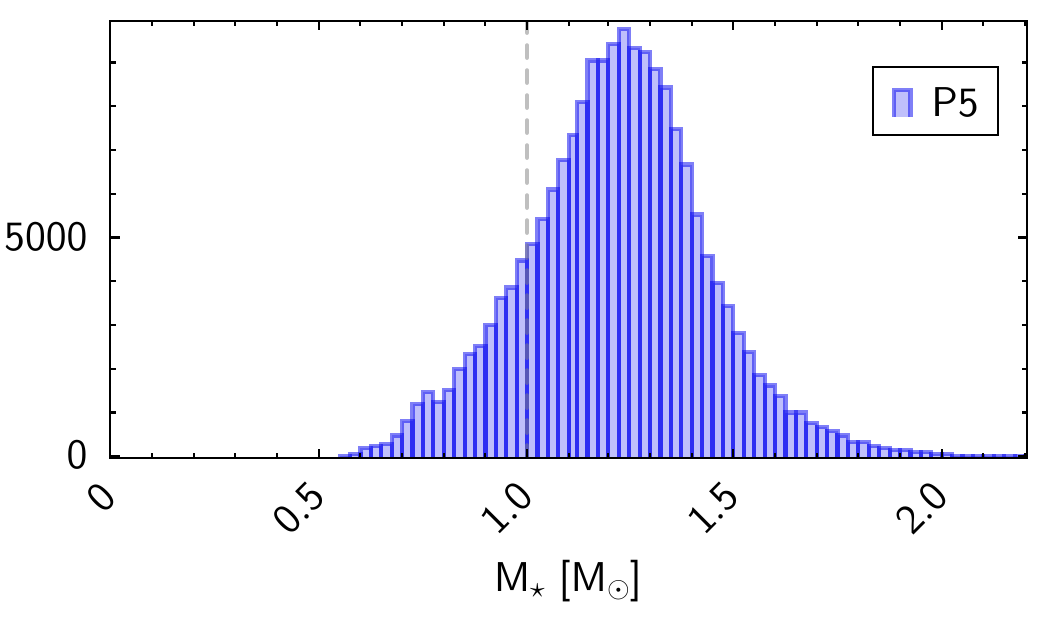}
	\includegraphics[width=0.66\columnwidth]{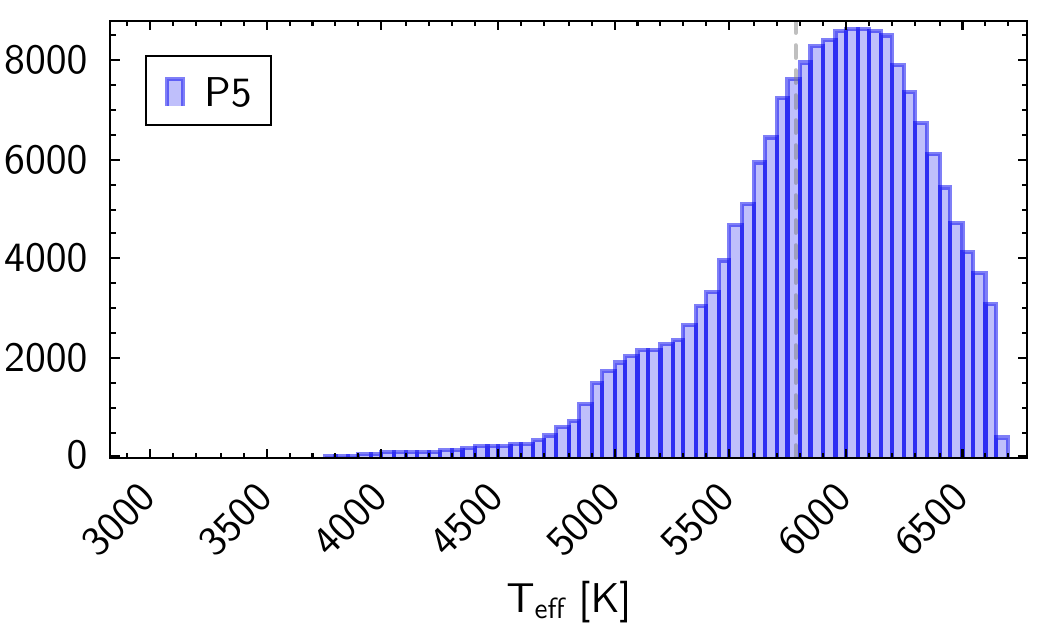}
	\caption{
    Distribution of stellar radii $R_\star$ (left panels), stellar mass $M_\star$ (middle column) and stellar effective temperature $T_\mathrm{eff}$ (right panels) for the P1, P2, P4, and P5 samples (green, magenta, red, and blue histograms, respectively). Sect.~\ref{sec:stellar_parameters} and Table~\ref{tab:counts} provide further details. The solar reference values of $1~R_\star$, $1~M_\star$, and 5778~K are marked with a vertical dashed line.
	}
	\label{fig:histograms}
\end{figure*}

Figure~\ref{fig:histograms} shows the distributions of stellar radii, masses, and effective temperatures for the P1, P2, P4, and P5 samples (excluding planet hosts; see Sect.~\ref{sec:confirmed_candidate_planet_hosts}). 
Table~\ref{tab:counts} lists the corresponding median, 10th, and 90th percentile of the distributions. Fig.~\ref{fig:histograms} shows the corresponding histograms. As expected from the Malmquist bias \citep{Butkevich2005}, our magnitude-limited sample is numerically dominated by early-type (F) and evolved stars. 
The expected PLATO exoplanet yield from tPIC targets is fully described in \citet{cabrera2026}

\begin{table}[]
    \centering
    \caption{Distribution of stellar samples parameters.}
    \begin{tabular}{lcccc} \hline \hline
    Parameter & P1 & P2 & P4 & P5 \\ \hline
    $T_\mathrm{eff}$ [K], median \rule{0pt}{16pt}  & 6\,049 & 6\,160 & 3\,654 & 5\,891 \\ 
    $T_\mathrm{eff}$ [K], $10\%$~percentile & 5\,308 & 5\,443 & 3\,397 & 5\,204 \\
    $T_\mathrm{eff}$ [K], $90\%$~percentile & 6\,455 & 6\,530 & 3\,797 & 6\,321 \\ 
    $R_\star$ [$R_\odot$], median  \rule{0pt}{16pt}& 1.55 & 1.59 &  0.52 & 1.44 \\
    $R_\star$ [$R_\odot$], $10\%$~percentile & 0.96 & 1.03 & 0.38 & 0.91 \\ 
    $R_\star$ [$R_\odot$], $90\%$~percentile & 2.84 & 2.93 & 0.63 & 2.55 \\
    $M_\star$ [$M_\odot$], median  \rule{0pt}{16pt} & 1.26 & 1.29 & 0.54 & 1.20 \\ 
    $M_\star$ [$M_\odot$], $10\%$~percentile & 0.97 & 1.02 & 0.39 & 0.92 \\
    $M_\star$ [$M_\odot$], $90\%$~percentile & 1.54 & 1.61 & 0.65 & 1.44 \\ \hline
    \end{tabular}
    \label{tab:counts}
    \tablefoot{Median, 10th and 90th percentiles of the distributions of effective temperature $T_\mathrm{eff}$, stellar radius $R_\odot$, and stellar mass $M_\odot$ for all the subsets of tPIC~2.2.}
\end{table}

\section{Confirmed and candidate planet hosts}
\label{sec:confirmed_candidate_planet_hosts}

We retrieved confirmed planets and planet candidates from the catalogs below using Exo-MerCat (The Exoplanet Merged Catalog; \citealt{alei2020,alei2025}), without any magnitude limit or constraint on the spectral type or luminosity class of the host star.
For planet host stars already contained in P1, P2, P4, or P, the stellar parameters ($T_\textrm{eff}$, mass, and radius), we calculated the procedures described above. Otherwise, we took the stellar parameters from either Gaia DR3 or from the following catalogs: Exoplanet Encyclopedia (EU)\footnote{ \url{https://exoplanet.eu/home/}}, NASA Exoplanet Archive (NASA)\footnote{\url{https://exoplanetarchive.ipac.caltech.edu/index.html}} (both planetary systems and TESS project candidate tables), and the Open Exoplanet Catalog (OEC)\footnote{ \url{https://www.openexoplanetcatalogue.com/}}.
Details on the implementation procedure used for the inclusion of planet host stars in the tPIC will be presented  by Marrese et al. (in prep.).
The confirmed and candidate planets in the present version of the PIC were updated on April 8, 2025. The next update is expected before the start of PLATO science operations. The list of confirmed  and candidate planet host stars will be updated at every spacecraft rotation (every three months).

\section{Comparisons with other catalogs}
\label{sec:comparisons}

\subsection{asPIC1.1.0}
\label{sec:aspic1.1.0}

\begin{figure*}[!t]
	\centering
 \includegraphics[width=0.24\textwidth]{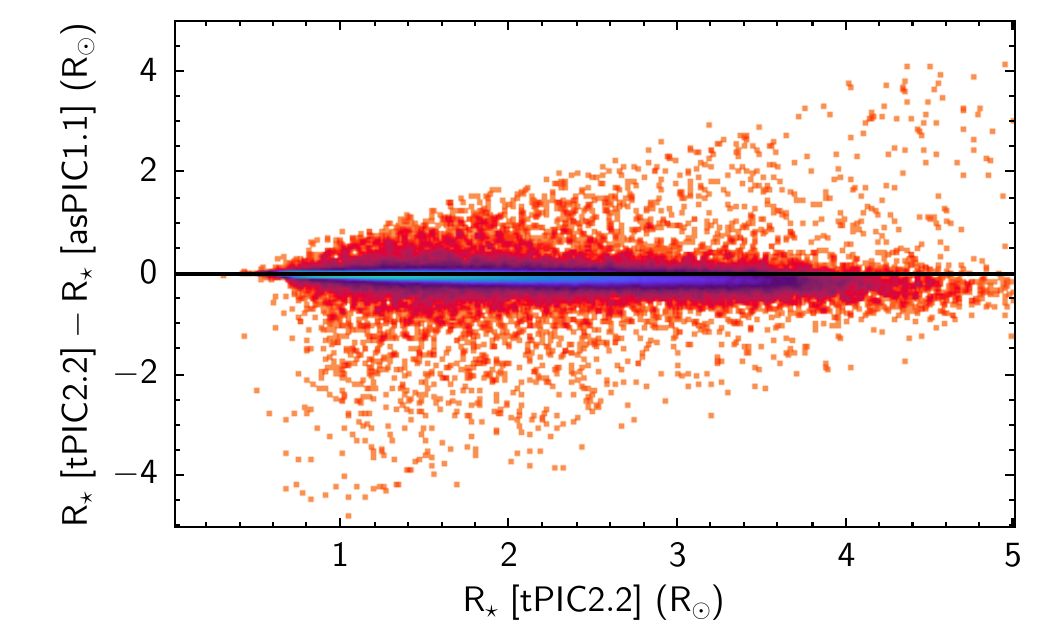}
 \includegraphics[width=0.24\textwidth]{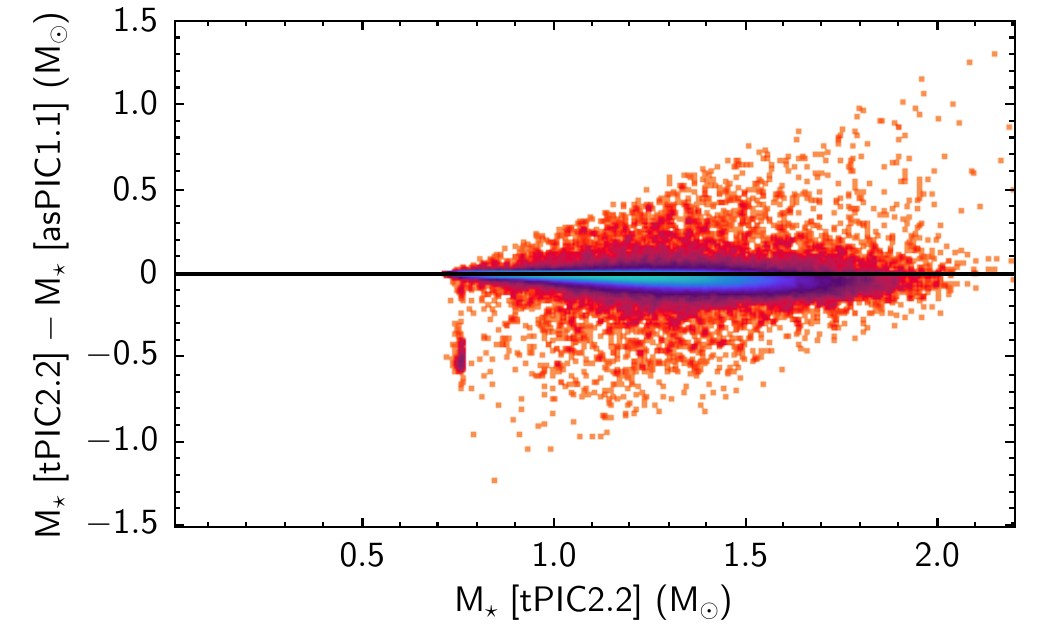}
  \includegraphics[width=0.24\textwidth]{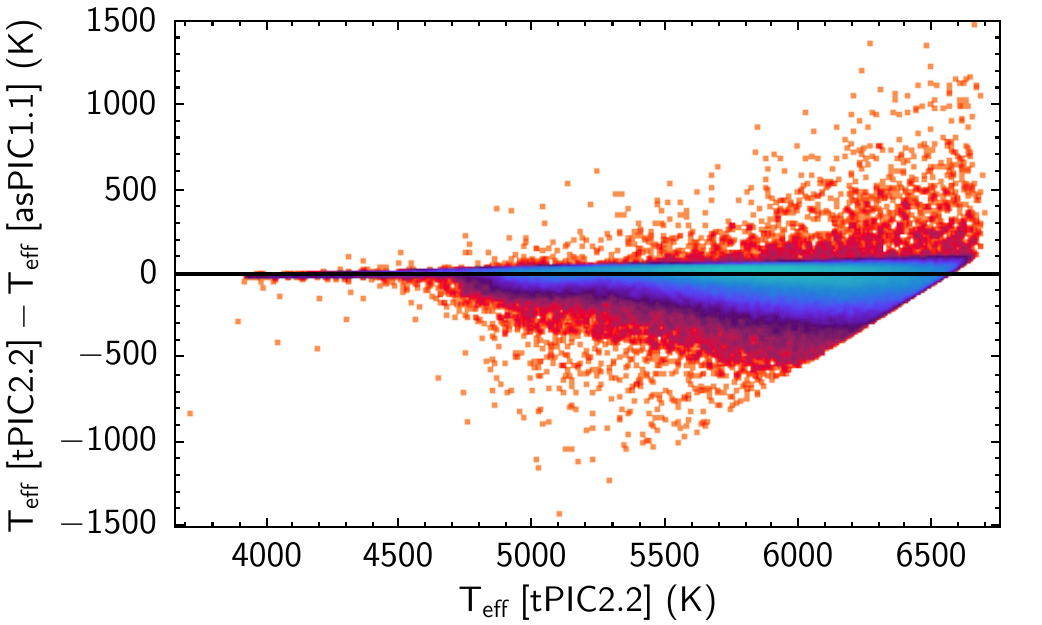}
	\includegraphics[width=0.24\textwidth]{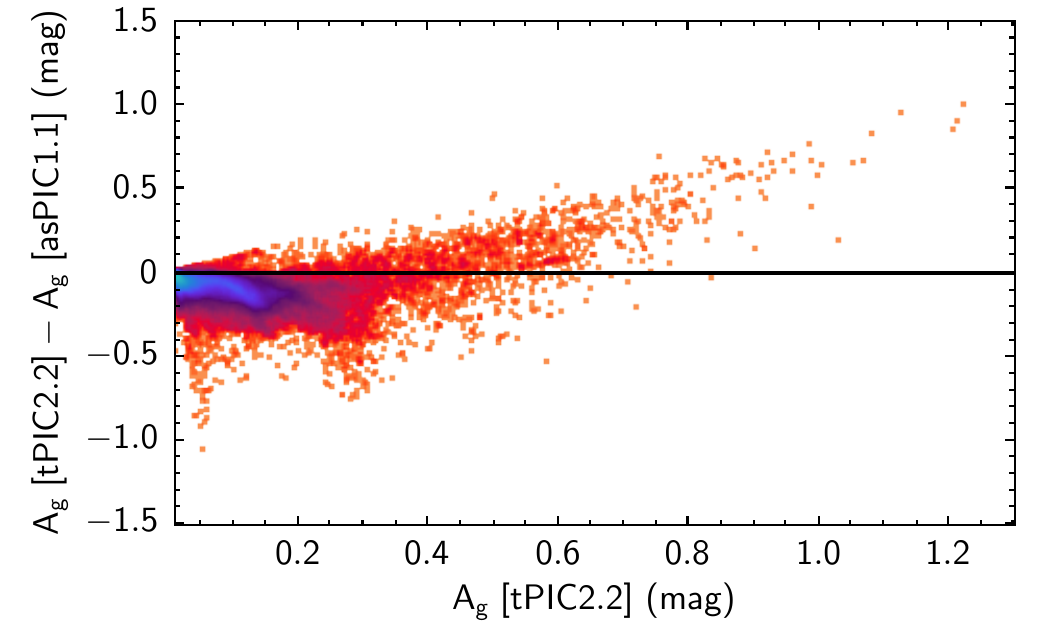}
 \includegraphics[width=0.24\textwidth]{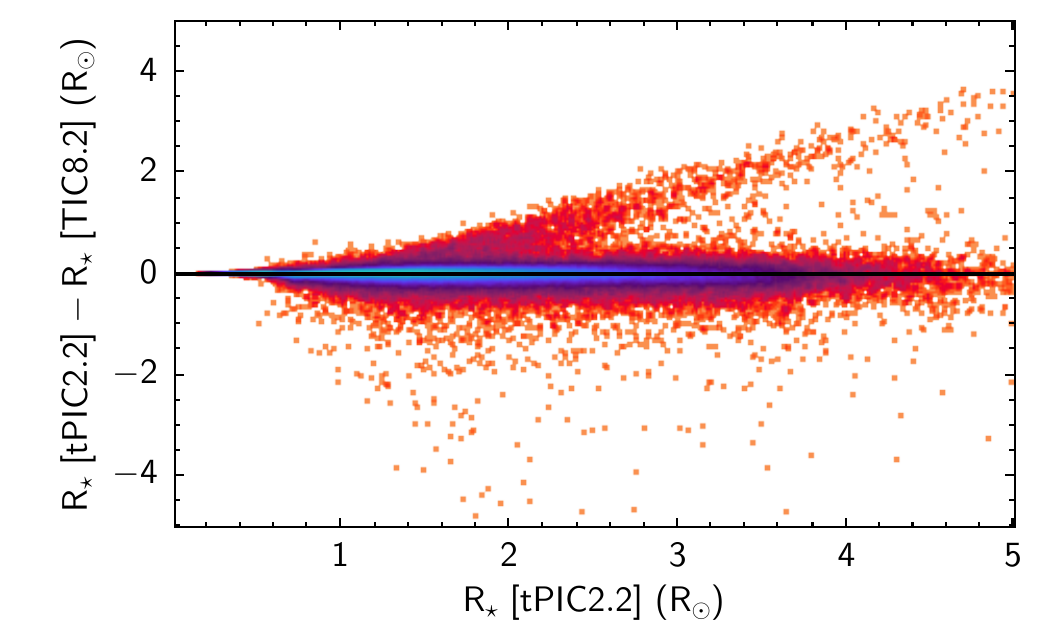}
 \includegraphics[width=0.24\textwidth]{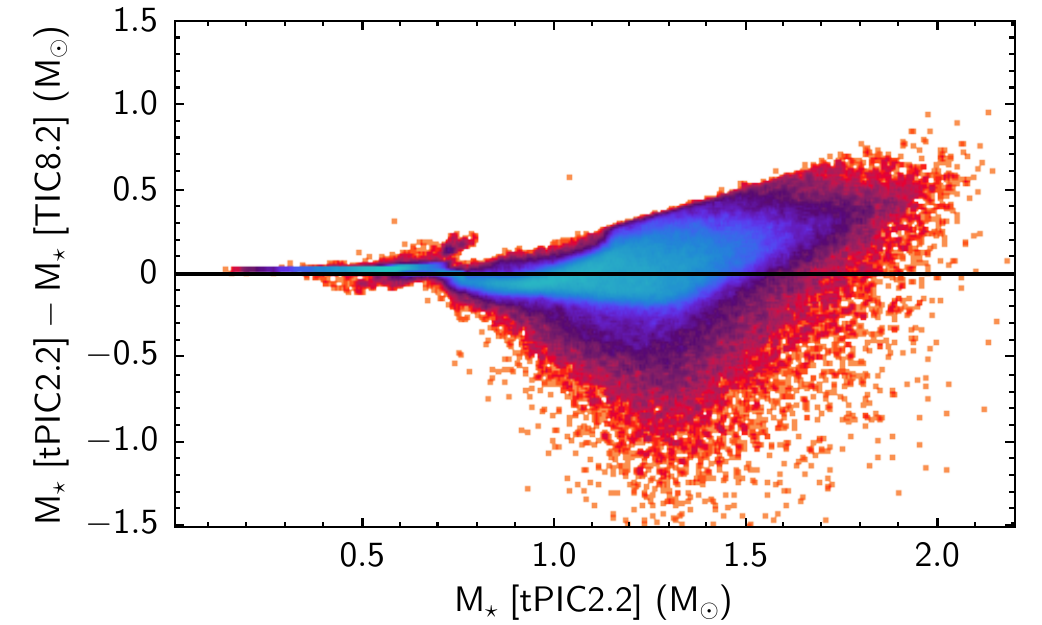}
  \includegraphics[width=0.24\textwidth]{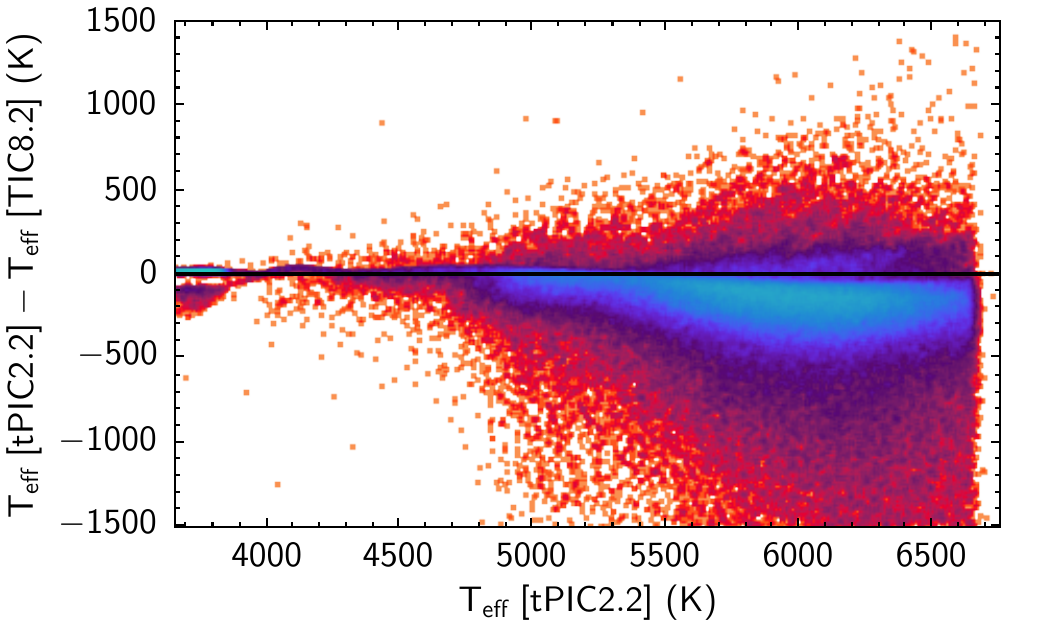}
	\includegraphics[width=0.24\textwidth]{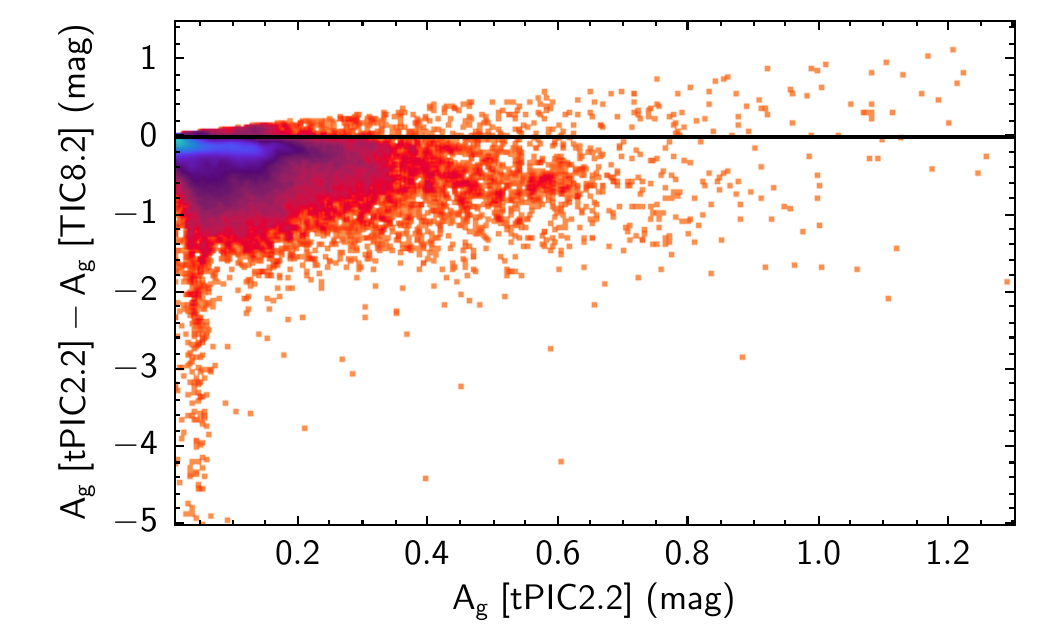}
	\caption{
    Comparison between tPIC (LOPS2) and asPIC1.1 (top panels) and TICv8.2 (bottom panels) for stellar parameters and extinction in the visible. The color scale (from red to blue) is proportional to the point density.
	}
	\label{fig:PIC2.1.0.1_CTLv8.01_new}
\end{figure*}

In the upper panels of Fig.~\ref{fig:PIC2.1.0.1_CTLv8.01_new}, we compare the stellar parameters and extinctions in 
tPIC and asPIC1.1 \citep{montalto2021}.
Table~\ref{tab:comparison_parameters} presents the average difference and their standard deviations between tPIC and asPIC1.1. The stellar parameters and extinctions in the two catalogs are consistent within their uncertainties. Extinctions are, on average, slightly smaller in tPIC than in asPIC1.1 in regions of low reddening, whereas they tend to be larger in regions of higher reddening. 
This trend may reflect differences in the spatial resolution and contrast of the adopted three-dimensional reddening maps, which can affect the reconstruction of small-scale extinction structures, particularly in regions of higher reddening. 

The overdensity around $M_\ast \simeq 0.6\,M_\odot$ in the upper middle panel of
Fig.~\ref{fig:PIC2.1.0.1_CTLv8.01_new} is associated with a very small number (127) of late-K subgiants located close to the lower limit of validity of the Moya et al. (2018) relation. For these stars, the empirical relation of Mann et al. (2015) is not applicable, and we estimated the masses using a 1 Gyr solar-metallicity isochrone. Since this approximation is representative of main-sequence stars, it tends to underestimate the masses of this small subset of evolved objects (see also Sect.~\ref{sec:stellar_masses}).

\begin{table*}[!ht]
	\centering
    \caption{Difference between stellar parameter values and extinctions in tPIC and other catalogs. The difference is defined as the tPIC value minus the value from the selected catalog.}
    \begin{tabular}{lcccc}
	\hline\hline
	  Catalog & $\Delta$Radius (R$\rm_{\odot}$) & $\Delta$Mass (M$\rm_{\odot}$) & $\Delta$T$\rm_{eff}$ (K) & $\Delta$A$\rm_V$ (mag)\\
	\hline
     asPIC1.1 & $-0.06\pm0.20$ & $-0.02\pm0.05$ & $-33\pm93$ & $-0.06\pm0.07$ \\
     TICv8.2 & $-0.007\pm0.225$ & $0.003\pm0.187$ & $-254\pm430$ & $-0.18\pm0.26$ \\
    \hline
	\end{tabular}
    \label{tab:comparison_parameters}
\end{table*}

\subsection{TIC}
\label{sec:tic}

The bottom panels of Figure~\ref{fig:PIC2.1.0.1_CTLv8.01_new} show a comparison of the stellar parameters and extinctions in tPIC and TICv8.2 \citep{paegert2021}.  Table~\ref{tab:comparison_parameters} presents the average differences and standard deviations between tPIC and TICv8.2. The stellar parameters and extinctions in tPIC and TICv8.2 are consistent within their uncertainties. Extinction and effective temperatures from TICv8.2 tends to be higher than those from tPIC. Stars with temperatures beyond $\sim 6\,500$ K belong to the sample of planet-host stars added to tPIC, for which the stellar parameters, including effective temperature, were not calculated using the methods described in this paper, but were instead adopted from external catalogs (see Sect.~\ref{sec:confirmed_candidate_planet_hosts}).

\section{Summary}
\label{sec:summary}

In this work we present the public release of the current PLATO Input Catalog of targets (tPIC) for the first Long-duration Observation Phase field  \citep[LOPS2;][]{nascimbeni2025}. We designed this catalog to fulfill the scientific objectives of the PLATO mission's core science program, which required a carefully preselected target sample because of telemetry constraints. The tPIC is part of a more complex catalog which also  includes cPIC (instrument calibration PIC), fgPIC (fine-guidance PIC), and scvPIC (science calibration and validation PIC), which are discussed in other papers. 

For almost all targets, we estimated interstellar reddening and derived fundamental stellar parameters - effective temperature, radius, and mass - based on astrometric and photometric observations. 
We constructed the tPIC using astrometric and photometric data from Gaia data release 3, complemented by three-dimensional maps of the local interstellar medium, to identify and  determine the main parameters of the stars belonging to the PLATO stellar samples. We used the Hipparcos catalog when Gaia data were not available. To divide the stars among the different tPIC samples, we used the beginning-of-life (BOL) NSR, corresponding to the as-built, as-designed for science configuration (BOLrandomSysNSRNCAM\_T). We also distinguish the counts between the ``on silicon'' and ``not on silicon'' cases because the tPIC includes stars located between and around the CCDs to account for possible spacecraft rotations or pointing inaccuracies. This means that if the pointing is perfect and the detector has the nominal rotation assumed by the simulations, the observations will not include stars classified as not on silicon.

The final catalog includes 217\,741 stars, including 202\,315  FGK dwarfs and subgiants, 15\,037  M dwarfs, and 789 known planet host stars. The two samples probe different distance regimes, with median distances of 512\,pc for FGK stars and 133\,pc for M dwarfs.

The tPIC is compliant with all PLATO requirements for the P1, P2, P4, and P5 samples (Sect.~\ref{sec:stellar_samples}) for a single field.
Table~\ref{tab:stellar_counts} summarizes the stellar counts and compares them with the required values for two LOP fields. 
In the tPIC, the P5 sample is disjoint from the P1 and P2 samples; that is, P5 stars satisfy the P5 sample selection criteria but do not belong to P1 or P2. The sample P2 is completely contained within P1.

As part of the main catalog, we release a dedicated target list containing all known confirmed and candidate exoplanet-host stars located within the LOPS2 field. Planet-host stars not present in the PIC will be added at the beginning of PLATO scientific observations and then at any spacecraft rotation (every three months).\\

\section*{Data availability}\label{sec:access}
The PIC2.2 catalog, which includes the tPIC, scvPIC, cPIC, and fgPIC subcatalogs can be retrieved from PLATO ArXiv\footnote{\url{https://pax.esac.esa.int/plato}}. The catalog is accompanied by  two PIC documents: LOPS2PIC2.2.0.1 Release Notes\footnote{\url{https://doi.org/10.5281/zenodo.19369481}}, which provide a detailed description of the catalog content, and the LOPS2PIC2.2.0.1 Data Definition\footnote{\url{https://doi.org/10.5281/zenodo.19369904}}, which describes the catalog structure and data models. The catalog release includes a catalog of target contaminants.

\begin{table}[!ht]
	\centering
    \caption{Comparison between tPIC stellar counts and required stellar counts for two long-duration pointings (Sect.~\ref{sec:stellar_samples}) for the P1, P2, P4, and P5 stellar samples.}
	\begin{tabular}{lcc}
	\hline
    \hline
    Sample & Number of targets & Required \\
           & Total & (in two LOPs)\\
    \hline
    P1 & 12\,900  & $\ge 15\,000$ \\
    P2 & 868 & $\ge 1\,000$ \\
    P4 & 15\,037 & $\ge 5\,000$ \\
    P5 & 189\,415 & $\ge 245\,000$ \\
    \hline
	\end{tabular}
    \label{tab:stellar_counts}
\end{table}

\begin{acknowledgements}
This work presents results from the European Space Agency (ESA) space mission PLATO. The PLATO payload, the PLATO Ground Segment and PLATO data processing are joint developments of ESA and the PLATO Mission Consortium (PMC). Funding for the PMC is provided at national levels, in particular by countries participating in the PLATO Multilateral Agreement (Austria, Belgium, Czech Republic, Denmark, France, Germany, Italy, Netherlands, Portugal, Spain, Sweden, Switzerland, Norway, and United Kingdom) and institutions from Brazil. Members of the PLATO Consortium can be found at \url{https://platomission.com/}. The ESA PLATO mission website is \url{https://www.cosmos.esa.int/plato}. We thank the teams working for PLATO for all their work. 
MM, GP, LP, VG, VN, SD, SO, EA, SB, DM, LM, IP, RR acknowledge support from PLATO ASI-INAF agreements n. 2022-28-HH.0. 
PM, SM, GA, MF acknowledge financial support from PLATO ASI-INAF agreements n. 2022-14-HH.0 and n. 2025-33.HH.0. We would like to acknowledge the financial support of INAF (Istituto Nazionale di Astrofisica), Osservatorio Astronomico di Roma, ASI (Agenzia Spaziale Italiana) under contract to INAF: ASI 2014-049-R.0 dedicated to SSDC. 
EA’s work has been partly carried out within the framework of the NCCR PlanetS supported by the Swiss National Science Foundation under grants 51NF40\_182901 and 51NF40\_205606. EA’s research was partly supported by an appointment to the NASA Postdoctoral Program at the NASA Goddard Space Research, administered by Oak Ridge Associated Universities under contract with NASA (ORAU-80HQTR21CA005).
V.~A. acknowledges support from FCT – Funda\c{c}\~ao para a Ci\^encia e Tecnologia through national funds and from FEDER via COMPETE2020 – Programa Operacional Competitividade e Internacionalização, under the grants UIDB/04434/2020 (DOI: 10.54499/UIDB/04434/2020) and UIDP/04434/2020 (DOI: 10.54499/UIDP/04434/2020), as well as through a work contract funded by the FCT Scientific Employment Stimulus program (reference 2023.06055.CEECIND/CP2839/CT0005, DOI: 10.54499/2023.06055.CEECIND/CP2839/CT0005)
CA acknowledges support from the BELgian Federal Science Policy Office (BELSPO) through PRODEX grants for Gaia and PLATO.
JMMH is funded by Spanish MICIU/AEI/10.13039/501100011033 and ERDF/EU grant PID2023-147338NB-C21.
This work has made use of data from the European Space Agency (ESA) mission
Gaia\footnote{\url{https://www.cosmos.esa.int/gaia}}, processed by the Gaia Data Processing and Analysis Consortium (DPAC\footnote{\url{https://www.cosmos.esa.int/web/gaia/dpac/consortium}}. Funding for the DPAC has been provided by national institutions, in particular the institutions participating in the Gaia Multilateral Agreement.
This paper includes data that has been provided by AAO Data Central\footnote{\url{datacentral.aao.gov.au}}.
The GALAH survey is based on observations made at the Australian Astronomical Observatory, under programmes A/2013B/13, A/2014A/25, A/2015A/19, A/2017A/18. We acknowledge the traditional owners of the land on which the AAT stands, the Gamilaraay people, and pay our respects to elders past and present.
The full catalog of known planets/candidates retrieved by Exo-MerCat is registered as a Virtual Observatory resource and it is available on TOPCAT \citep{taylor2005}. Exo-MerCat is open-source and accessible via a public GitHub repository \footnote{\url{https://github.com/Exo-MerCat/Exo-MerCat}.}. Some of the results in this paper have been derived using the HEALPix \citep{gorski2005} package.
\end{acknowledgements}

%
\bibliographystyle{aa} 
\bibliography{bibexample} 
%

\begin{appendix}

\section{Libraries of synthetic stellar spectra}
\label{sec:StellarLibraries}

The calibration relations presented in this work were derived using synthetic stellar spectra from several 
libraries described in this section.

\subsection*{MPS-ATLAS}
\label{sec:MPS-ATLAS}

MPS-ATLAS is a library\footnote{\url{https://edmond.mpdl.mpg.de/dataset.xhtml?persistentId=doi:10.17617/3.NJ56TR}} of stellar specific intensity spectra, disk-integrated flux spectra, and model atmospheres computed with MPS-ATLAS code on an extensive and fine grid of stellar parameters \citep{witzke2021,kostogryz2022}. The library provides low resolution synthetic spectra (R = 55) and the wavelength range spans the interval between 90 nm and 1\,600 $\mu$m. 
The calculations assume the same setting as for the Kurucz's grid modeled fluxes: convection is turned on without overshoot, the mixing length is set to 1.25 and microturbulence is equal to 2 km s$^{-1}$. The database provides
two data sets which differ for the adopted chemical abundances and the treatment of limb darkening.
The first set adopts the chemical abundances from
\citet{grevessesauval1998}, the second set adopts chemical abundances from \citet{asplund2009} and the mixing length parameters which depend on stellar effective temperature, metallicity, and surface gravity from \citet{viani2018}. 
We used the first set consistent with the elemental abundances in Kurucz's calculations \citep{witzke2021}. For dwarf stars, we retrieved 56 synthetic spectra with effective temperatures ranging from 3\,500 K to 9\,000 K, $\log g = 4.5$, 
[M/H] = 0.0. For giants we retrieved 56 synthetic spectra with effective temperatures ranging from 3500 K to 9000 K, $\log g = 3.0$, [M/H] = 0.0.

\subsection*{MARCS}
\label{sec:MARCS}

MARCS (Model Atmospheres with a Radiative and Convective Scheme) is a grid of one-dimensional, hydrostatic, plane-parallel and spherical LTE model atmospheres \citep{gustafsson2008}. These may be used together with atomic and molecular spectral line data and software for radiative transfer to generate synthetic stellar spectra. 
We used the .flx files supplied with the models which are statistical samples of the model surface fluxes derived in the model calculations. Despite the fact that these are rough estimates of the surface fluxes and are not synthetic spectra, they are sufficiently accurate for our purpose of deriving broadband colors. The  wavelengths range from 1\,300~\AA$\,$ to 200\,000~\AA$\,$ in vacuum and the wavelength step is not constant but has a constant resolving power of $R=20\,000$. The stellar atmospheric model parameters range in effective temperature from 2\,500 K to 8\,000 K. For dwarf models we retrieved 31 stellar spectra using plane parallel atmospheric models and set the gravity to $\log g = 4.5$ for $\rm T_{\rm eff}>5\,000$ K and $\log g = 5.0$ for $\rm T_{\rm eff}\le 5\,000$ K
and  microturbulence $\xi_t =1 \, {\rm km\, s} ^{-1}$.
For giants, we retrieved 31 models using spherical symmetry for 1 $M_{\odot}$,  
set the  microturbulence to $\xi_t = 2 \, {\rm km\, s}^{-1}$ and the gravity to $\log g = 3.0 $
for models with $\rm T_{\rm eff} > 5\,000$ K and set $\log g$ accordingly to the relation $\log g = 0.00155 \times  \rm T_{\rm eff} -4.964$ for $\rm T_{\rm eff}\le 5\,000$ K. Such a relation was obtained modeling the red giant branch of a 1 Gyr, solar metallicity isochrone from the Padova database. 
The metallicity and alpha elements were set to [Fe/H] = 0.0 and [$\alpha$/Fe] = 0.0 both for dwarf and giant models.

\subsection*{POLLUX}
\label{sec:POLLUX}

POLLUX \citep{palacios2010} is a database\footnote{\url{http://npollux.lupm.univ-montp2.fr/}} of stellar spectra developed at the Laboratoire Univers et Particules de Montpellier (LUPM - University of Montpellier - CNRS). It provides a  library of
theoretical synthetic stellar spectra at high resolution with a broad coverage of the atmospheric parameters (effective temperature $T_{\rm eff}$, gravity $\log g$ and metallicity [Fe/H]) as well as spectral types across the Hertzsprung-Russell Diagram.

From this database, we retrieved spectra in the visible domain (3\,000~\AA$\,$ to 12\,000~\AA) from the AMBRE collection which uses MARCS\footnote{\url{https://marcs.astro.uu.se/}} \citep{gustafsson2008} models and the TURBOSPECTRUM \citep{alvarezplez1998} radiative code to produce synthetic spectra with resolution $R > 150\,000$.  For dwarfs, we retrieved 18 spectra ranging in effective temperature between 3\,500 K to 8\,000 K. The gravity was set to $\log g = 4.5$,
the metallicity and alpha elements were set to [Fe/H] = 0 and [$\alpha$/Fe] = 0. The  microturbulence parameter was set to $\xi_t =1 \, {\rm km\, s} ^{-1}$.
For giants, we retrieved 17 spectra ranging in temperature between 3\,500 K to 8\,000 K. The gravity was set to $\log g = 3.0 $
for models with $\rm T_{\rm eff}\ge 5\,000$ K and accordingly to the equation $\log g = 0.00155 \times \rm T_{\rm eff}-4.96411$ for $\rm T_{\rm eff} \le 5\,000$ K. We used spherical  symmetry models for $1\, M_{\odot}$. For $\rm T_{\rm eff} \ge 5\,000$ K metallicity and alpha elements were set to [Fe/H] = 0.2 and [$\alpha$/Fe] = 0.2 and  microturbulence to $\xi_t = 2 \, {\rm km\, s}^{-1}$. For $\rm T_{\rm eff} < 5\,000$ K 
metallicity and alpha elements were fixed to [Fe/H] = 0.0 and [$\alpha$/Fe] = 0.0 and  microturbulence to $\xi_t =1 \, {\rm km} ^{-1}$.
From the POLLUX database we also retrieved visible spectra
(3\,000~\AA$\,$ to 12\,000~\AA) from the BT-Dusty collection
which uses the PHOENIX radiative transfer code and produced spectra with resolution $R>100\,000$. 
For dwarfs, we retrieved 39 spectra with effective temperature ranging from 2\,100 K and 6\,000 K and $\log g = 4.5$ for $\rm T_{\rm eff}> 3\,000$ K and $\log g = 5.0$ for $\rm T_{\rm eff}\le 3\,000$ K. For giants, we retrieved 41 spectra with effective temperature ranging from 2100 K and 6000 K and $\log g = 3.0$ for $T_{\rm eff}\ge 5\,000$ K and $\log g < 3.0$ for $T_{\rm eff}< 5\,000$ K. Metallicity and $\alpha$ elements were set to
[Fe/H] = 0, [$\alpha$/Fe]=0, respectively and  microturbulence varied with the temperature and was automatically set in the models.

Finally, from the POLLUX database we also retrieved visible spectra (3\,000~\AA$\,$ to 12\,000~\AA) from the CMFGEN collection
of hot stars which uses the CMFGEN\footnote{\url{http://kookaburra.phyast.pitt.edu/hillier/web/CMFGEN.htm}}
radiative transfer code and produced spectra with resolution $R=150000$. In this case, we used the same set of 35 spectra both for dwarfs and giants. 
The effective temperatures were comprised between 12\,020 K and 63\,880 K, the gravity between $\log g=2.2$ and $\log g=4.2$, metallicity and $\alpha$ elements were set to
[Fe/H] = 0.05 and [$\alpha$/Fe] = 0, respectively and  microturbulence to $\xi_t=5-10\, {\rm km\,s}^{-1}$.

\subsection*{Coelho}
\label{sec:Coelho}

\citet{coelho2005} presented a library\footnote{\url{https://cdsarc.cds.unistra.fr/viz-bin/cat/VI/120}} of high resolution synthetic stellar spectra from 300 nm to 1.8 $\mu$m at a resolution of $R=275\,000$. The library spans all the stellar types that are relevant to the integrated light of old and intermediate-age stellar populations in the involved spectral region (spectral types F through M and all luminosity classes).
For dwarfs, we  retrieved 15 models with effective temperature between 3500 K and 7\,000 K, $\log g=4.5$, [Fe/H] = 0.0, [$\alpha$/Fe] = 0.0. The  microturbulence velocity was set to $\xi_t=1.0 \,{\rm km\,s}^{-1}$. For giants, we  retrieved 15 models with effective temperature between 3\,500 K and 7\,000 K. The gravity was set to log g=3.0 for stars with $T\rm_{eff}> 5\,000$ K and accordingly to the relation $\log g = 0.00155 \times T\rm_{eff}-4.96411$
for stars with $T\rm_{eff}\le 5\,000$. The metallicity was set to
[Fe/H] = 0.0 and the $\alpha$ elements abundance to [$\alpha$/Fe] = 0.0.

\section{PLATO response function}
\label{sec:PLATOresponseFunction}

The response function of PLATO was presented   by \citet{marchiori2019} with an instrument design that was representative for the phase B of the project. We used at that time performance parameters based on end-of-life requirements for the normal cameras. This scenario was required to show that the mission would fulfill its science goals using conservative assumptions. At the moment of writing this paper, we are in phase D  and we need to review the assumptions used in phase B. We  must re-define the PLATO magnitude in \citet{marchiori2019} to include performance
parameters which are representative of the Beginning of Life performance of the PLATO normal and fast cameras. This scenario is required for the selection of the targets of the PLATO Input Catalog \citep{montalto2021}.

Here we  provide a short summary based on the more detailed description  that will appear in \citet{cabrera2026}.

For every camera (the blue fast camera F-CAM$_{\rm b}$,  the red fast camera F-CAM$_{\rm r}$ and the normal camera N-CAM) the as designed (blue,red,normal) camera effective transmission was calculated as the product of four components: (i) the as designed filter, (ii) the transmission loss due to molecular contamination,
(iii) the transmission loss due to particulate contamination, (iv) the as designed transmission of the camera.

In short, the impact of contamination  was  calculated as per requirements Beginning of Life  The filters are as designed \citep{cabrera2026}. The transmission values of the lenses  were  defined by the system team based on measurements on the TOU \citep[Telescope Optical Unit,][]{munari2022} Proto-Flight Model. The assumed CCD quantum efficiency is the mean value of the flight model distribution.

\begin{figure}[!t]
	\centering
	\includegraphics[width=\columnwidth]{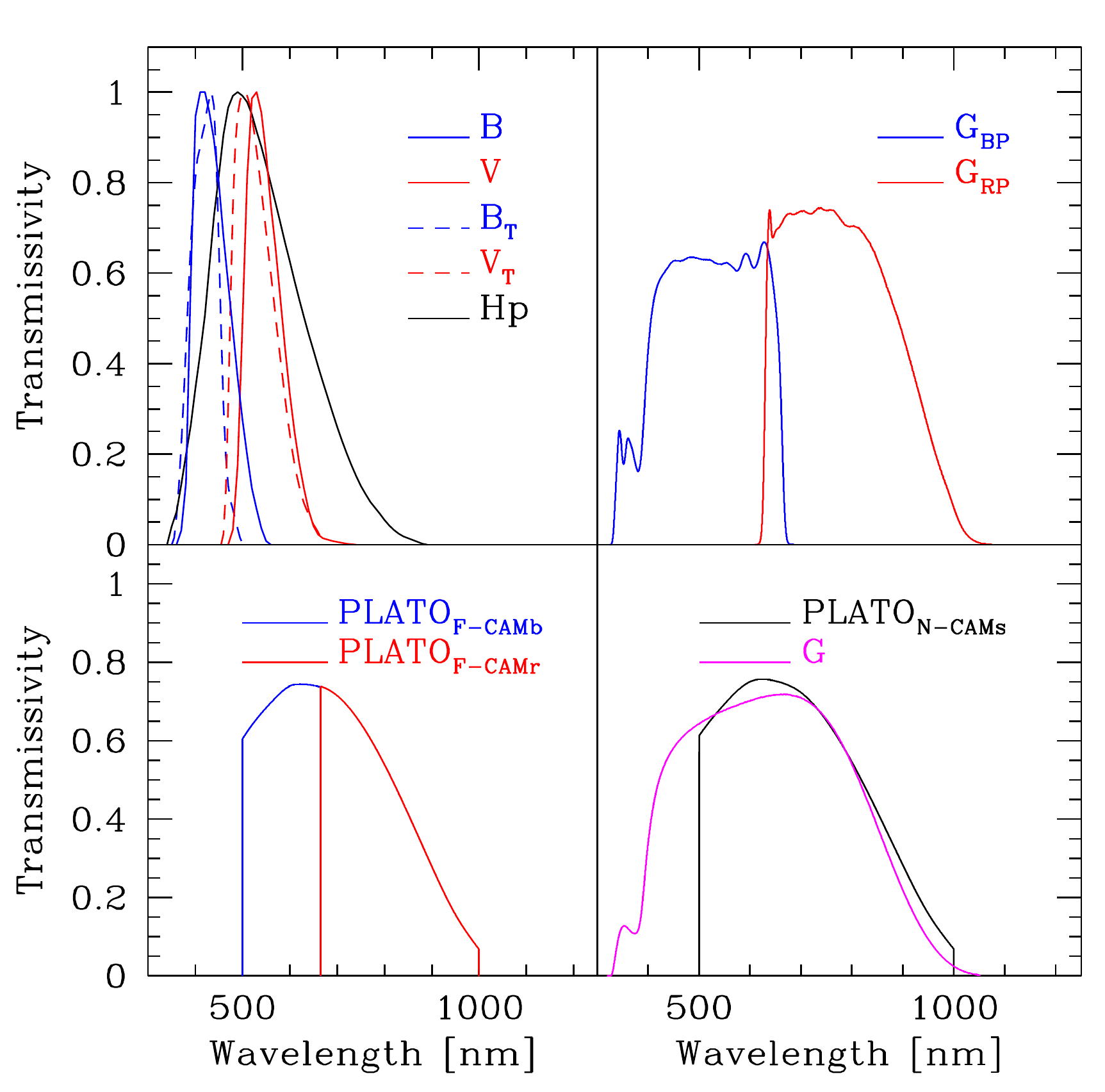}
	\caption{
     {\it Upper left:} spectral response function of Hipparcos (Hp, solid black), Tycho-2 $B\rm_T$ (dashed blue, Tycho-2 $V\rm_T$ (dashed red), Johnson $B$ (solid blue), Johnson $V$ (solid red).
{\it Upper right:}
     Gaia DR3
     $G_{\rm BP}$ (solid blue) and $G_{\rm RP}$ (solid red). 
     {\it Bottom left:}
     PLATO blue fast camera (F-CAM$\rm_b$, solid blue) and PLATO red fast camera (F-CAM$\rm_r$, solid red).
      {\it Bottom right:} PLATO N-CAMs (solid black),
     Gaia DR3 $G$ (solid magenta), 
	}
	\label{fig:transmission_curves}
\end{figure}

The transmission and the quantum efficiency curves are tabulated as a function of wavelength in the files from the MPDB (Mission Parameter Database\footnote{\url{https://www.researchgate.net/publication/365800892_PLATO_Mission_Parameter_Database}}). 
In all calculations that follow we used spline interpolation to interpolate the response curve at the spectral resolution of a given input spectrum. Note that the response function was set identically equal to zero outside the [500 nm, 1000 nm] interval for the normal cameras and
outside the interval [500 nm, 665 nm] and [665 nm, 1000 nm] for the blue and red cameras, respectively. Fig. \ref{fig:transmission_curves} shows the response function of PLATO N-CAMs, PLATO blue fast camera (F-CAMb), PLATO red fast camera (F-CAMr) Gaia DR3 $G$, $G_{\rm BP}$ and $G_{\rm RP}$\footnote{\url{https://www.cosmos.esa.int/web/gaia/edr3-passbands}}, Hipparcos (Hp), Tycho-2 
$B_{\rm T}$, Tycho-2 $V_{\rm T}$,
Johnson $B$, Johnson $V$ \citep{bessell2012}, as specified in the panels.

\section{PLATO magnitude}
\label{sec:PLATOmagnitude}

We determined the PLATO magnitude in the VEGA system. We used the spectrum alpha\_lyr\_stis\_011.fits retrieved from the CALSPEC library\footnote{\url{https://archive.stsci.edu/hlsps/reference-atlases/cdbs/current_calspec/}} as reference spectrum of Vega. We then calculated the PLATO magnitude zero point in this system ($P_{ZP}$) with

\begin{equation}
P\rm_{ZP}=2.5\log_{10}\Big(\frac{\Theta}{hc}{\int_{\lambda_{min}}^{\lambda_{max}}\,f_\textrm{Vega}(\lambda)\,S(\lambda)\,\lambda\,d\lambda}\Big)+P_\textrm{Vega},
\end{equation}

\noindent
where $f_\textrm{Vega}(\lambda)$ is the Vega spectral flux at Earth which we expressed
in units of W m$^{-2}$ nm$^{-1}$, $\Theta$ is the instrument pupil area ($\Theta$=0.01131 m$^{2}$, for one camera), $h$ is the Planck constant and $c$ the speed of light ($hc=$1.98644586$\times\,10^{-16}$ J nm),

$P_\textrm{Vega}$ is the PLATO magnitude assigned to Vega. We set $P_\textrm{Vega}=0.023$ mag, equal to the measured Johnson $V$ magnitude of Vega \citep{bohlin2007}, in order to anchor the PLATO photometric zero point to the observed flux-calibrated spectrum of Vega.

$S(\lambda)$ is the PLATO response function described in Sec.\ref{sec:PLATOresponseFunction} 
and $\rm\lambda_{min}$, $\rm\lambda_{max}$ denote
the extreme wavelengths of each filter.
For the normal cameras $P\rm_{ZP,NCAMs} = 20.78$. For the blue fast camera $P\rm_{ZP,FCAM_b} = 20.18$ and for the red fast camera we obtained $P\rm_{ZP,FCAM_r} = 19.81$.

Given the PLATO magnitude zero point $P_{ZP_X}$ and the total flux ($F$) measured by PLATO (in photons s$^{-1}$), the PLATO magnitude ($P_X$) is obtained from:

\begin{equation}
P\rm_{X}=-2.5\log_{10}(F)+P\rm_{ZP_X},
\end{equation}

\noindent
where X = NCAMs,FCAM$\rm_b$,FCAM$\rm_r$. By convention it was decided to calculate the PLATO magnitude for one camera and assuming that each star is observed at the center of the camera.

\subsection{PLATO magnitude from calibration relations}
\label{sec:PLATOmagnitudeFromCalibrationRelations}

We derive some relations between PLATO magnitudes and
magnitudes in other photometric systems which allow us to calculate
the PLATO magnitude of a given source from the magnitude
of the same source in other photometric systems. The relations involve colors in different bandpasses, reddening free (intrinsic). Moreover, we will distinguish between relationships which should be applied to dwarf stars and relationships which should be applied to giant stars, as well as relationships for normal and fast cameras.

\subsection{Synthetic Magnitudes}
\label{sec:SyntheticMagnitudes}

To establish the relations between magnitudes in different photometric systems, we calculated synthetic magnitudes in the various bandpasses using synthetic stellar spectra from the atmosphere model libraries described in Appendix~\ref{sec:MPS-ATLAS} (MPS-ATLAS, MARCS, POLLUX, and Coelho). Multiple libraries were considered to ensure coverage of the full range of stellar parameters relevant to the PLATO target samples and to mitigate possible systematic differences among individual atmosphere models.
Below we describe the photometric systems considered in this work.

\subsubsection{{\it Gaia DR3 G, \texorpdfstring{$G_{\rm BP}$}{GBP} and \texorpdfstring{$G_{\rm RP}$}{GRP}} synthetic magnitudes}
\label{sec:GaiaDR3SyntheticMagnitudes}

The Gaia DR3 bandpasses (S$_X$) have been obtained from the Gaia DR3 documentation\footnote{\url{https://www.cosmos.esa.int/web/gaia/edr3-passbands}}
We derived the zero points in the VEGA system using the same Vega spectrum we adopted for the calculation of the PLATO magnitude (Sec.\ref{sec:PLATOmagnitude})
and following the definition of the zero point in the Gaia documentation\footnote{\url{https://gea.esac.esa.int/archive/documentation/GEDR3/Data_processing/chap_cu5pho/cu5pho_sec_photProc/cu5pho_ssec_photCal.html}}.
The zero point ($G\rm_{X}$) was obtained from 

\begin{equation}
G_\textrm{X}=2.5\log_{10}\left (
\frac{
{\int_{\lambda_{min}}^{\lambda_{max}}\,f_\textrm{Vega}(\lambda)\,S_{X}(\lambda)\,\lambda\,d\lambda}
}
{
{\int_{\lambda_{min}}^{\lambda_{max}}\,S_{X}(\lambda)\,\lambda\,d\lambda}
}
\right ) +G\rm_{X, Vega}
\end{equation}

\noindent
where X$=G$, $G_{\rm BP}$ and $G_{\rm RP}$, G$_{X, \rm Vega}=0.023$ and $f_\textrm{Vega}$ is in W m$^{-2}$ nm$^{-1}$. Here $\lambda_\textrm{min}$ and $\lambda_\textrm{max}$  denote the extreme wavelengths of each bandpass:
$\lambda_\textrm{min}=321.6$ nm and $\lambda_\textrm{max}=1\,051$ nm for the $G$ bandpass,
$\lambda_\textrm{min}=324$ nm and $\lambda_\textrm{max}=724.4$ nm for the $G_{\rm BP}$ bandpass,
$\lambda_\textrm{min}=609$ nm and $\lambda_\textrm{max}=1081$ nm for the $G_{\rm RP}$ bandpass.

The resulting zero point were $G\rm_{ZP}=-26.48$, $G\rm_{BP, ZP}=-25.95$ and $G\rm_{RP, ZP}=-27.22$. These zero points are all compatible within 0.016 mag with the zero points for synthetic 
photometry computed by the Gaia team.

\subsubsection{Hipparcos {\it Hp} and Tycho-2 {\it B$_T$, V$_T$} synthetic magnitudes}
\label{sec:HipparcosHpTychoBTVTSyntheticMagnitudes}

The normalized Hipparcos $Hp$ bandpass and the Tycho-2  $B_T$ and $V_T$ bandpasses have been obtained from \citet{bessell2012}. We derived the zero points in the VEGA system using the same Vega spectrum we adopted for the calculation of the PLATO magnitude (Sec.\ref{sec:PLATOmagnitude}). The zero point ($X_\textrm{ZP}$) was obtained from 

\begin{equation}
X_\textrm{ZP}=2.5\log_{10}\left (
\frac{
{\int_{\lambda_{min}}^{\lambda_{max}}\,f_\textrm{Vega}(\lambda)\,S_{X}(\lambda)\,\lambda\,d\lambda}
}
{
{\int_{\lambda_{min}}^{\lambda_{max}}\,S_{X}(\lambda)\,\lambda\,d\lambda}
}
\right )+X_\textrm{Vega},
\end{equation}

\noindent
where $X=Hp,B_T,V_T$ and $X_\textrm{Vega}=0.023$, $\lambda_\textrm{min}$ and $\lambda_\textrm{max}$ denote the extreme wavelength of each bandpass:
$\lambda_\textrm{min}=340$ nm and $\lambda_\textrm{max}=890$ nm for the $Hp$ bandpass,
$\lambda_\textrm{min}=350$ nm and $\lambda_\textrm{max}=505$ nm for the $B_T$ bandpass,
$\lambda_\textrm{min}=455$ nm and $\lambda_\textrm{max}=675$ nm for the $V_T$ bandpass.
The resulting zero points were $Hp_\textrm{ZP}=-26.03$, $B_\textrm{T,ZP}=-25.42$ and $V_\textrm{T,ZP}=-25.99$.

\subsubsection{Johnson {\it B, V} synthetic magnitudes}
\label{sec:JohnsonVSyntheticMagnitudes}

The normalized Johnson's {\it B} bandpass ({\it S$_{B}$}) has been obtained from \citet{bessell2012}. We derived the zero points in the VEGA system using the same Vega spectrum we adopted for the calculation of the PLATO magnitude (Sec.\ref{sec:PLATOmagnitude}). The zero point ($B_\textrm{ZP}$) was obtained from 

\begin{equation}
B_\textrm{ZP}=2.5\log_{10}\left (
\frac{
{\int_{\lambda_{min}}^{\lambda_{max}}\,f_\textrm{Vega}(\lambda)\,S_{B}(\lambda)\,\lambda\,d\lambda}
}
{
{\int_{\lambda_{min}}^{\lambda_{max}}\,S_{B}(\lambda)\,\lambda\,d\lambda}
}
\right )+B_\textrm{Vega},
\end{equation}

\noindent
where $B_{\textrm{Vega}}=0.023$, $\lambda_\textrm{min}=360$ nm and $\lambda_\textrm{max}=560$ nm. The resulting zero point was $B_\textrm{ZP}=-25.47$.

\noindent
The normalized Johnson's $V$ bandpass ({\it S$_{V}$}) has been obtained from \citet{bessell2012}. We derived the zero points in the VEGA system using the same Vega spectrum we adopted for the calculation of the PLATO magnitude (Sec.\ref{sec:PLATOmagnitude}). The zero point ($V_\textrm{ZP}$) was obtained from 

\begin{equation}
V_\textrm{ZP}=2.5\log_{10}\left (
\frac{
{\int_{\lambda_{min}}^{\lambda_{max}}\,f_\textrm{Vega}(\lambda)\,S_{V}(\lambda)\,\lambda\,d\lambda}
}
{
{\int_{\lambda_{min}}^{\lambda_{max}}\,S_{V}(\lambda)\,\lambda\,d\lambda}
}
\right)+V_\textrm{Vega}
\end{equation}

\noindent where $V_\textrm{Vega}=0.023$, $\lambda_\textrm{min}=470$ nm and $\lambda_\textrm{max}=740$ nm. The resulting zero point was $V_\textrm{ZP}=-26.08$.

\subsection{Extinction corrected PLATO magnitude ($P_0$)}
\label{sec:extcorrpmag}

\begin{figure*}[!ht]
	\centering
	\includegraphics[width=0.39\textwidth]{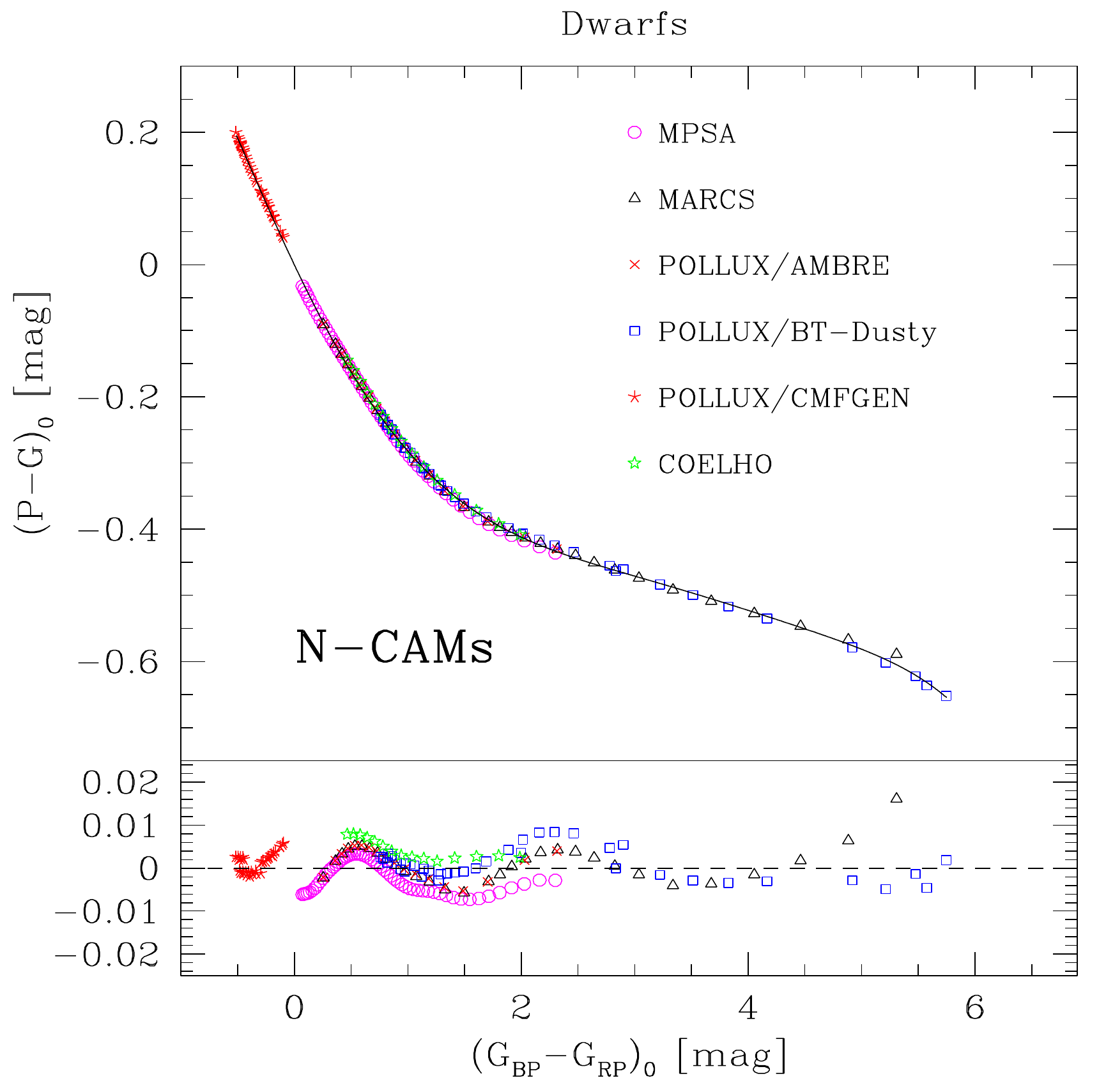}    
	\includegraphics[width=0.39\textwidth]{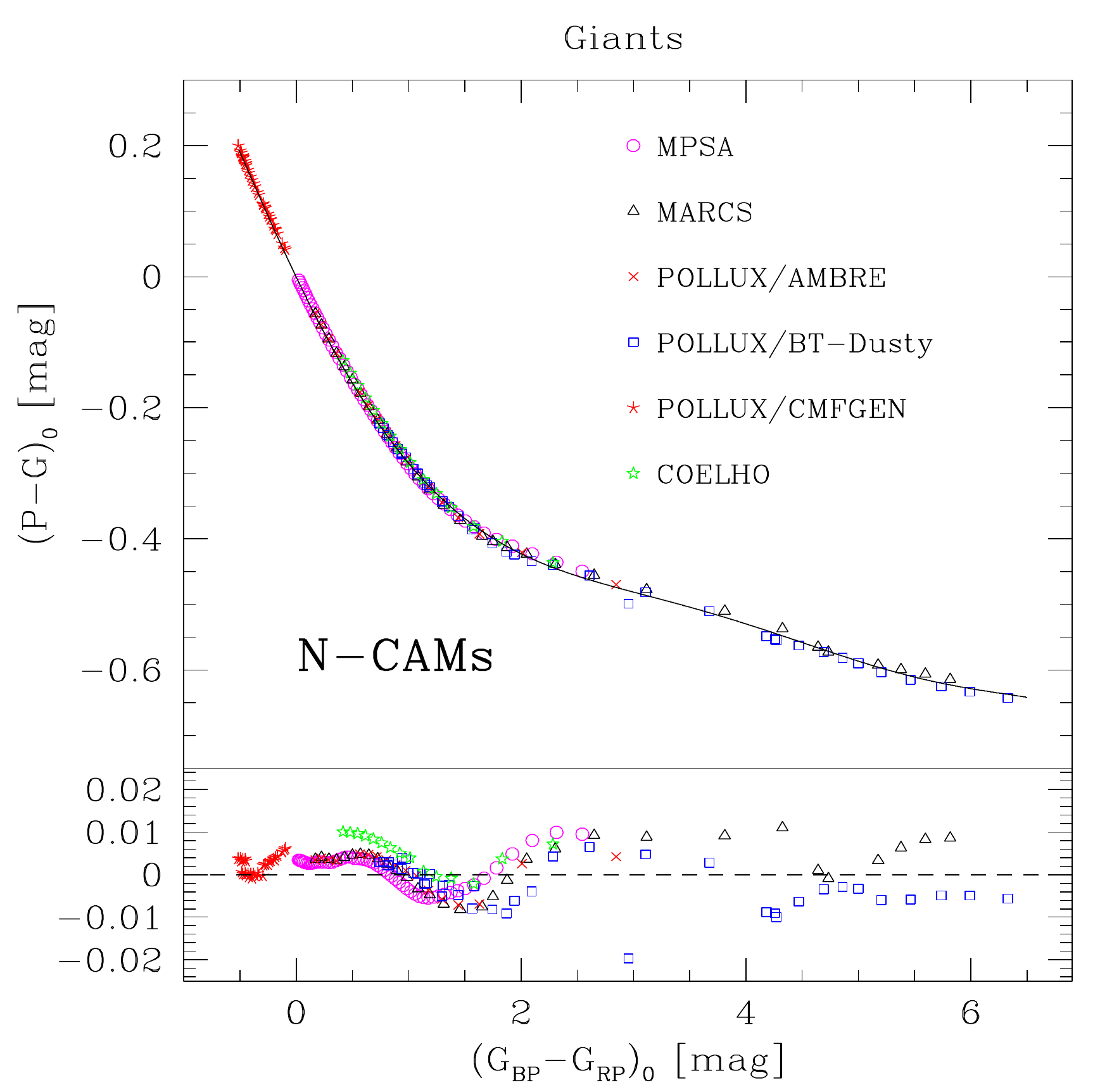}
	\includegraphics[width=0.39\textwidth]{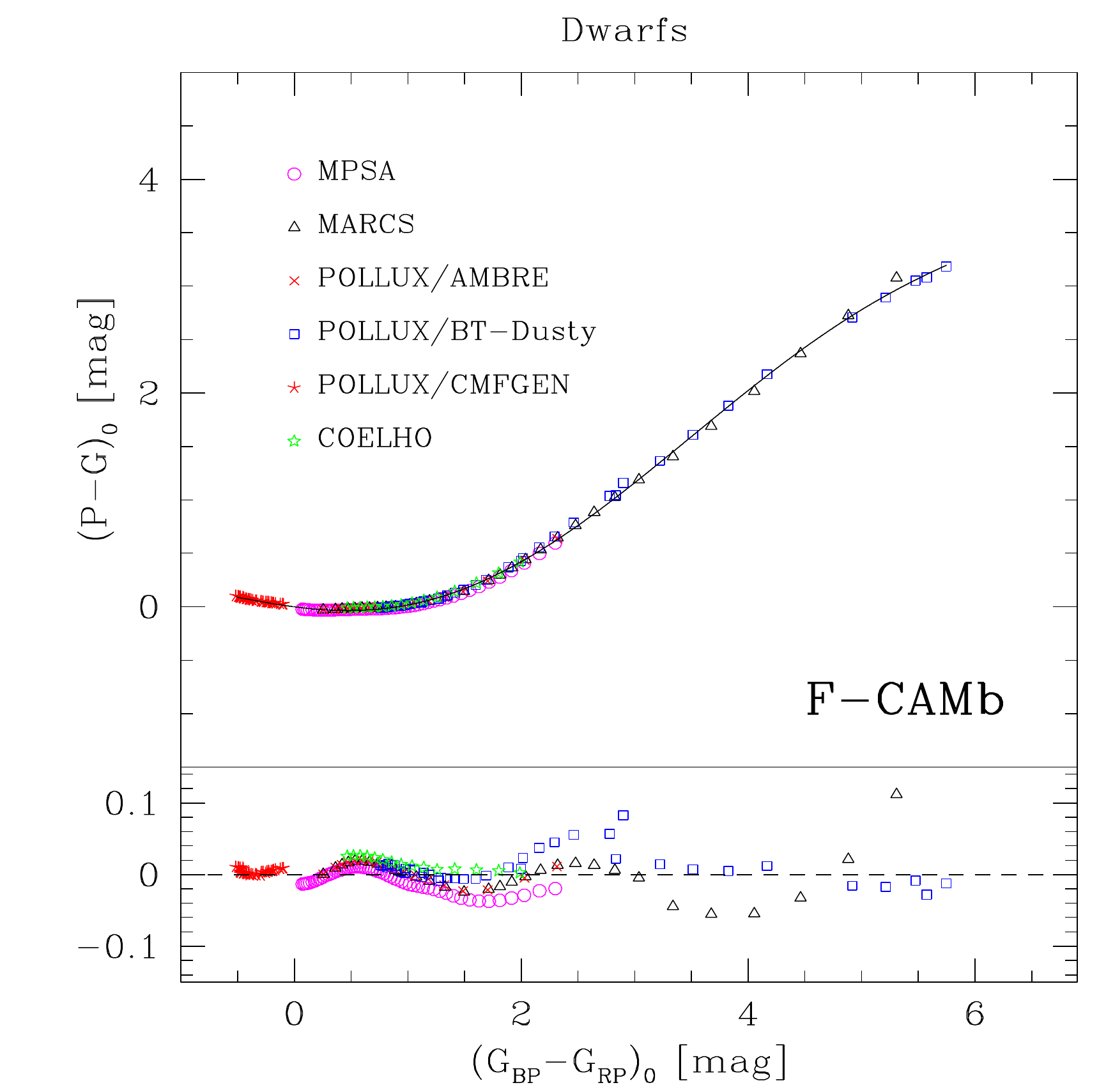}
	\includegraphics[width=0.39\textwidth]{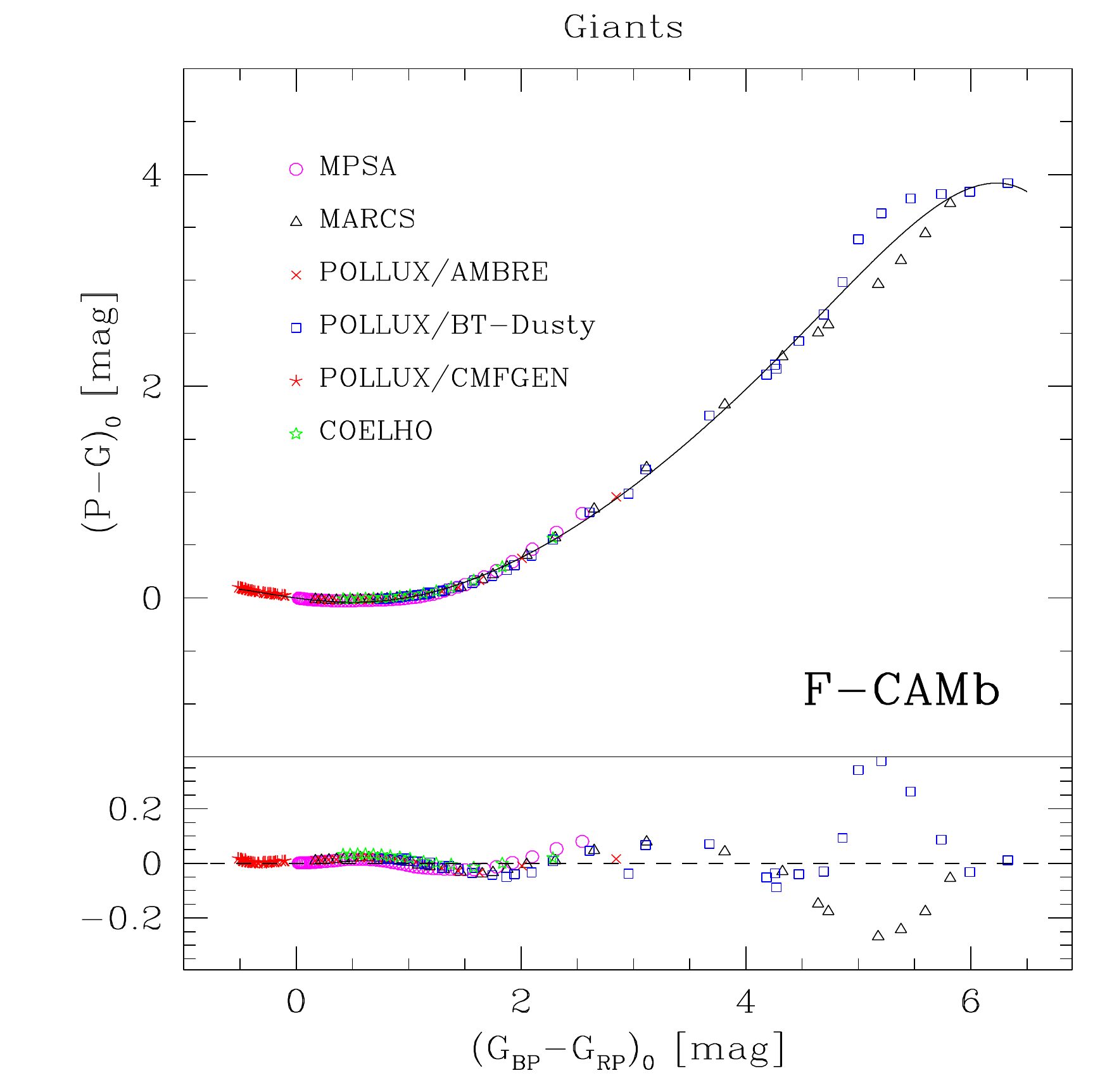}
\includegraphics[width=0.39\textwidth]
    {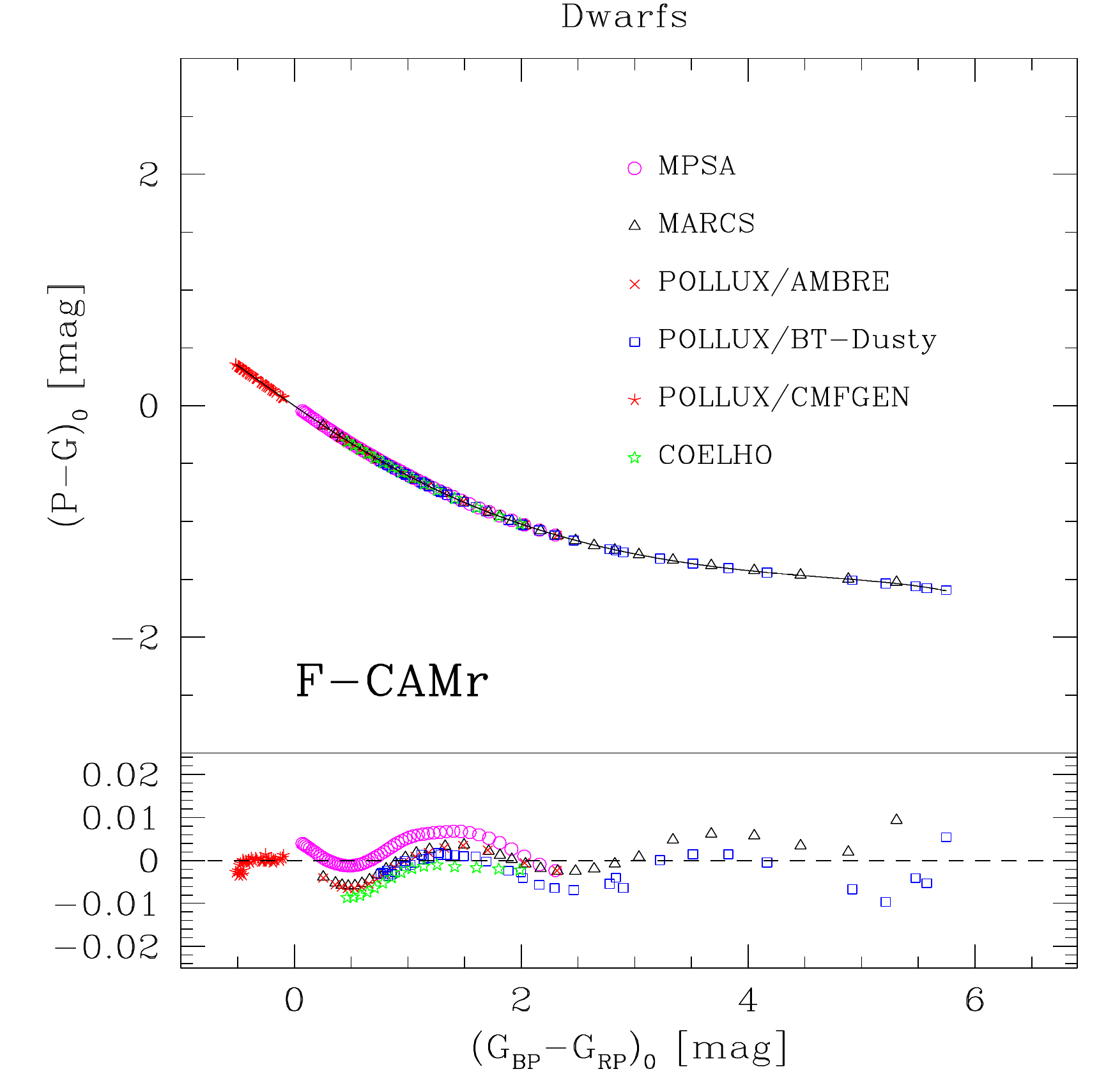}
    \includegraphics[width=0.39\textwidth]
    {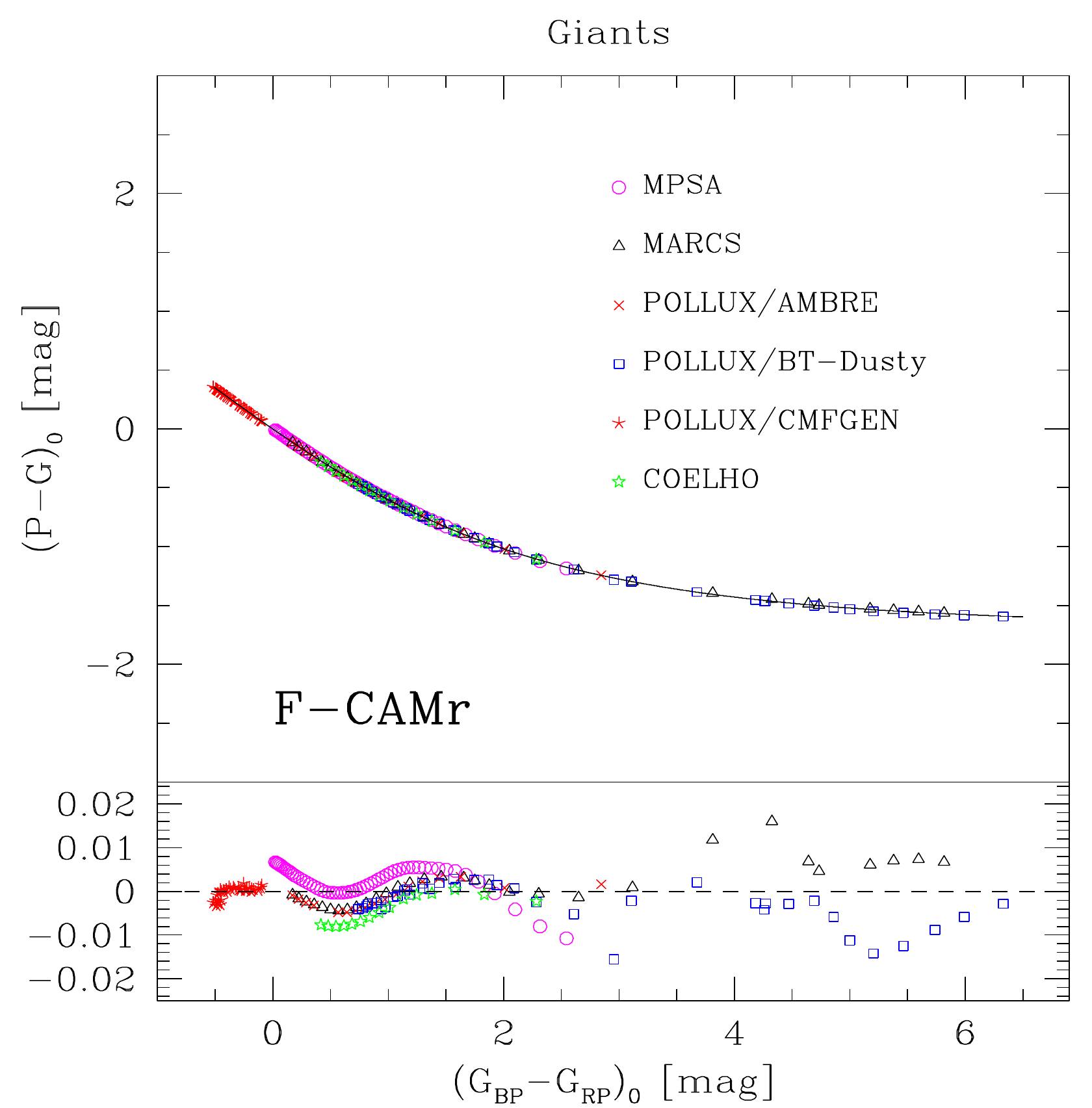}
	\caption{
    \textit{Top left}: $(P-G)_0$ vs. $(G\rm_{BP}-G\rm_{RP})_0$ synthetic two color diagram for dwarf stars models and normal cameras (N-CAMs) calculated from the MPSA, MARCS, POLLUX/AMBRE, POLLUX/BT-Dusty, POLLUX/CMFGEN and COELHO stellar libraries, as indicated by the legend. The best fit of the synthetic colors is shown with the continuous solid line.
    \textit{Bottom panel}: residuals between the synthetic color $(P-G)_0$ and the best fit interpolation model as a function of the synthetic $(G\rm_{BP}-G\rm_{RP})_0$ color.
\textit{Top right:} same as top left panel for giant stars and N-CAMs.    
\textit{Middle left:} same as top left panel for F-CAMb and dwarf stars.
\textit{Middle right:} same as top left panel for F-CAMb and giant stars.
\textit{Bottom left:} same as top left panel for F-CAMr and dwarf stars.
\textit{Bottom right:} same as top left panel for F-CAMr and giant stars.
    }
	\label{fig:PG_BPRP_DWARFS_NCAMS}
\end{figure*}

\begin{figure*}[!ht]
	\centering
	\includegraphics[width=0.39\textwidth]{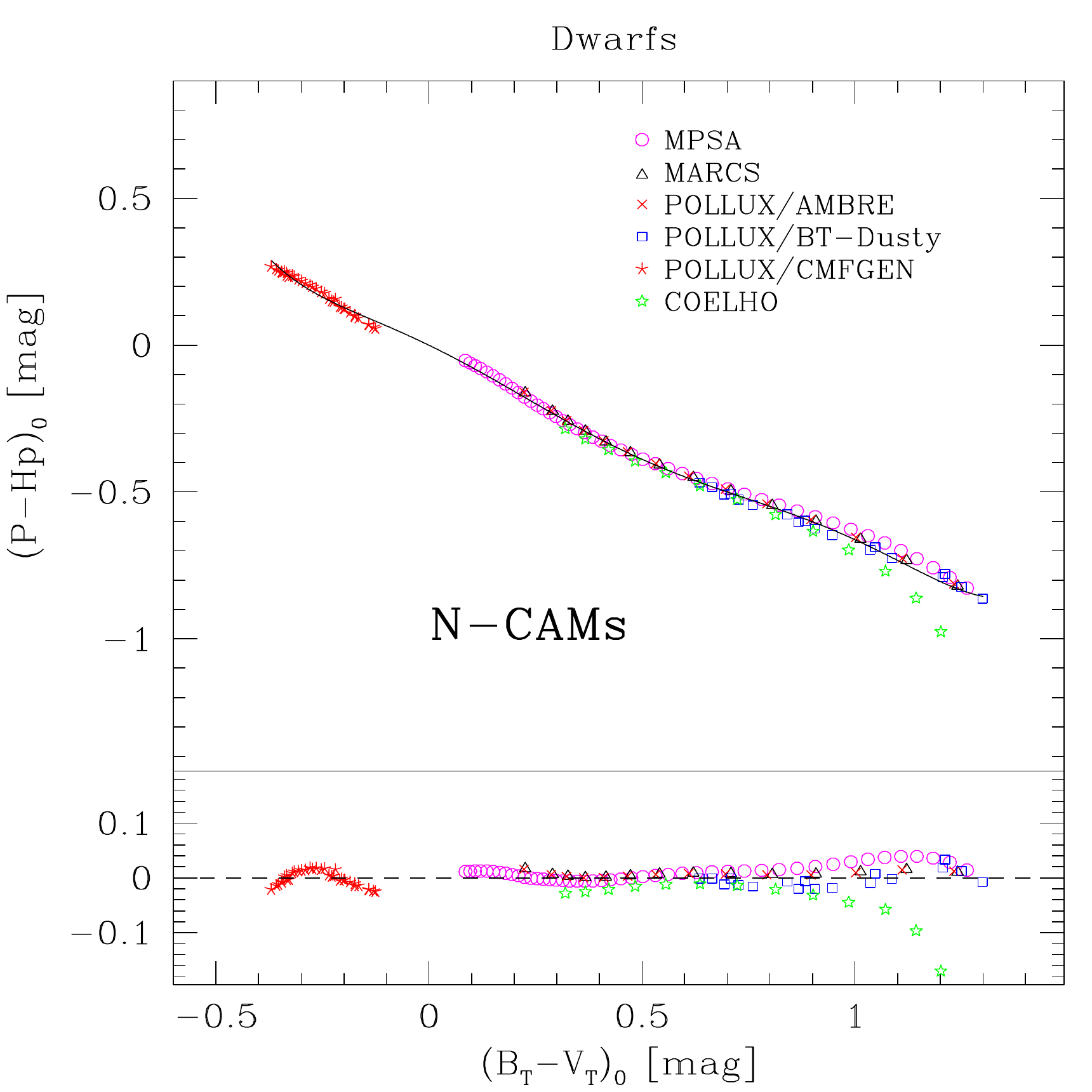}
\includegraphics[width=0.39\textwidth]{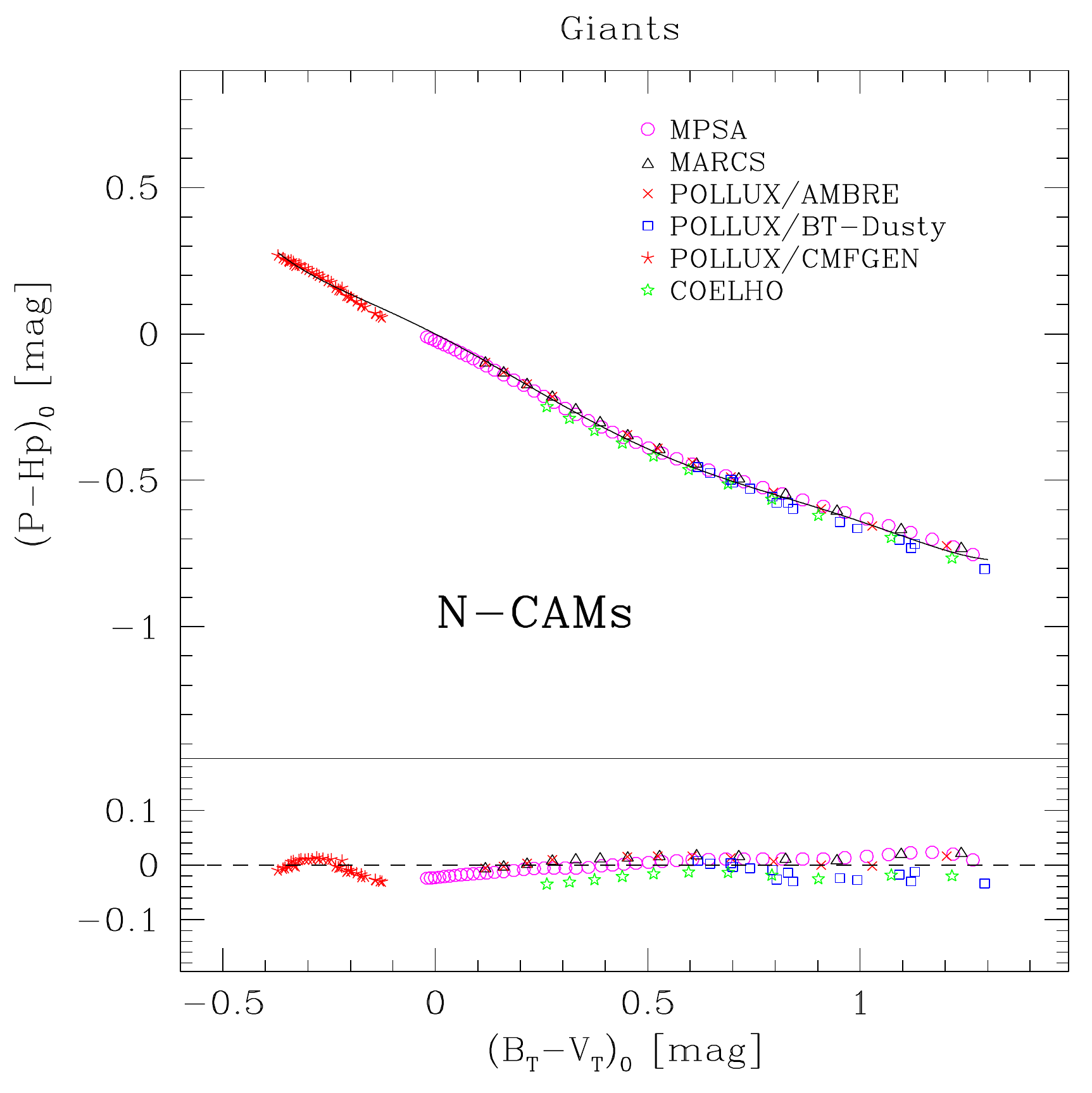}
    \includegraphics[width=0.39\textwidth]{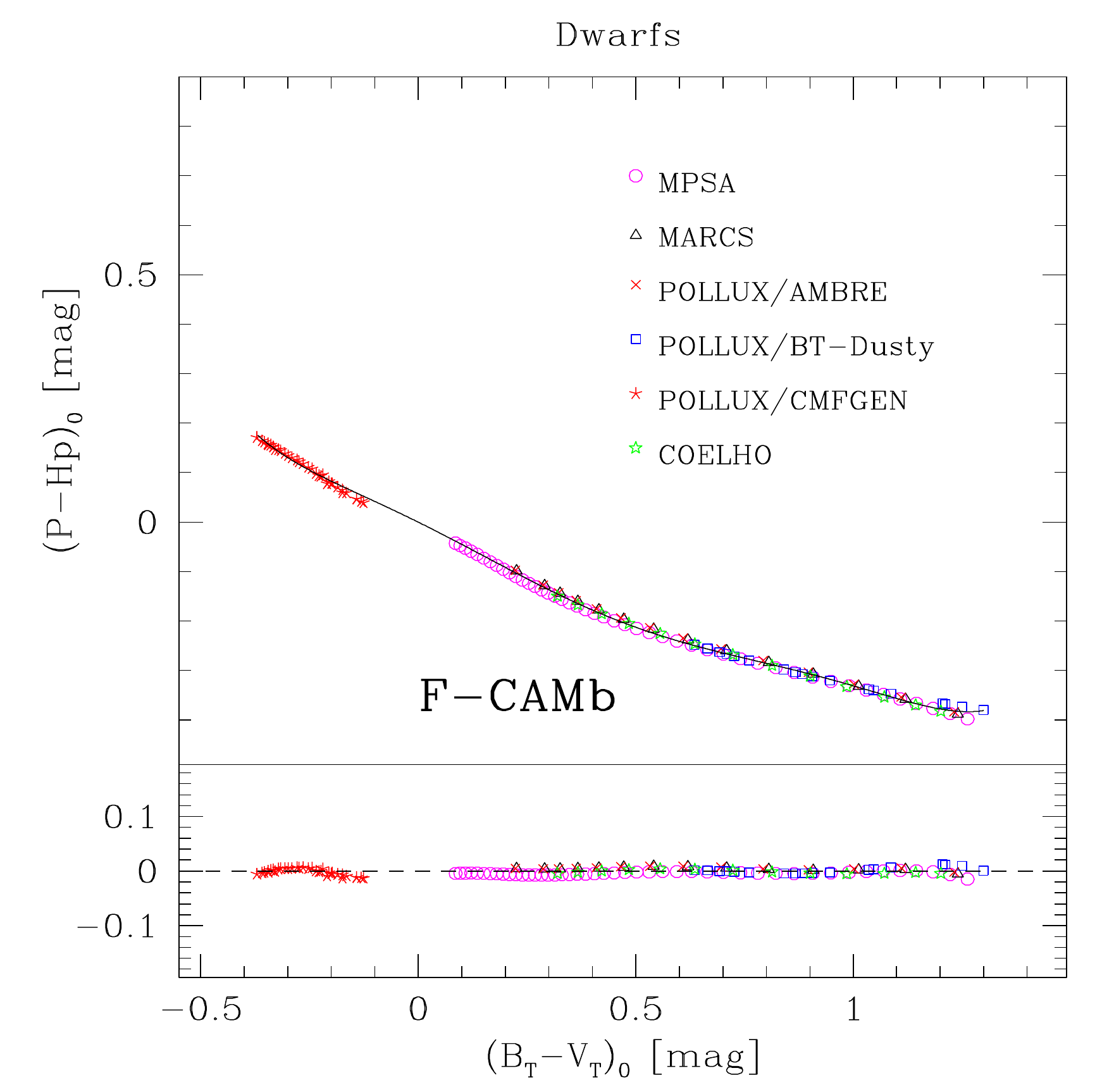}
    \includegraphics[width=0.39\textwidth]{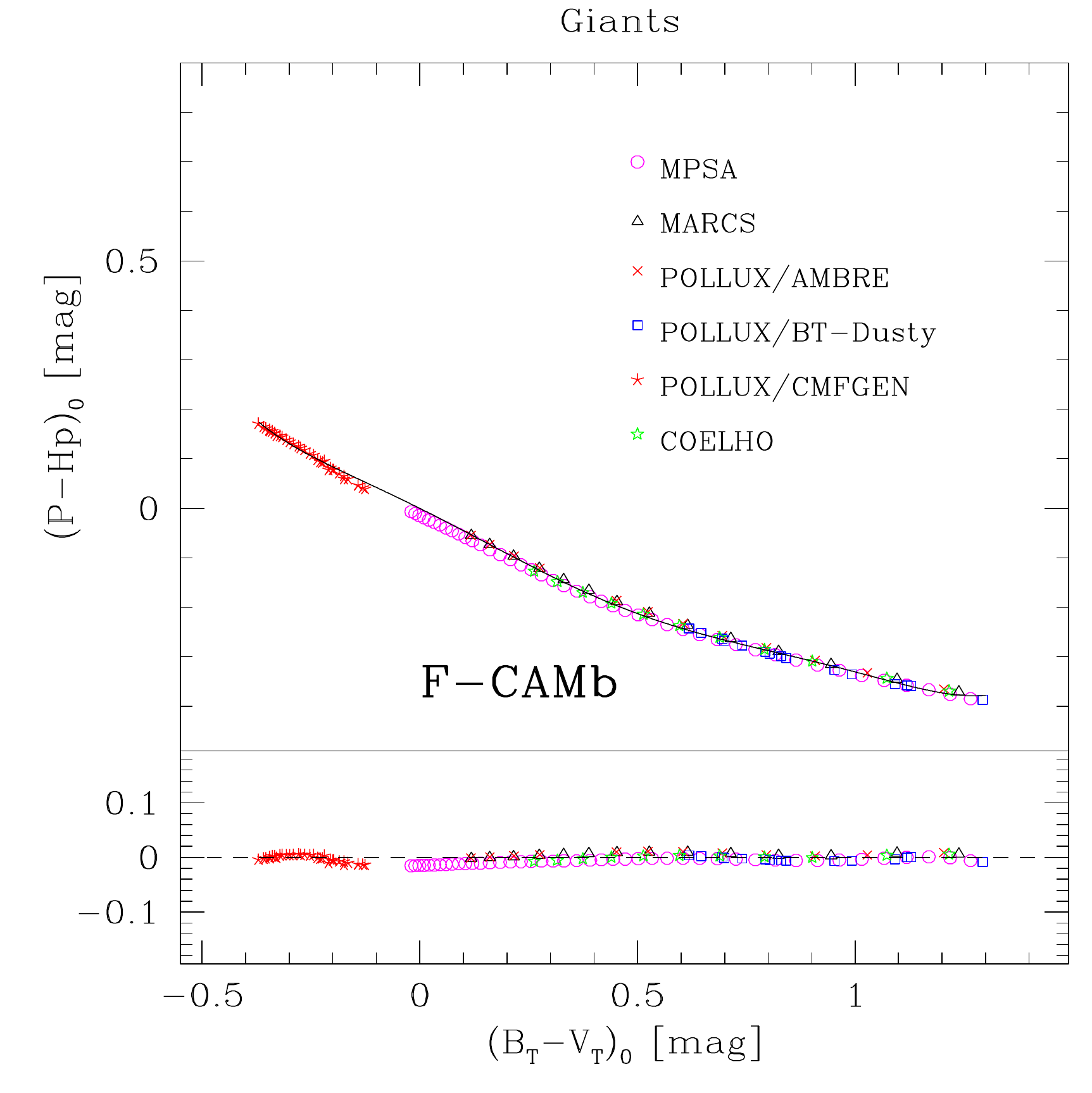}
    \includegraphics[width=0.39\textwidth]{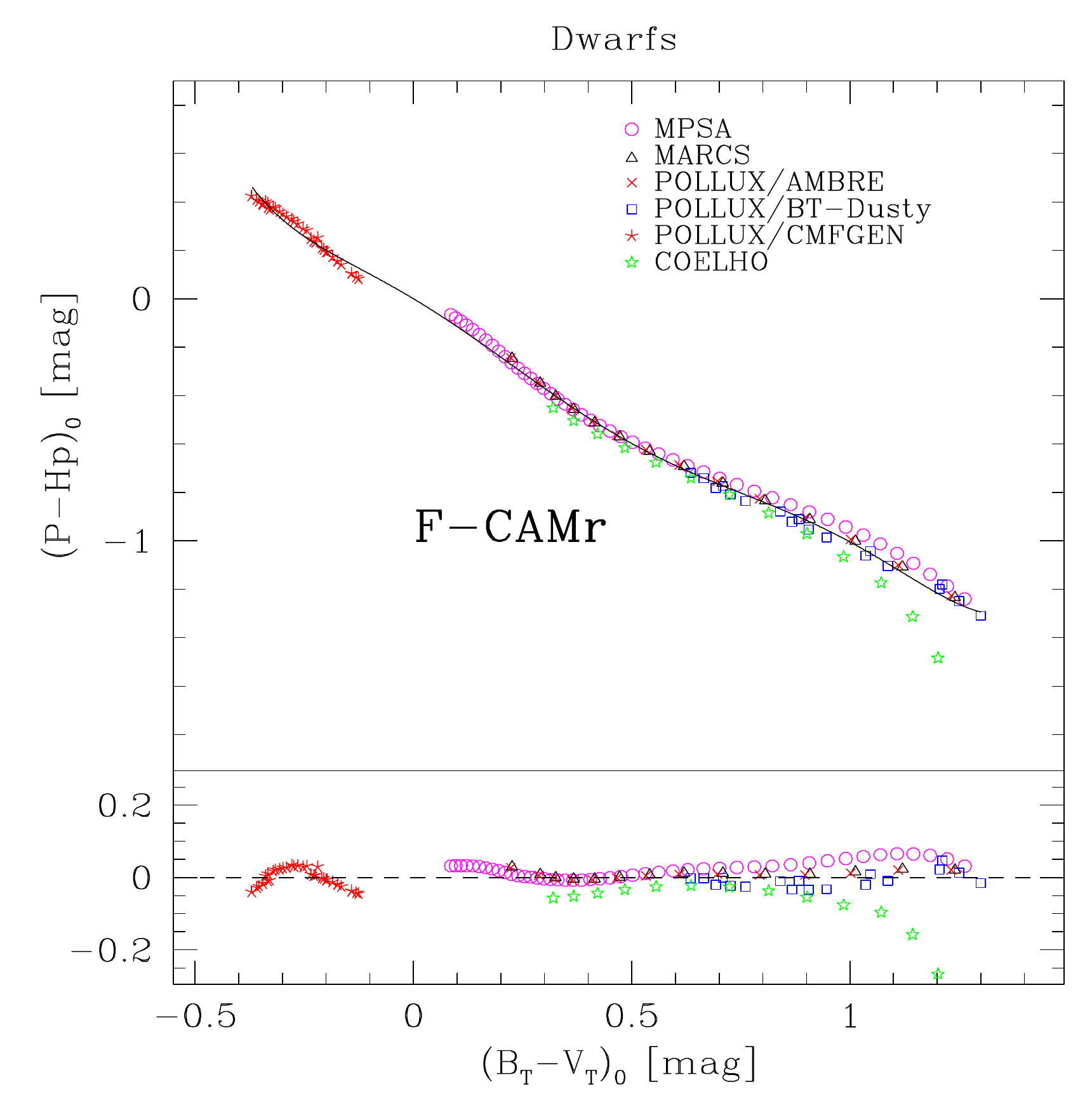}
    \includegraphics[width=0.39\textwidth]{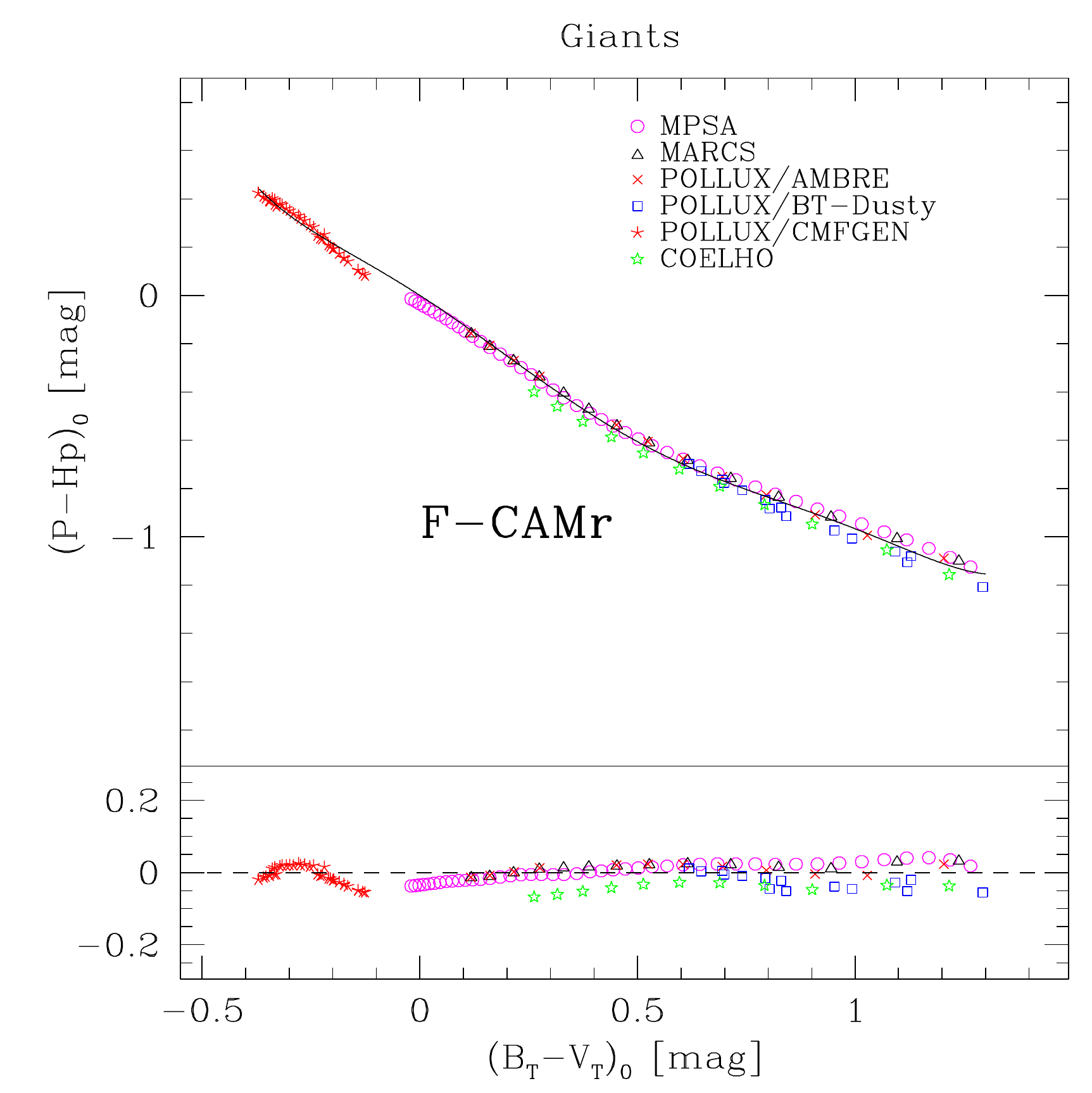}   
	\caption{
     Same as Fig.~\ref{fig:PG_BPRP_DWARFS_NCAMS}
    for the (P-Hp)$_0$ vs. ($\rm B_T$-$\rm V_T$)$_0$ synthetic two color diagram. 
	}
	\label{fig:PHp_BTVT_DWARFS_NCAMS}
\end{figure*}

\begin{figure*}[!ht]
	\centering
	\includegraphics[width=0.39\textwidth]{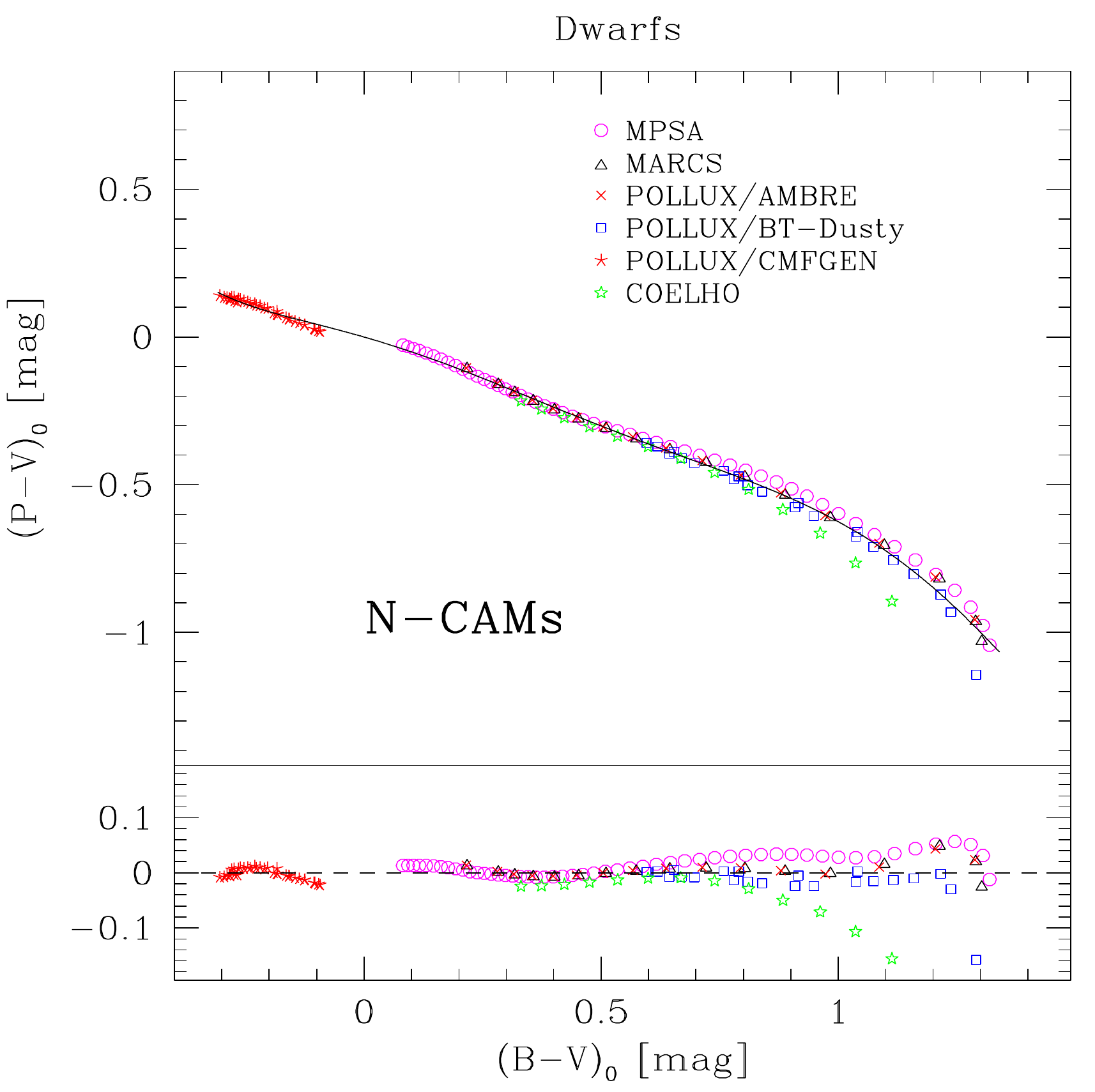}
    	\includegraphics[width=0.39\textwidth]{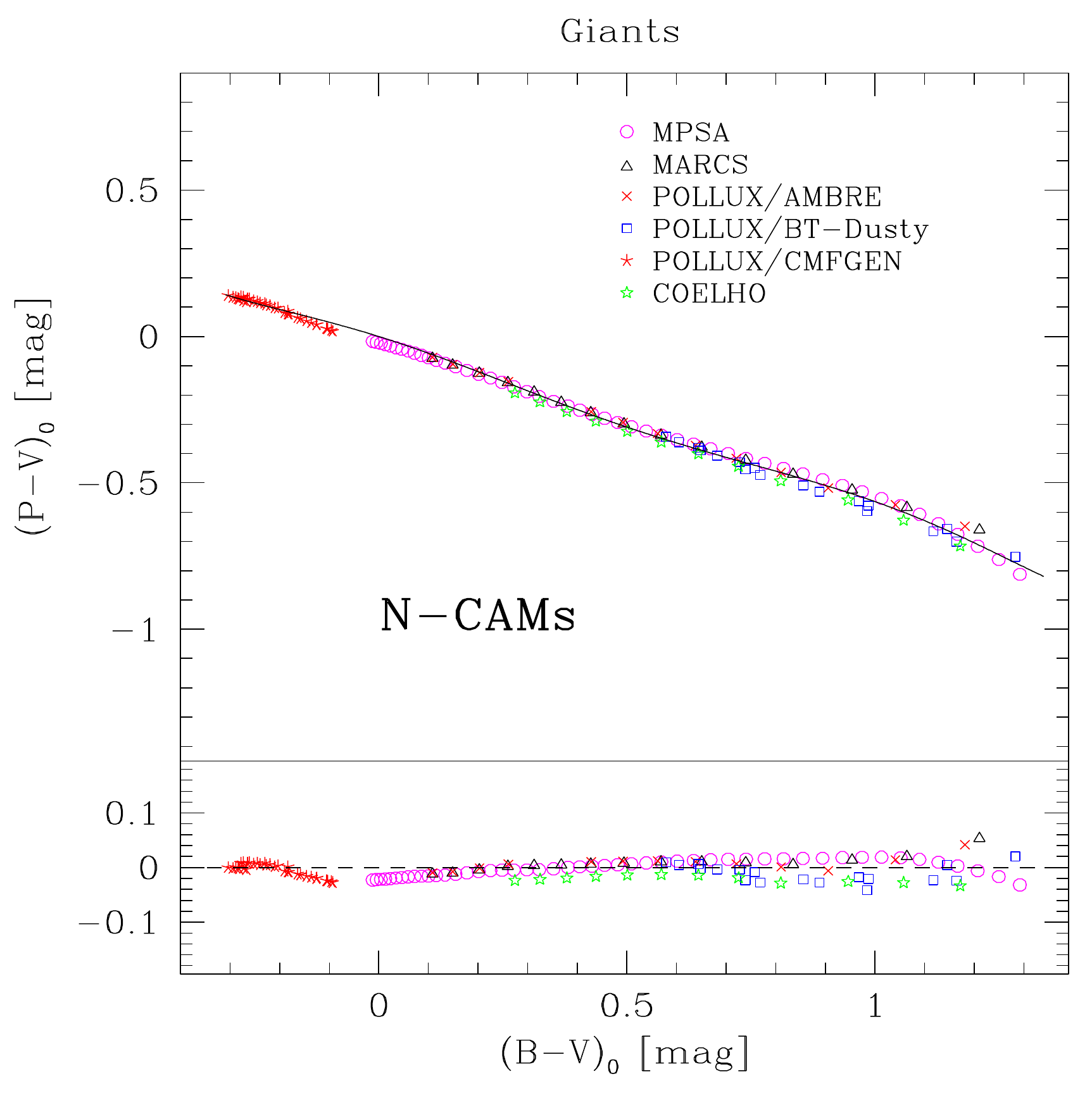}
    \includegraphics[width=0.39\textwidth]{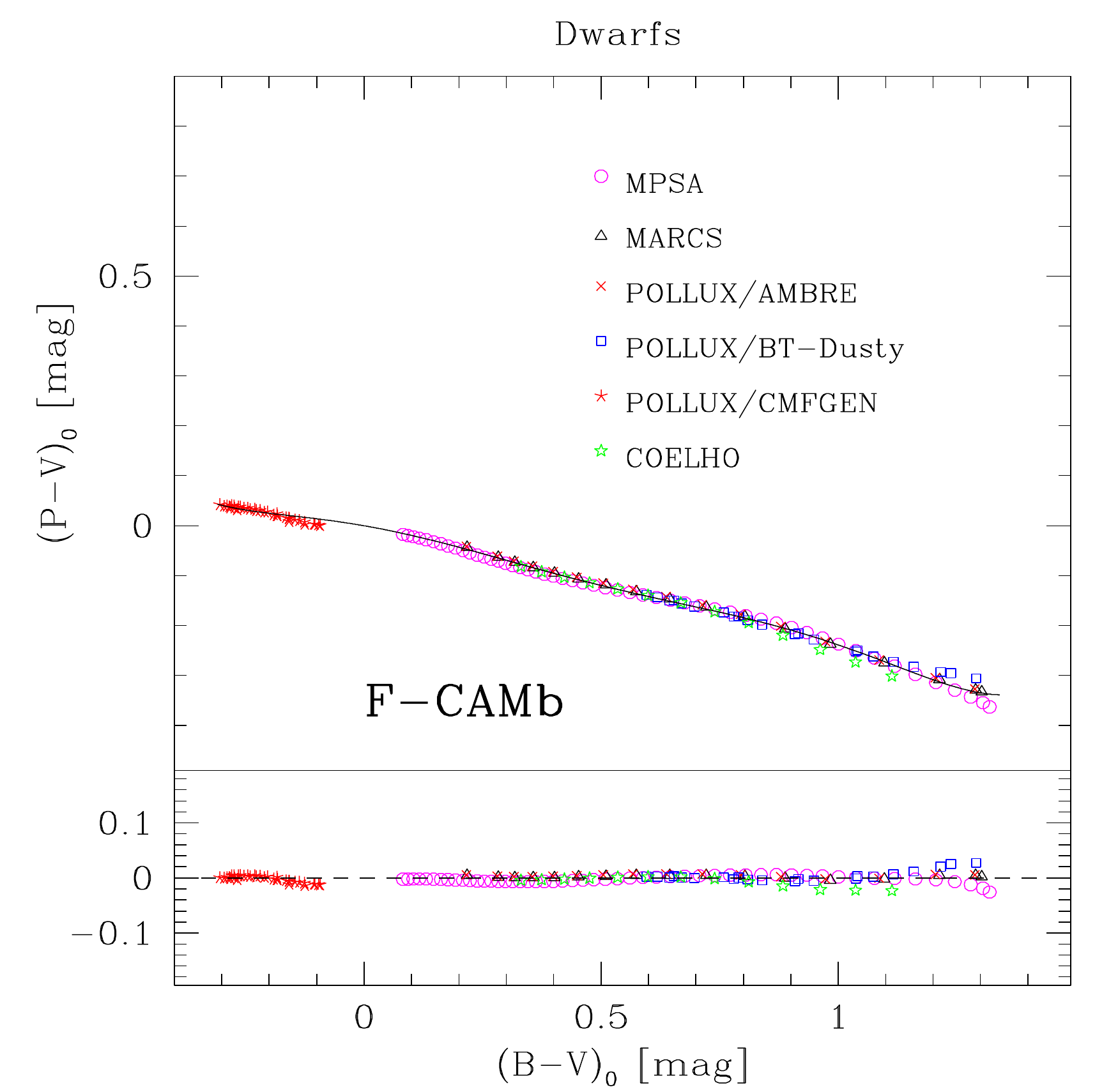}
    \includegraphics[width=0.39\textwidth]{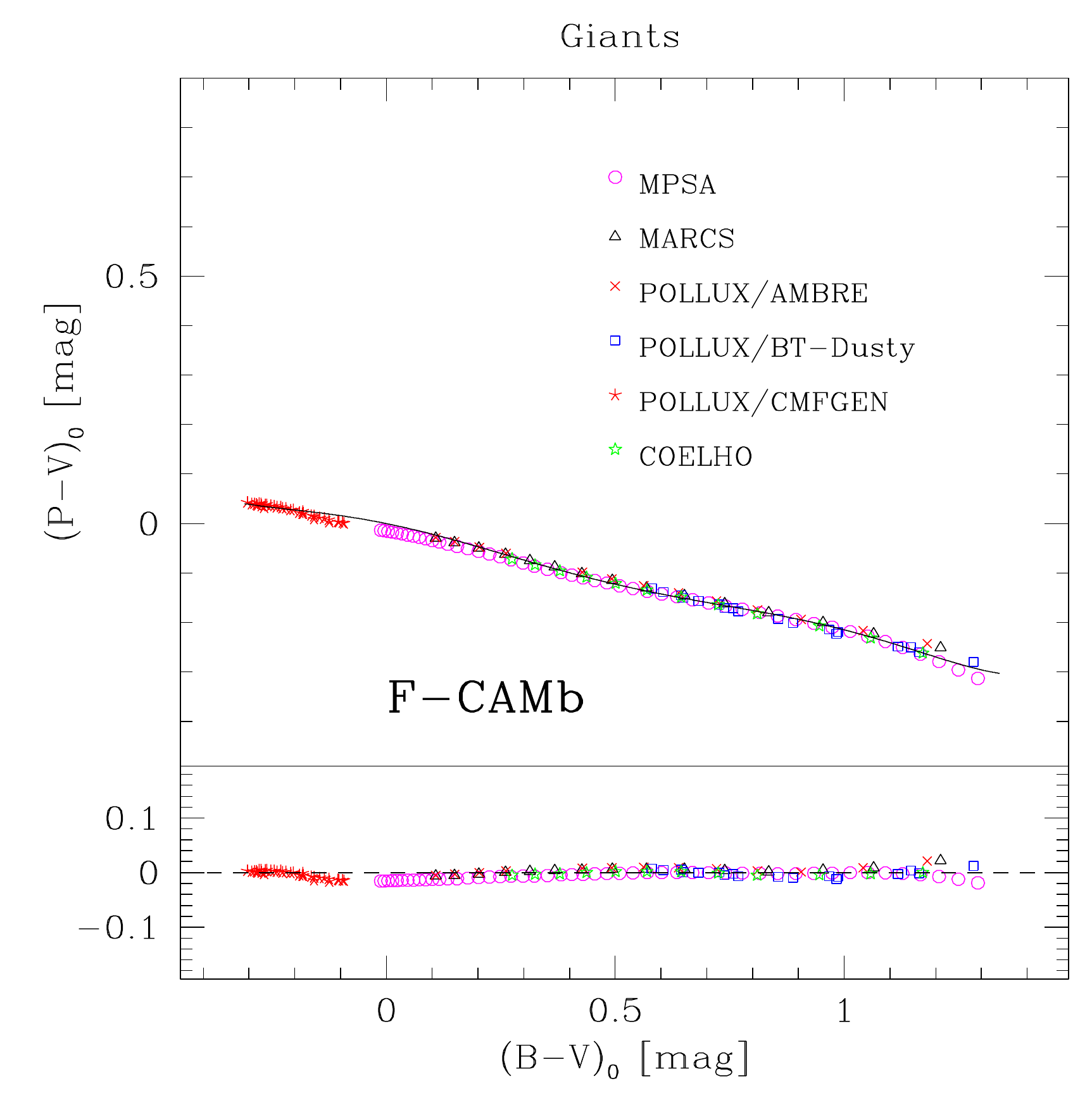}
    \includegraphics[width=0.39\textwidth]{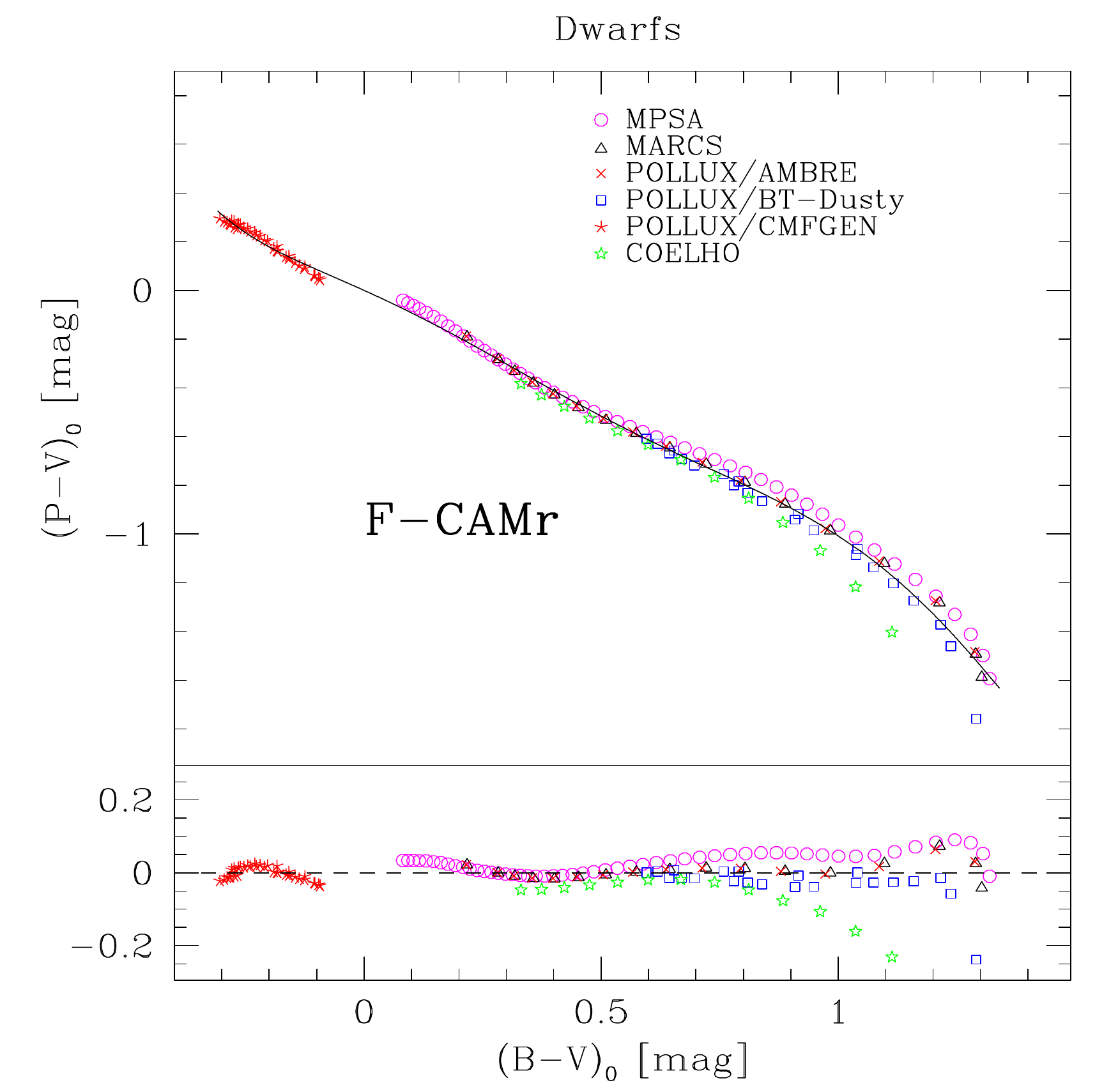}
    \includegraphics[width=0.39\textwidth]{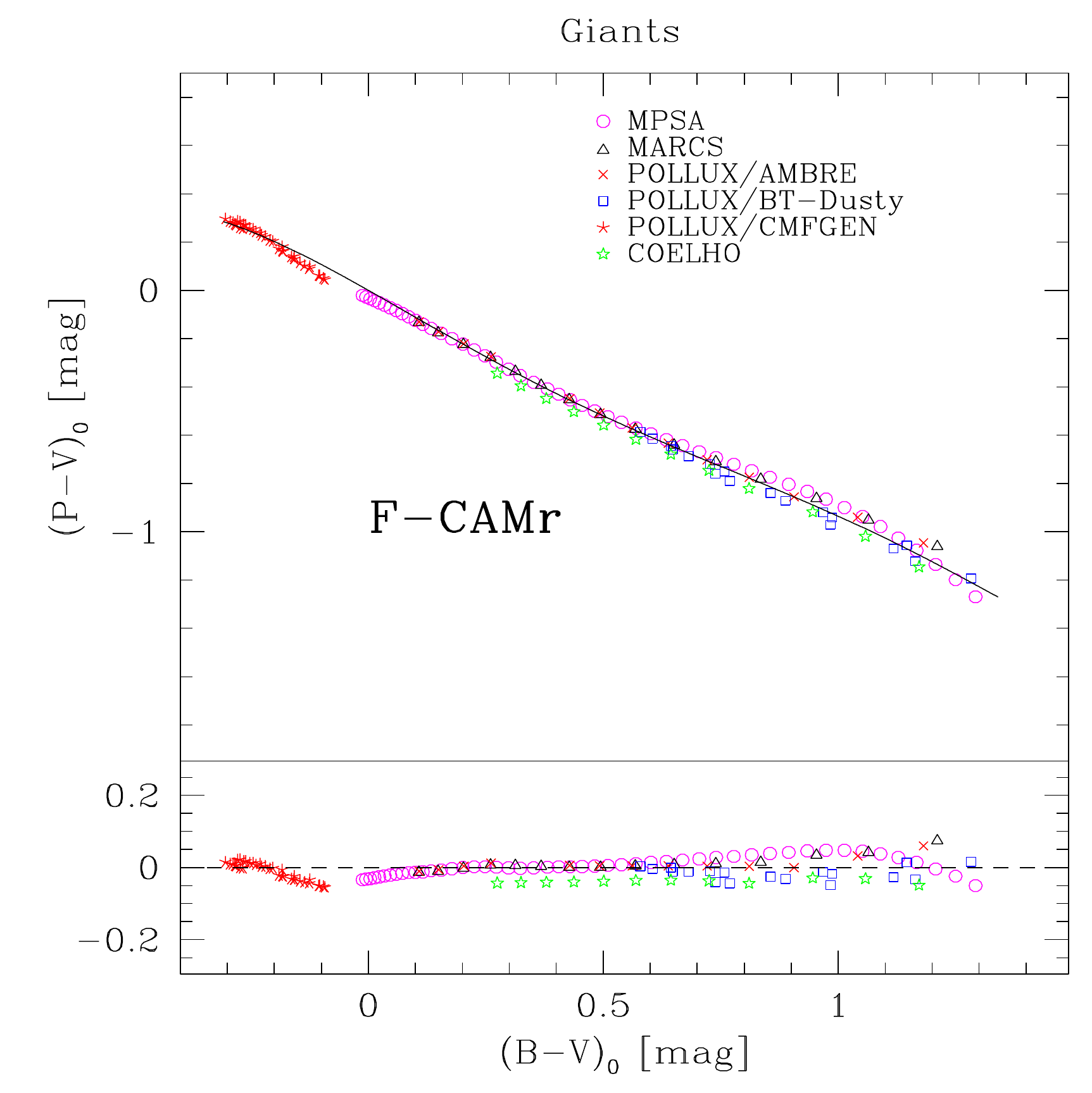}
	\caption{
    Same as  Fig.~\ref{fig:PG_BPRP_DWARFS_NCAMS} for the  (P-V)$_0$ vs. ($\rm B$-$\rm V$)$_0$ synthetic two color diagram. 
	}
	\label{fig:PV_BV_DWARFS_NCAMS}
\end{figure*}

The extinction corrected (intrinsic) PLATO band photometry ($P_0$) is therefore obtained from:

\begin{equation}
\label{eq:p0}
    P_0=X_0+\sum_{i=1}^{i=6} b_i(Color_0)^i,
\end{equation}

\noindent where $X_0$ and $Color_0$ are the adopted intrinsic photometric band and color.
Figures~\ref{fig:PG_BPRP_DWARFS_NCAMS}-\ref{fig:PV_BV_DWARFS_NCAMS} represent the different calibration relations we calculated to derive the PLATO magnitude and Tables~\ref{tab:best_fit_coefficients_dwarfs_ncams}-\ref{tab:best_fit_coefficients_dwarfs_gv_bprp} report the corresponding best-fit coefficients.

\subsection{Apparent PLATO magnitude}
\noindent
The apparent PLATO magnitude ($P$) is obtained from:

\begin{equation}\label{eq:vphot}
    P=P_0+A_P,
\end{equation}

\noindent where $A_P$ is the extinction in the PLATO band, calculated from the monochromatic extinction 
and the effective temperature.

\subsection{Uncertainty on the PLATO magnitude ($\delta P_0$)}
\label{sec:uncertaintypmag}

The uncertainty on the  intrinsic PLATO magnitude ($\delta P_0$) is obtained by error propagation with the following equation:

\begin{equation}
    \delta P_0=\delta X_0+\sum_{i=1}^{i=6} i \times b_i \times (Color)_0^{i-1} \times \delta (Color)_0,
\end{equation}

\noindent where

\begin{equation}
  \delta X_0=\sqrt{(\delta X)^2+(\delta A_X)^2}
\end{equation}

\noindent and $\delta X$ is the uncertainty in the adopted photometric band and $\delta A_X$ is the extinction uncertainty  in that photometric band and $\delta (Color)_0$ is the uncertainty in the adopted intrinsic color which, for example, in the case of the Gaia color, is given by

\begin{equation}
    \delta (G_{\rm BP}-G_{\rm RP})_0=\sqrt{(\delta G_{\rm BP})^2+(\delta G_{\rm RP})^2+(\delta E(G_{\rm BP}-G_{\rm RP}))^2}
\end{equation}

\noindent
where  $\delta G_{\rm BP}$ and $\delta G_{\rm RP}$ are the uncertainties in the Gaia DR3 photometry and  $\delta E(G_{\rm BP}-G_{\rm RP})$  is  the uncertainty  in the reddening in the $(G_{\rm BP}-G_{\rm RP})$ color, respectively.

\noindent
The uncertainty on the apparent PLATO magnitude ($\delta P$) is obtained from:

\begin{equation}
    \delta P=\sqrt{(\delta P_0)^2+(\delta A_P)^2}
\end{equation}

\noindent
where $\delta A_P$ is the uncertainty in the PLATO band extinction (Sect.~\ref{sec:uncertainty}).

\begin{table*}   
	\centering
     \caption{Best-fit coefficients of the  $(P-G)_0$ vs. $(G_{\rm BP}-G_{\rm RP})_0$ relations. 
   }
	\begin{tabular}{c c c c c c c c c}
	\hline
    \hline
 CAMS & type & $b_1$ & $b_2$ & $b_3$ & $b_4$ & $b_5$ & $b_6$ & $(G_{\rm BP}-G_{\rm RP})_0$ \\
 & & -- & (mag$^{-1}$) & (mag$^{-2}$) & (mag$^{-3}$) & (mag$^{-4}$) & (mag$^{-5}$) & min, max (mag)\\
    \hline
 N-CAMs & dwarfs & -0.3613390    &    0.0632494   &     0.0301607   &  -0.0163962   &     0.0027984   &    -0.0001679 & -0.51, 5.75 \\
 F-CAMb & dwarfs & -0.1386193   &     0.1103836   &     0.0582385    &   -0.0144120   &     0.0006554     &   0.0000251 & -0.51, 5.75 \\
F-CAMr & dwarfs & -0.6795686   &     0.0539941   &     0.0331913   &    -0.0123407    &    0.0019006    &   -0.0001174 & -0.51, 5.75\\ 
  N-CAMs & giants & -0.3586933    &    0.0598219   &    0.0244786    &  -0.0119261    &    0.0017487   &    -0.0000870 & -0.51, 6.33\\
  F-CAMb & giants & -0.1448778    &    0.0940880   &     0.0880533   &    -0.0369099    &    0.0064762    &   -0.0004327  & -0.51, 6.33\\
   F-CAMr & giants & -0.6773443   &     0.0550916    &    0.0294329    &   -0.0095837    &    0.0011505   &    -0.0000510 & -0.51, 6.33\\
   \hline
   \end{tabular}
   \label{tab:best_fit_coefficients_dwarfs_ncams}
\end{table*}

\begin{table*}    
	\centering
    \caption{Best-fit coefficients of the $(P-Hp)_0$ vs. $(B_T-V_T)_0$ relationship. }
	\begin{tabular}{c c c c c c c c c}
	\hline
    \hline
 CAMS & type & $b_1$ & $b_2$ & $b_3$ & $b_4$ & $b_5$ & $b_6$ & $(B_{T}-V_{T})_0$ \\  
  & & -- & (mag$^{-1}$) & (mag$^{-2}$) & (mag$^{-3}$) & (mag$^{-4}$) & (mag$^{-5}$) & min, max (mag)\\
     \hline
   N-CAMs & dwarfs & -0.7037266 & -0.4784983 & -0.1113304 & 3.1211357 & -3.7857743  & 1.2958674 &
   -0.37, 1.3\\
   F-CAMb & dwarfs & -0.4346966 & -0.1644953  & 0.0492126 & 1.3866888 & -1.8215308 & 0.6540379 & -0.37, 1.3 \\ 
   F-CAMr & dwarfs & -1.0811309 & -0.7144800 & -0.3403424 & 5.1274203 & -6.0248121 & 2.0286010 & -0.37, 1.3\\
    N-CAMs & giants & -0.7414054  & -0.3938972  &  0.0830866 & 2.1720965  & -2.6647386 &  0.9047329 & -0.37, 1.3 \\
    F-CAMb & giants & -0.4374622 & -0.1402522 & 0.0529479 & 1.1891877 & -1.5414390 &  0.5466525 & -0.37, 1.3\\
     F-CAMr & giants &  -1.1623179 & -0.5898588  &   0.1205326   &  3.5051465   & -4.2977908   &  1.4575598 & -0.37, 1.3 \\
   \hline
  \end{tabular}
   \label{tab:best_fit_coefficients_hipparcos}
\end{table*}

\begin{table*}  
	\centering
    \caption{Best-fit coefficients of the $(P-V)_0$ vs. $(B-V)_0$ relation.}  
	\begin{tabular}{c c c c c c c c c}
	\hline
    \hline
 CAMS & type & $b_1$ & $b_2$ & $b_3$ & $b_4$ & $b_5$ & $b_6$ & $(B_{T}-V)_0 $ \\
 & & -- & (mag$^{-1}$) & (mag$^{-2}$) & (mag$^{-3}$) & (mag$^{-4}$) & (mag$^{-5}$) & min, max (mag) \\
     \hline
 N-CAMs & dwarfs & -0.4603150  & -0.3788919 & -0.5312191 & 2.4784103 & -2.4168036 & 0.6840207 & -0.3 1.3 \\
F-CAMb & dwarfs & -0.1651463 & -0.2767556  & -0.0625853 & 1.2844708 & -1.5560036 &  0.5374649 & -0.3, 1.3 \\ 
  F-CAMr & dwarfs &  -0.8750680 & -0.3658863 & -1.2473050 & 4.5093262 & -4.2180117 & 1.1888158 & -0.3, 1.3\\ 
 N-CAMs & giants & -0.5297327  & -0.4101526  &  0.0586109  &  1.6082424   &  -1.8874444  &  0.5968343  & -0.3, 1.3 \\  
  F-CAMb & giants & -0.1883739   &   -0.2787896   &  0.1182966     &   0.9240725     &   -1.1974594     &   0.4076061 & -0.3 1.3 \\
   F-CAMr & giants & -1.0984071  &  -0.2152440  &  0.9783525    &   -0.6119324   &  -0.0957097  &   0.1072434 & -0.3 1.3\\
   \hline
  \end{tabular}
   \label{tab:best_fit_johnson}
\end{table*}

\begin{table*}[!h]
	\centering
    \caption{Best-fit coefficients of the  $(G-V)_0$ vs. $(G_{\rm BP}-G_{\rm RP})_0$ relation.}
	\begin{tabular}{c c c c c c c c}
	\hline
    \hline
  type & $b_1$ & $b_2$ & $b_3$ & $b_4$ & $b_5$ & $b_6$ & $(G_{\rm BP}-G_{\rm RP})_0$ \\
    & -- & (mag$^{-1}$) & (mag$^{-2}$) & (mag$^{-3}$) & (mag$^{-4}$) & (mag$^{-5}$) & min, max (mag)\\
    \hline
  dwarfs & -0.0085337 & -0.2422638 & 0.0493499 & -0.0161423 & 0.0038450 & -0.0003356 & -0.51, 5.75\\
 giants & 0.0034037 & -0.2155382 & 0.0017512 & 0.0123098 & -0.0031739 & 0.0002454 & -0.51, 6.33\\
    \hline
    \end{tabular}
   \label{tab:best_fit_coefficients_dwarfs_gv_bprp}
\end{table*}

\begin{figure}
	\centering
\includegraphics[width=\columnwidth]{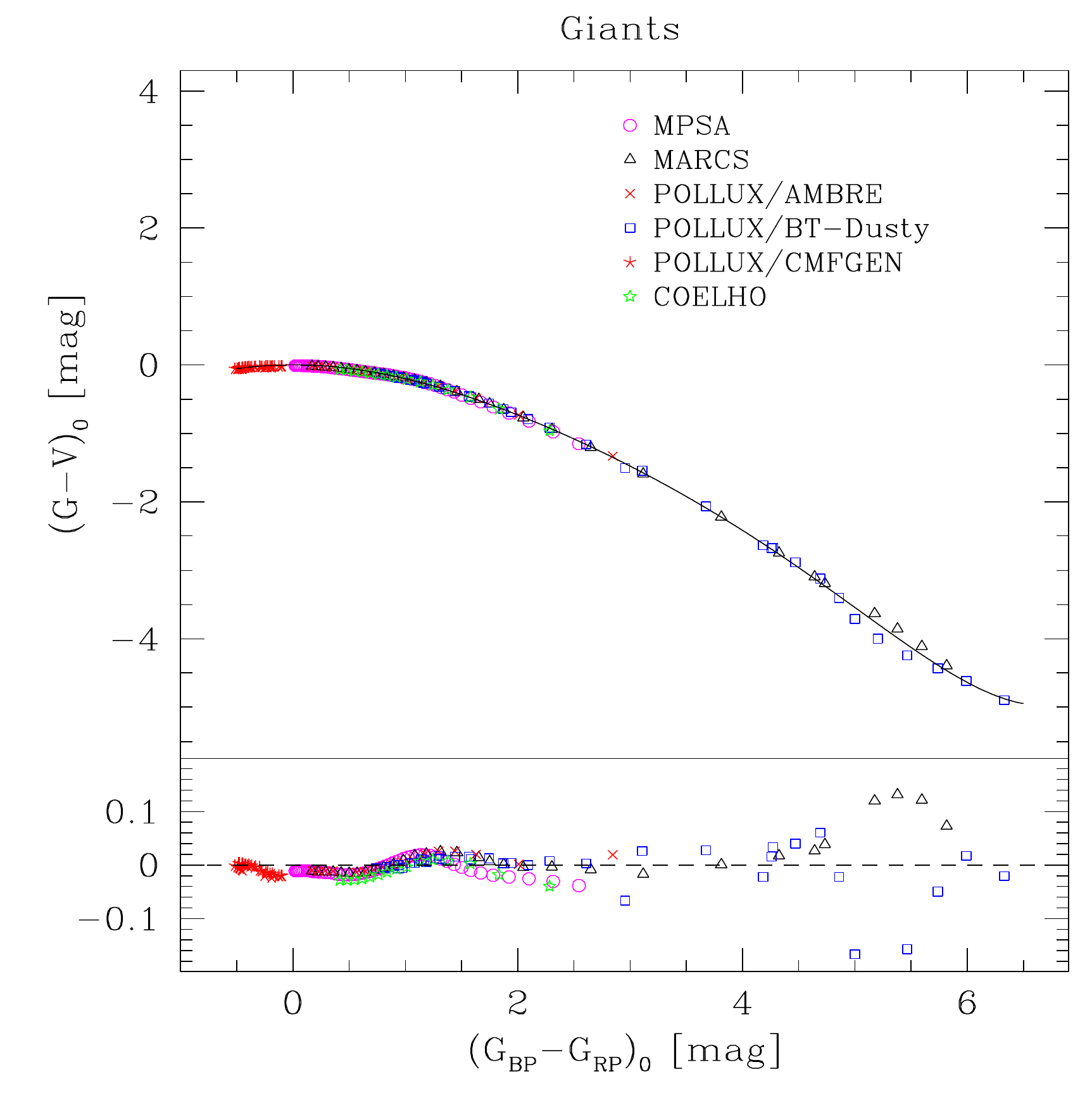}
	\caption{
    \textit{Top panel}: 
    $(G-V)_0$ vs. $(G_\mathrm{BP}-G_\mathrm{RP})_0$ synthetic two colors diagram for giant stars models calculated from the MPSA, MARCS, POLLUX/AMBRE, POLLUX/BT-Dusty, POLLUX/CMFGEN and COELHO stellar libraries, as indicated by the legend. The best fit of the synthetic colors is shown with the continuous solid line. \textit{Bottom panel}: Residuals between the synthetic color $(G-V)_0$ and the best-fit interpolation model as a function of the synthetic $(G_\mathrm{BP}-G_\mathrm{RP})_0$ color.}
	\label{fig:GV_BPRP_GIANTS}
\end{figure}

\section{Flags in PIC}
\label{sec:flags}

We introduced some flags in tPIC which are useful for identifying stellar samples or properties of the objects we selected. The definition of these flags is reported below, and a comprehensive overview including all the bitmask values defined in Sects.~\ref{sec:main_source_flag}, \ref{sec:source_flag}, \ref{sec:nss_flag}, \ref{sec:quality_flag}, and \ref{sec:planet_flag} are reported in Table~\ref{tab:pic_bitmask}.

\subsection{Main source flag}
\label{sec:main_source_flag}

The main source flag (PICmainSourceFlagBOL) is a bitmask (Table~\ref{tab:pic_bitmask}) indicating to which subPIC an object belongs to and if it is part of the proprietary sample. The proprietary sample will be defined only after launch, so its corresponding flag is null in the present version of the PIC.
 In this case the NSR is calculated at Beginning of Life (BOL).

\subsection{Source flag}
\label{sec:source_flag}

The source flag (tPICsourceFlagNCAM\_BOL) is a bitmask (Table~\ref{tab:pic_bitmask}) indicating which tPIC sample an object belongs to. The NSR used to define the P1, P2 and P5 samples is calculated at the Beginning of Life (BOL).
The PIC includes also the NSR calculated at the End of Life (EOL).

\subsection{Non-Single Star flag}
\label{sec:nss_flag}

The Non-Single Star flag (NSSflag) is a bitmask (Table~\ref{tab:pic_bitmask}) indicating the presence of binary/multiple stellar systems.

\subsection{Quality flag}
\label{sec:quality_flag}

The quality flag (qualityFlag) is a bitmask (Table~\ref{tab:pic_bitmask})  containing indicators for issues in Gaia astrometry and photometry. The numerical thresholds we adopted correspond to the 95$\rm ^{th}$ percentile of the cumulative distribution of each parameter for all sources with $G\le13$.  For the definition of the Gaia astrometry and photometry quality indicators we invite the reader to consult the 
Gaia documentation\footnote{\url{https://gea.esac.esa.int/archive/documentation/GDR3/index.html}}.
The parameter beta is the blending fraction defined as in Sect. 9.3 of
\citet{riello2021} and $C^{*}$, $\sigma_{C^{*}}$ are the corrected BP and RP flux excess and the 1-$\rm\sigma$ scatter as described in Sect. 9.4 of \citet{riello2021}.

\subsection{Planet flag}
\label{sec:planet_flag}
The planet flag (tPICplanetFlag) is a bitmask (Table~\ref{tab:pic_bitmask}) indicating the presence of at least one confirmed or candidate planet according to the discovery method. 
The caseFlag is obtained for all stars regardless of the presence of any planet. The tPICplanetFlag is calculated only for planet host stars and includes information only on the discovery method of a given confirmed/candidate planet.
If a planet is detected via multiple methods only the discovery method is reported in the tPICplanetFlag. If, for the same host, different planets are discovered by different methods all the corresponding bits are set in the tPICplanetFlag.

\begin{table}
    \centering   
   \caption{tPIC bitmask values.}
    \begin{tabular}{ll}
    \hline
    \hline
    \\
    \multicolumn{2}{c}{PICmainSourceFlagBOL}\\
     Bitmask value & Meaning  \\
    \hline
    $0000000_2$ = 0 & reset all bits \\
    $0000001_2$ = 1 & Prime sample star \\
    $0000010_2$ = 2 & Proprietary Sample Star \\
    $0000100_2$ = 4 & tPIC star \\
    $0001000_2$ = 8 & fgPIC star \\
    $0010000_2$ = 16 & cPIC star \\
    $0100000_2$ = 32 & scvPIC star \\
    $1000000_2$ = 64 & colorsample star \\
    \hline
     & \\
    \multicolumn{2}{c}{tPICsourceFlagNCAM\_BOL}\\
     Bitmask value & Meaning  \\
    \hline    
    $0000000_2$ = 0 & reset all bits \\
    $0000001_2$ = 1 & P1 star \\
    $0000010_2$ = 2 & P2 star \\
    $0000100_2$ = 4 & P5 star \\
    $0001000_2$ = 8 & P4 star \\
    $0010000_2$ = 16 & Candidate / confirmed planet(s) host star \\
    $0100000_2$ = 32 & Prime Sample \\
    $1000000_2$ = 64 & Proprietary Sample star \\
    \hline
         & \\
    \multicolumn{2}{c}{NSSflag}\\
     Bitmask value & Meaning  \\
    \hline    
    $000000_2 = 0$ & no NSS info available \\
    $000001_2 =  1$ & NSS \\
    $000010_2 =  2$ & wide binaries \\
    $000100_2 =  4$ & eclipsing \\
    $001000_2 =  8$ & astrometric \\
    $010000_2 = 16$ & spectroscopic \\
    \hline
         & \\
    \multicolumn{2}{c}{qualityFlag}\\
     Bitmask value & Meaning  \\
    \hline 
    $000000000_2$ = 0 & no issues \\
    $000000001_2$ =  1 & Gaia ruwe $>$ 3.6 \\
    $000000010_2$ =  2 & astrometric\_excess\_noise$>$0.61 AND astrometric\_excess\_noise\_sig $>$ 2 \\
    $000000100_2$ =  4 & ipd\_frac\_multi\_peak $>$ 14 \\
    $000001000_2$ =  8 & ipd\_frac\_harmonic\_amplitude $>$ 0.09 \\
    $000010000_2$ =  16 & ipd\_frac\_odd\_win $>$ 1 \\
    $000100000_2$ =  32 & beta $>$ 0.1 \citep{riello2021}\\
    $001000000_2$ =  64 & $\rm |C^{*}|>3\sigma_{C^{*}}$ \\
    $010000000_2$ =  128 &  The saturation correction in Appendix C.1 of \citet{riello2021} was applied \\
    $100000000_2$ =  256 & The correction in Sect. 8.4 of \citet{riello2021} was applied \\    
    \hline
         & \\
    \multicolumn{2}{c}{tPICplanetFlag}\\
     Bitmask value & Meaning  \\
    \hline 
    $000000000000_2$ = 0 & reset all bits \\
    $000000000001_2$ = 1 & At least one confirmed planet discovered by transit \\
    $000000000010_2$ = 2 & At least one candidate planet discovered by transit \\
    $000000000100_2$ = 4 & At least one confirmed planet discovered by radial velocity \\
    $000000001000_2$ = 8 & At least one candidate planet discovered by radial velocity \\
    $000000010000_2$ = 16 & At least one confirmed planet discovered by imaging \\
    $000000100000_2$ = 32 & At least one candidate planet discovered by imaging \\
    $000001000000_2$ = 64 & At least one confirmed planet discovered by  astrometry \\
    $000010000000_2$ = 128 & At least one candidate planet discovered by astrometry \\
    $000100000000_2$ = 256 & At least one confirmed planet discovered by timing \\
    $001000000000_2$ = 512 & At least one candidate planet discovered by timing \\
    $010000000000_2$ = 1024 & At least one confirmed planet discovered by disk kinematics \\
    $100000000000_2$ = 2048 & At least one candidate planet discovered by disk kinematics \\
    \hline
    \end{tabular}
   \label{tab:pic_bitmask}
   
\end{table}

\subsection{Case flag}
\label{sec:case_flag}

The flag (caseFlag) is an integer number that can take a value between [1,27]. It is used to classify the stars into dwarfs or giants and FGK or M stars based on the location on the CMD.
In some cases (e.\,g. case 7, 13, 19, 21, 22, 23, 24, 25, 26), where some crucial information is missing (e.\,g. distance  or color) we considered the star as a dwarf and  
imposed an average color. This was necessary to permit the calculation of important quantities (e.\,g. the PLATO magnitude in any possible situation).
The detailed description of each value of the flag is explained in the Table~\ref{tab:CaseFlag}.

\begin{table*}
\centering
   \caption{Description of the situations corresponding to the different values assumed by the caseFlag.}
	\begin{tabular}{cl}
	\hline
    \hline
     CaseFlag & Description \\
    \hline
      1 & FGK dwarf/subgiant selected using Gaia DR3 photometry with extinction \\
      2 & M dwarf selected using Gaia DR3 photometry with extinction \\
      3 & Dwarf/Subgiant (not FGKM) selected using Gaia DR3 photometry with extinction \\
      4 & Giant selected using Gaia DR3 photometry with extinction \\
      5 & Dwarf/Subgiant based on absolute apparent  Gaia DR3 photometry \\
      6 & Giant based on absolute apparent  Gaia DR3 photometry \\
      7 & Using only apparent G magnitude (imposed dwarf/subgiant with average color) \\
      8 & FGK dwarf/subgiant selected Hipparcos photometry and extinction \\
      9 & Dwarf/subgiant (not FGK) selected Hipparcos photometry and extinction \\
      10 & Giant selected Hipparcos photometry and extinction \\
      11 & Dwarf/Subgiant based on absolute apparent Hipparcos photometry \\
      12 & Giant based on absolute apparent Hipparcos photometry \\
      13 & Using only apparent {\it Hp} magnitude (imposed dwarf with average color) \\
      14 & FGK dwarf/subgiant selected using Johnson photometry and extinction \\
      15 & Dwarf/Subgiant (not FGK) selected using Johnson photometry with extinction  \\
      16 & Giant selected using Johnson photometry with extinction \\
      17 & Dwarf/Subgiant based on absolute apparent Johnson photometry  \\
      18 & Giant based on absolute apparent Johnson photometry  \\
      19 & Using only apparent {\it V} magnitude (imposed dwarf with average color)  \\
      20 & No input photometry in particular on {\it G}, {\it Hp} or {\it V} band\\
      21 & Distance not defined, we use Gaia DR3 apparent magnitude and color(imposed dwarf) \\
      22 & Distance and color not defined, we use apparent {\it G} magnitude (imposed dwarf) \\
      23 & Distance not defined, we use Hipparcos apparent magnitude and {\it B$_T$-V$_T$} (imposed dwarf) \\
      24 & Distance and color not defined, we use apparent {\it Hp} magnitude (imposed dwarf) \\
      25 & Distance not defined, we use Johnosn apparent {\it V} magnitude and {\it B-V} color(imposed dwarf) \\
      26 & Distance and color not defined, we use apparent {\it V} magnitude (imposed dwarf)  \\
      27 & No distance and input photometry \\   
    \hline
    \end{tabular}
   \label{tab:CaseFlag}
\end{table*}

\end{appendix}

\end{document}